\documentclass[pre,reprint]{revtex4-1}
\usepackage{graphics,graphicx}
\usepackage{amsmath}
\usepackage{amssymb}
\usepackage{subfigure}
\usepackage{dcolumn}
\usepackage{bm}
\usepackage{hyperref}

\begin{document}

\preprint{for submission to PRF}

\title{The Lagrangian Deformation Structure of Three-Dimensional Steady Flow}

\author{Daniel R. Lester}
\affiliation{School of Civil, Environmental and Chemical Engineering, Royal Melbourne Institute of Technology, Melbourne, Victoria 3001, Australia}
\email{daniel.lester@rmit.edu.au}
\author{Marco Dentz}
\affiliation{Spanish National Research Council (IDAEA-CSIC), 08034 Barcelona, Spain}
\author{Tanguy Le Borgne}
\affiliation{Geosciences Rennes, UMR 6118, Universit\'{e} de Rennes 1, CNRS, 35042 Rennes, France}
\author{Felipe P.J. de Barros}
\affiliation{Sonny Astani Department of Civil and Environmental Engineering, University of Southern California, Los Angeles, USA}

\date{\today}

\begin{abstract}
Fluid deformation and strain history are central to wide range of fluid phenomena ranging from mixing and particle transport to stress development in complex fluids and the formation of Lagrangian coherent structures (LCSs). To understand and model these processes it is necessary to quantify Lagrangian deformation in terms of Eulerian flow properties, currently an open problem. Whilst this problem has received much attention in the context of unsteady three-dimensional (3D) turbulent flow, there also exist several important classes of steady 3D flow such as chaotic, non-Newtonian and porous media flows. For steady 3D flows we develop a Protean (streamline) coordinate transform which renders both the velocity gradient and deformation gradient upper triangular. This frame not only simplifies computation of fluid deformation metrics such as finite-time Lyapunov exponents (FTLEs) and elucidates the deformation structure of the flow, but moreover explicitly recovers kinematic and topological constraints upon deformation evolution included those related to steady flow, helicity density and the Poincar\'{e}-Bendixson theorem. We apply this transform to several classes of steady 3D flow, including helical and non-helical, compressible and incompressible flows, and find random flows exhibit remarkably simple (Gaussian) deformation structure. As such this technique provides the basis for the development of stochastic models of fluid deformation in random flows which adhere to the kinematic constraints inherent to various flow classes.
\end{abstract}

\maketitle

\section{Introduction}\label{sec:intro}
Fluid deformation and strain history are central to wide range of fluid-borne phenomena, including fluid mixing and transport phenomena~\cite{Villermaux:2003ab}, identification of Lagrangian coherent structures and transport barriers~\cite{Haller2001248}, prediction of particle clustering or dispersion~\cite{Bec20082037}, alignment of material elements and scalar gradients~\cite{Klein2000246}, understanding and prediction of pair dispersion~\cite{Thalabard:2014aa}, and the development of stress in complex fluids~\cite{Truesdell:1992aa}. In many of these applications, the ability to link Eulerian flow features such as the spatial velocity gradient to Lagrangian evolution of the deformation gradient tensor provides significant insights into the deformation structure of the flow. Whilst this problem is well-studied in the context of turbulent flows~\cite{Girimaji:1990aa,Meneveau:2011aa, Ilyn:2015aa} there also exist several classes of steady 3D flow including chaotic and porous media flow which exhibit complex deformation behaviours. The steady nature of these flows imposes important constraints upon the evolution dynamics of both the velocity and deformation gradient tensors and so simplifies the link between flow structure and deformation. Such insights are of relevance to studies of fluid mixing, scalar dissipation or pair dispersion as they allow statistically quantified flow properties to be linked to evolution of fluid deformation which directly controls the associated fluid-borne phenomena. For example, in the context of fluid mixing, the distribution of Lagrangian fluid stretching rates serves as a quantitative input for lamellar mixing models~\cite{Villermaux:2003ab, Wiggins:2004aa, Barros:2012aa, Le-Borgne:2015aa} based upon evolution and coalescence of concentration inhomogeneities which evolve as interacting lamellae. Similarly, the evolution of material surfaces and interfaces  governs chemical reaction and front propagation. Whilst fluid deformation plays a pivotal role in these problems, it is often difficult to correlate Lagrangian deformation to Eulerian properties of the flow field. This is particularly challenging in the case of highly heterogeneous flow fields where perturbation methods are not appropriate.

Fluid deformation and strain history in non-Newtonian fluids controls the development of material stress and molecular or fibre orientation, and so is central to the constitutive modelling of complex fluids~\cite{Wineman:2009aa}. Many such constitutive models are posed in terms of a memory integral representation for the stress tensor~\cite{Wineman:2009aa}, the solution of which can be very computationally intensive. Several works~\cite{Court:1981aa, Malkus:1981aa, Soulages:2008aa,Feigl:1996aa} have devised Lagrangian numerical methods to alleviate the computational overhead associated with tracking and resolving particle trajectories and calculation of associated strain histories and convolutions thereof.  This computational overhead is significantly reduced when the system is in transformed in streamline coordinates~\cite{Duda:1967aa, Finnigan:1983aa}, resulting in an upper triangular rate of strain tensor. For steady flows tracking of Lagrangian trajectories and free surface flows is greatly simplified as particles are confined to fluid streamlines~\cite{Duda:1967aa}. This approach also simplifies calculation of fluid deformation and strain history, as the deformation gradient tensor admits closed-form solution consisting of a definite integrals evaluated along the streamline~\cite{Adachi:1986aa,Winter:1982aa}. This solution permits much more efficient calculation of both fluid deformation, strain history and associated convolutions.

Reorientation into streamline coordinates is also closely related to QR decomposition methods~\cite{Dieci:2008aa, Dieci:1997aa} for accurate computation of Lyapunov exponents in continuous dynamical systems (posed as linear first order ODEs). This approach is based upon continuous decomposition of the fundamental solution matrix into orthogonal $\mathcal{Q}$ and upper triangular $\mathcal{R}$ components which satisfy individual auxiliary evolution equations (ODEs) and possess beneficial qualities with respect to numerical approximation of Lyapunov spectra and adherence to constraints such as regularity and preservation of phase space. QR decomposition of 2 degree-of-freedom (d.o.f) autonomous linear first-order continuous dynamical systems is directly analogous to that of 2D fluid deformation in streamline coordinates, whereby the orthogonal matrix $\mathcal{Q}$ represents a rotation into the ``streamline'' coordinates of the dynamical system, and the upper triangular matrix $\mathcal{R}$ is given by closed-form integrals of the reoriented systems along a ``streamline''. Previously, QR decomposition methods have been considered as primarily numerical techniques for efficient and accurate computation, rather than to elucidate the governing dynamics and constraints of the dynamical system. 
 
As such, reorientation of deformation evolution in 2D flows into streamline coordinates not only provides benefits with respect to numerical analysis, but also basic physical properties of the flow field (e.g. incompressibility, mass conservation) are naturally preserved by this formulation. Moreover, additional topological constraints associated with steady flow of continua (or autonomous continuous dynamical systems in general) are naturally recovered without need for ad-hoc enforcement. These constraints are particularly relevant as they directly limit evolution of the deformation gradient tensor.

One such constraint is imposed by the Poincar\'{e}-Bendixson theorem~\cite{Teschl:2012aa} which arises due to the fact that 1D streamlines cannot ``cross'' each other in a steady 2D flow. A direct consequence of this topological constraint is that fluid deformation in steady 2D flows on topologically simple manifolds can only grow at most algebraically in time. This constraint has important implications for the construction of models of fluid deformation in such flows, as naive implementations inherently generate exponential fluid stretching due to the homogeneous linear form of the temporal deformation evolution equation (\ref{eqn:deform}). Although sub-exponential stretching can be imposed via \emph{ad-hoc} means such as the imposition of exponential waiting times between stretching events~\cite{Villermaux:2003ab}, such dynamics are not supported by experimental observations. To properly capture the underlying deformation dynamics, it is necessary that the modelling framework naturally recovers constraints on the deformation dynamics. Such constraints both simplify the modelling process and more clearly elucidate the role of flow structure in controlling fluid deformation.

Whilst the Poincar\'{e}-Bendixson theorem applies to 2D steady flow fields (or autonomous continuous dynamical systems on manifolds of zero topological genus in general), it also has significant implications for deformation in some steady 3D flows. As shown in \cite{Sposito:2001aa,Sposito:1997aa}, flows with zero helicity density also must exhibit sub-exponential stretching as a consequence. The local helicity density $h(\mathbf{x})$ introduced by Moffatt~\cite{Moffatt:1969aa} is the product of the fluid velocity $\mathbf{v}(\mathbf{x})$ and vorticity $\boldsymbol\omega(\mathbf{x})$
\begin{equation}
h(\mathbf{x}):=\mathbf{v}(\mathbf{x})\cdot\boldsymbol\omega(\mathbf{x}),\label{eqn:helicity}
\end{equation}
and the total helicity $H$ given by the volume integral $H:=\int_{\mathcal{D}}h(\mathbf{x})d^n\mathbf{x}$ over the closed flow domain $\mathcal{D}$ is a measure of topological complexity of the flow field, related to the total knottedness of vortex lines within $\mathcal{D}$. There exist a wide class of zero helicity density flows, including all 2D flows; e.g. those defined by a Stokes streamfunction $\psi$, $\mathbf{v}=\nabla\times \psi \hat{\mathbf{z}}$, irrotational 2D and 3D flows $\mathbf{v}=\nabla\phi$ in terms of the velocity potential $\phi$, isotropic heterogeneous Darcy flow $\mathbf{v}=k\nabla\phi$ (where $k$ is the heterogeneous hydraulic conductivity), and toroidal flows $\mathbf{v}=\nabla\times\nabla\times \varphi \hat{\mathbf{z}}$ where $\varphi$ is the toroidal potential such as the Arter flow~\cite{Holm:1991aa}. Note that in general tensorial Darcy flows have non-zero helicity density \cite{Ye:2015aa}.

Spositio~\cite{Sposito:2001aa} shows that fluid streamlines in zero helicity density flows are confined to topologically simple 2D surfaces to which the Poincar\'{e}-Bendixson locally applies, enforcing sub-exponential fluid deformation under steady flow conditions. As per 2D steady flow, an appropriate modelling framework must enforce this constraint to avoid spurious non-physical behaviour. The use of streamline coordinates for steady 3D flows is analogous to the QR method for 3 d.o.f. systems, however in 3D the QR method does not necessarily conform to streamline coordinates and so does not recover these topological constraints. Such constraints not only have significant implications for models of fluid deformation in random flows, but also the fluid mechanics of complex fluids, as the rate of fluid stretching is integral to many constitutive models.

As outlined above, transformation into streamline coordinates in steady 2D flow greatly simplifies the evolution equations for the deformation tensor and recovers constraints upon the deformation dynamics. In this paper we extend this concept to steady 3D flows and uncover the deformation structure of several classes of such flows. Whilst streamline coordinates are unique in steady 2D flows, such a frame is not unique in steady 3D flows due arbitrary rotation of frame about a streamline, and do not necessarily render the transformed velocity gradient tensor upper triangular. In this study we develop a transform which aligns with 3D streamline coordinates and also renders the velocity gradient tensor upper triangular. Following Adachi~\cite{Adachi:1983aa} we term a system with these properties as a \emph{Protean} coordinate system, which inherits the advantages of both the QR decomposition and streamline coordinate systems with respect to both numerical computation and physical insights into the dynamics which govern fluid deformation in 2D and 3D steady flows.

We present an alternative derivation to previous studies~\cite{Adachi:1983aa,Adachi:1986aa,Clermont:1993aa, Winter:1982aa} for the streamline coordinate transform into 2D flow, and then extend this to steady 3D flows to generate integral solutions of the full strain tensor. Whilst it was previously considered~\cite{Clermont:1993aa} that such analysis was too complex to be undertaken in 3D flow, we show this is possible for all steady 3D flow as the QR decomposition may be applied to all real square matrices.

Due to the simplicity of deformation evolution equation in the Protean frame, this transformation elucidates the link between Eulerian flow features and fluid deformation along Lagrangian trajectories and moreover, naturally enforces physical constraints upon deformation evolution. This structure naturally recovers the inherent kinematic and topological constraints associated with the Poincar\'{e}-Bendixson theorem and zero helicity flow, and so provides a framework for the development of a kinematically-consistent Continuous Time Random Walk (CTRW) model~\cite{Dentz:2015aa} of deformation in steady 3D flow. By way of example we apply this approach to a series of model flows, both compressible and incompressible, helical and non-helical and assess the numerical accuracy of the method and implications for understanding fluid deformation structure in complex 3D flows.

The remainder of the paper is organised as follows; in the following Section we briefly review the evolution of finite strain tensors in continuous media, and in Section~\ref{sec:transform} we consider objective transformation of the deformation gradient tensor into the Protean frame. We apply this transformation in Section~\ref{sec:2D} to derive the core results for steady 2D flow via a more transparent route than previous studies, and then extend application to 3D flow in Section~\ref{sec:3D}. This method is then applied to the example flow classes in Section~\ref{sec:examples} and concluding remarks are provided Section~\ref{sec:conclusions}.

\section{Evolution of Finite Strain Tensors}\label{sec:strain_history}

Fluid deformation and strain history are typically couched in terms of finite-strain tensors such as the right Cauchy-Green $\mathbf{C}=\mathbf{F}^T\mathbf{F}$, left Cauchy-Green (Finger) $\mathbf{B}=\mathbf{F}\mathbf{F}^T$, Green-Lagrangian $\mathbf{E}=\frac{1}{2}(\mathbf{C}-\mathbf{1})$ and Hencky $\mathbf{H}=\frac{1}{2}\ln \mathbf{C}$ strain tensors \cite{Truesdell:1992aa}, all of which are derived directly from the deformation gradient tensor $\mathbf{F}$, defined as
\begin{equation}
\mathbf{F}(t):=\frac{\partial \mathbf{x}}{\partial \mathbf{X}},\,\,\,\,F_{ij}(t):=\frac{\partial x_i}{\partial X_j} \label{eqn:Fdefn}
\end{equation}
where $\mathbf{x}$ are the $\mathbf{X}$ respectively are reference material vectors in the Eulerian and Lagrangian frames. As such the deformation gradient tensor $\mathbf{F}$ quantifies how the infinitesimal vector $d\mathbf{x}$ deforms from its reference state $d\mathbf{X}$ as $d\mathbf{x}=\mathbf{F}(t)\cdot d\mathbf{X}$. The evolution of an infinitesimal line element $d\mathbf{l}(t)$ then evolves as
\begin{equation}
d\mathbf{l}(t)=\mathbf{F}(t)\cdot d\mathbf{l}(0),
\end{equation}
and similarly the the infinitesimal areal element $d\mathbf{A}(t)$ spanned by the line elements $d\mathbf{X}_1(t)$, $d\mathbf{X}_2(t)$ evolves as
\begin{equation}
d\mathbf{A}(t)=\det[\mathbf{F}(t)](\mathbf{F}^{-1}(t))^T\cdot d\mathbf{A}(0),
\end{equation}
an the fluid volume $V(t)$ evolves as
\begin{equation}
V(t)=\det[\mathbf{F}(t)]V(0).
\end{equation}
Following the definition (\ref{eqn:Fdefn}), the deformation gradient tensor $\mathbf{F}(t)$ evolves with travel time $t$ along a Lagrangian trajectory as
\begin{equation}
\frac{d\mathbf{F}}{dt}=\boldsymbol\epsilon(t)\cdot\mathbf{F}(t),\quad\mathbf{F}(0)=\mathbf{1},\label{eqn:deform}
\end{equation}
where the velocity gradient:= tensor $\boldsymbol\epsilon(t):=\nabla\mathbf{v}(\mathbf{x}(t))^T$, and the operator $d/dt$ denotes differentiation along the Lagrangian trajectory $\mathbf{x}(t)$.

The finite-strain tensors $\mathbf{C}$, $\mathbf{B}$, $\mathbf{E}$, $\mathbf{H}$, are all are objective (frame-indifferent) and hence provide appropriate strain measures for constitutive modelling~\cite{Truesdell:1992aa} of non-Newtonian fluids. The stress-strain relationships for viscoelastic materials are often encoded via memory-integral constitutive models of the form~\cite{Wineman:2009aa}
\begin{equation}
\boldsymbol\sigma(t)=\mathcal{F}[\mathbf{F}(t-s)^\infty_{s=0}],\label{eqn:strainhistory}
\end{equation}
where $\boldsymbol\sigma(t)$ is the fluid stress, $s$ denotes historical time and $\mathcal{F}$ represents a tensorial-valued functional which is dependant upon the entire strain history of the material. Explicitly, many memory integral equations are based upon convolution of the strain history with tensorial kernels $\mathbf{K}[\mathbf{C}(t),s]$ which decay monotonically with increasing history $s$, such as is given the Pipkin-Rogers (or similar) constitutive theory
\begin{equation}
\boldsymbol\sigma(t)=\mathbf{F}(t)\cdot\left\{\mathbf{K}[\mathbf{C}(t),0]+\int_0^t \frac{\partial \mathbf{K}[\mathbf{C}(s),t-s]}{\partial(t-s)} ds \right\}\cdot\mathbf{F}(t)^T.\label{eqn:PipkinRogers}
\end{equation}
In general, calculation of the stress evolution and general fluid mechanics of multidimensional viscoelastic flows via integral equations such as (\ref{eqn:PipkinRogers}) can be very computationally expensive. Such computations may be significantly simplified via transformation of the deformation gradient $\mathbf{F}(t)$ into the Protean coordinate frame~\cite{Winter:1982aa}.

\section{Transformation of the Deformation Tensor into the Protean Frame}\label{sec:transform}

Whilst a formal description of fluid deformation in an arbitrary orthogonal curvilinear coordinate frame requires an excursion into differential geometry and tensor calculus, to clarify exposition and physical interpretation we consider transform of an orthogonal Cartesian coordinate system, and note that the formulation herein may be extended to general curvilinear coordinates via the standard tools of differential geometry. We denote spatial coordinates in the Cartesian coordinate system as $\mathbf{x}=\{x_1, x_2, x_3\}$, and the reoriented Protean (streamline) coordinate system by $\mathbf{x}^\prime=\{x_1^\prime, x_2^\prime, x_3^\prime\}$, such that the arbitrary velocity vector $\mathbf{v}(\mathbf{x})=\{v_1,v_2,v_3\}$ in the Cartesian frame transforms to $\mathbf{v}^\prime(\mathbf{x}^\prime)=\{v,0,0\}$ in the Protean frame, where $v=|\mathbf{v}|$. These two frames are related by the local objective transform~\cite{Truesdell:1992aa}
\begin{equation}
\mathbf{x}^\prime=\mathbf{x}_0(t)+\mathbf{Q}^T(t)\cdot\mathbf{x},\label{eqn:objective_xform}
\end{equation}
where $\mathbf{x}_0(t)$ is an arbitrary translation vector, and $\mathbf{Q}(t)$ is a proper orthogonal transformation such as a rotation, hence $\mathbf{Q}^T(t)\cdot\mathbf{Q}(t)=\mathbf{1}$ and $\det[\mathbf{Q}(t)]=1$. The differential element $d\mathbf{x}$ transforms as
\begin{align}
d\mathbf{x}^\prime = \mathbf{Q}^T(t)\cdot d\mathbf{x},\label{eqn:dxtransform}
\end{align}
and from (\ref{eqn:Fdefn}) the deformation tensor then transforms as
\begin{align}
\mathbf{F}^\prime(t)=\mathbf{Q}^T(t)\cdot\mathbf{F}(t)\cdot\mathbf{Q}(0).\label{eqn:Ftransform}
\end{align}
Whilst $\mathbf{F}(t)$ is often referred to as the deformation gradient \emph{tensor}, it is not objective (as per (\ref{eqn:Ftransform}), $\mathbf{F}^\prime(t)\neq\mathbf{P}^T(t)\cdot\mathbf{F}(t)\cdot\mathbf{P}(t)$ for some orthogonal matrix $\mathbf{P}(t)$), and so is not a tensor in the formal sense \cite{Truesdell:1992aa}, however the finite strain tensors $\mathbf{C}$, $\mathbf{B}$, $\mathbf{E}$, $\mathbf{H}$ are objective, and form a suitable basis for constitutive modelling. Differentiating (\ref{eqn:Ftransform}) with respect to the Lagrangian travel time $t$ yields
\begin{equation}
\begin{split}
\frac{d\mathbf{F}^\prime}{dt}=
%&\dot{\mathbf{Q}}^T(t)\cdot\mathbf{F}(t)\cdot\mathbf{Q}(0)+\mathbf{Q}^T(t)\cdot\boldsymbol\epsilon(t)\cdot\mathbf{F}(t)\cdot\mathbf{Q}(0),\\
=&\boldsymbol\epsilon^\prime(t)\cdot\mathbf{F}^\prime(t),\label{eqn:rotF}
\end{split}
\end{equation}
where the transformed rate of strain tensor $\boldsymbol\epsilon^\prime(t)$ is then
\begin{equation}
\begin{split}
\boldsymbol\epsilon^\prime(t)=&\mathbf{Q}^T(t)\cdot\boldsymbol\epsilon(t)\cdot\mathbf{Q}(t)+\mathbf{A}(t),\\
:=&\tilde{\boldsymbol\epsilon}(t)+\mathbf{A}(t),\label{eqn:epsilon}
\end{split}
\end{equation}
and the contribution due to a moving coordinate frame is $\mathbf{A}(t):=\dot{\mathbf{Q}}^T(t)\cdot\mathbf{Q}(t)$. As $\mathbf{Q}(t)$ is orthogonal, then $\dot{\mathbf{Q}}^T(t)\cdot\mathbf{Q}(t)+\mathbf{Q}^T(t)\cdot\dot{\mathbf{Q}}(t)=0$ and so $\mathbf{A}(t)$ is skew-symmetric. From (\ref{eqn:dxtransform}), the velocity vector $\mathbf{v}(t)=d\mathbf{x}/dt$ also transforms as
\begin{equation}
\mathbf{v}^\prime(t)=\mathbf{Q}^T(t)\cdot\mathbf{v}(t).
\end{equation}
The basic idea regarding use of Protean coordinates is to find an appropriate local reorientation $\mathbf{Q}(t)$ to render the transformed velocity gradient $\boldsymbol\epsilon^\prime(t)$ upper triangular, yielding explicit solution of (\ref{eqn:rotF}).

\section{2D Deformation Gradient Tensor in Protean Coordinates}\label{sec:2D}

In two spatial dimensions  the reorientation matrix $\mathbf{Q}(t)$ is then
\begin{equation}
\mathbf{Q}(t)=\frac{1}{v}
\left(
\begin{array}{cc}
v_1 & -v_2 \\
v_2 & v_1 \\
\end{array}
\right),\label{eqn:Qrotn}
\end{equation}
%\begin{equation}
%\mathbf{e}_1^\prime=\frac{1}{v}
%\left(\begin{array}{c}
%v_1\\
%v_2\\
%\end{array}\right),\,\,
%\mathbf{e}_2^\prime=\frac{1}{v}
%\left(\begin{array}{c}
%-v_2\\
%v_1\\
%\end{array}\right),
%\end{equation}
%so $\mathbf{Q}(t)=\{\mathbf{e}_1^\prime,\mathbf{e}_2^\prime\}$
and the basis vectors $\mathbf{e}_1^\prime, \mathbf{e}_2^\prime$ in the Protean frame are explicitly $\mathbf{Q}(t)=\{\mathbf{e}_1^\prime,\mathbf{e}_2^\prime\}^T$. The moving coordinate frame contribution $\mathbf{A}(t)$ may be expressed as
\begin{equation}
\mathbf{A}(t)=
\left(\begin{array}{cc}
0 & \mathbf{e}_2^\prime\cdot\dot{\mathbf{e}}_1^\prime \\
-\mathbf{e}_2^\prime\cdot\dot{\mathbf{e}}_1^\prime & 0 \\
\end{array}\right).\label{eqn:Qtrans}
\end{equation}
Note that $\mathbf{e}_1^\prime=\mathbf{v}/v$, and since
\begin{equation}
\dot{\mathbf{v}}(t)=\boldsymbol\epsilon(t)\cdot\mathbf{v}(t),\label{eqn:velevolve}
\end{equation}
 for the travel time $t$ along a streamline, then the basis vector $\mathbf{e}_1^\prime$ evolves with Lagrangian time $t$ as
\begin{equation}
\dot{\mathbf{e}}_1^\prime=\frac{d\ln v}{dt}\mathbf{e}_1^\prime+\boldsymbol\epsilon(t)\cdot\mathbf{e}_1^\prime,
\end{equation}
and as the basis vectors $\mathbf{e}_1^\prime$, $\mathbf{e}_2^\prime$ are orthogonal
\begin{equation}
\mathbf{e}_2^\prime\cdot\dot{\mathbf{e}}_1^\prime=\mathbf{e}_2^\prime\cdot\boldsymbol\epsilon(t)\cdot\dot{\mathbf{e}}_1^\prime.
\end{equation}
Since $\mathbf{Q}(t)=\{\mathbf{e}^\prime_1, \mathbf{e}^\prime_2\}^T$, the elements of $\tilde{\boldsymbol\epsilon}(t)$ in (\ref{eqn:epsilon}) are then
\begin{equation}
\tilde{\epsilon}_{ij}=[\mathbf{Q}^T(t)\cdot\boldsymbol\epsilon(t)\cdot\mathbf{Q}(t)]_{ij}=\mathbf{e}_i^\prime\cdot\boldsymbol\epsilon\cdot\mathbf{e}_j^\prime,
\end{equation}
and so
\begin{equation}
\mathbf{e}_2^\prime\cdot\dot{\mathbf{e}}_1^\prime=\tilde{\epsilon}_{21}.
\end{equation}
From (\ref{eqn:epsilon}), (\ref{eqn:Qtrans}) the reoriented rate of strain tensor $\boldsymbol\epsilon^\prime(t)$ is then upper triangular:
\begin{equation}
\boldsymbol\epsilon^\prime(t)=
\left(\begin{array}{cc}
\tilde{\epsilon}_{11} & \tilde{\epsilon}_{12}+\tilde{\epsilon}_{21} \\
0 & \tilde{\epsilon}_{22} \\
\end{array}\right).\label{eqn:F2D}
\end{equation}
Hence in two dimensional steady flow reorientation into Protean
coordinates automatically renders the transformed velocity gradient
tensor $\boldsymbol\epsilon^\prime(t)$ upper triangular. Note that
this method is analogous to the continuous QR decomposition method for
a $d$--dimensional autonomous linear system, as outlined in
Appendix~\ref{App:QR}. 

Due to the upper triangular form of $\boldsymbol\epsilon^\prime(t)$,  solution of the evolution equation (\ref{eqn:rotF}) for the Protean deformation gradient tensor $\mathbf{F}^\prime(t)$  is particularly simple (via Gaussian elimination and the initial condition $\mathbf{F}^\prime(0)=\mathbf{Q}(0)^T\cdot\mathbf{F}(0)\cdot\mathbf{Q}(0)=\mathbf{1}$)
\begin{align}
&F_{21}^\prime(t)=0,\\
&F_{11}^\prime(t)=\frac{v(t)}{v(0)},\\
&F_{22}^\prime(t)=\exp\left(\int_0^t dt^\prime \epsilon^\prime_{22}(t^\prime)\right),\\
&F_{12}^\prime(t)=v(t)\int_0^t dt^\prime \frac{\epsilon^\prime_{12}(t^\prime)F_{22}^\prime(t^\prime)}{v(t^\prime)}.
\end{align}
For incompressible flow $\epsilon^\prime_{22}=-\epsilon^\prime_{11}$ these expressions simplify to 
\begin{align}
&F_{21}^\prime(t)=0,\\
&F_{11}^\prime(t)=\frac{v(t)}{v(0)},\\
&F_{22}^\prime(t)=\frac{v(0)}{v(t)},\\
&F_{12}^\prime(t)=v(t)v(0)\int_0^t dt^\prime \frac{\epsilon^\prime_{12}(t^\prime)}{v(t^\prime)^2}.\label{eqn:F12_2D}
\end{align}
This result has previously been obtained by several workers \cite{Winter:1982aa, Adachi:1983aa,Adachi:1986aa, Clermont:1993aa} to solve $\mathbf{F}^\prime(t)$. As the reorientation matrix $\mathbf{Q}(t)$ is known along a streamline, the Cartesian deformation tensor can then be directly calculated as $\mathbf{F}(t)=\mathbf{Q}(t)\cdot\mathbf{F}^\prime(t)\cdot\mathbf{Q}^T(0)$.

The use of streamline coordinates in 2D steady continuous flow naturally imposes the topological constraints associated with these flows, namely that streamlines cannot cross on smooth 2D manifolds. In turn this constraint limits the rate of strain tensor along as streamline (\ref{eqn:velevolve}), as formalised by the Poincar\'{e}-Bendixson theorem. One implication of this theorem is that fluid stretching in steady 2D flow both parallel and transverse to the flow direction (quantified respectively by $\epsilon^\prime_{11}(t)$, $\epsilon^\prime_{22}(t)$) must converge toward zero mean over long times for both incompressible and compressible flow. As such, the deformation components $F_{11}^\prime(t)$, $F_{22}^\prime(t)$ cannot grow or decay without bound over long times. This is reflected by the explicit solution $F_{11}^\prime(t)=v(t)/v(0)$, where fluid stretching along the streamline coordinate is governed by fluctuations in the advective velocity $v(t)$. Similarly transverse stretching in incompressible flow is also bound as $F_{22}^\prime(t)=v(0)/v(t)$, and for compressible flow transverse stretching is limited by volumetric compression which must be finite due to mass conservation.

Therefore persistent fluid deformation in steady 2D flow can only arise via the off-diagonal term $\epsilon^\prime_{12}(t)$, and the solution for $F_{12}^\prime(t)$ in (\ref{eqn:F12_2D}) algebraic growth for $F_{12}^\prime(t)$ as per the Poincar\'{e}-Bendixson theorem. From (\ref{eqn:F2D}), the off-diagonal term $\epsilon^\prime_{12}(t)$=$\tilde{\epsilon}_{12}+\tilde{\epsilon}_{21}$
consists of contributions from shear deformation $\tilde{\epsilon}_{12}$ between streamlines and curvature of a streamline $\tilde{\epsilon}_{21}$ in the Protean frame. For open streamlines (and in particular flows such as Darcy and potential flows which do not admit closed streamlines~\cite{Bear:1972aa}), the average of the curvature $\tilde{\epsilon}_{21}$ must average to zero over long times, whereas for closed streamlines, periodicity of the integral (\ref{eqn:F12_2D}) yields fluid stretching which is linear in time. For flows which are ergodic in the Lagrangian sense (i.e. either mixing flows which admit ergodic particle trajectories or random flow fields), the long time average of the shear contribution $\tilde{\epsilon}_{12}$ must also have zero mean due to stationarity.
%MARCO CHANGED THE FOLLOWING: DAN Agreed
% Hence from  (\ref{eqn:F12_2D}) persistent fluid
% stretching in such flows may only arise if the net correlation between
% these contributions and $v^{\prime-2}$ is non-zero.

As the elongation $\rho(t)$ of a material line $\mathbf z(t)$ is given
by 
\begin{equation}
\begin{split}
\rho(t) &= \sqrt{\mathbf z(0) \mathbf {F^\prime}^T(t) \mathbf {F^\prime}(t) \mathbf
  z(0)} \\
&= \sqrt{z_2(0)^2 F^\prime_{22}(t)^2+[z_2(0) F^\prime_{12}(t) + z_{1}(0)
  F^\prime_{11}(t)]^2},\\
  &\approx |z_2{0}F_{12}^\prime(t)|
\end{split}
\end{equation}
then persistent fluid stretching in such flows arises from the fact that the elongation $\rho(t)$ is governed by the absolute value of $F_{12}^\prime(t)$. In fact, episodes of low streamwise velocity $v(t)$ add up to a strong increase of elongation as evidenced by the presence of $v(t^\prime)^2$ in the denominator of the integrand
in~\eqref{eqn:F12_2D}. The detailed link between transverse shear, velocity fluctuations and fluid deformation
is studied by Dentz et al~\cite{Dentz:2015ab,Dentz:2015aa}, who show via a stretching continuous time random walk (CTRW) that the rate of fluid deformation in stationary random 2D steady flows can be modelled as a L\'{e}vy walk directly from these correlations. Hence the simple structure of the deformation gradient tensor in Protean coordinates clearly identifies the flow properties and constraints which govern fluid deformation.

%This simplicity offered by the Protean frame also facilitates development of highly efficient numerical procedures to solve the convolution of the strain history with memory kernel in integral constitutive models (such as (\ref{eqn:PipkinRogers})), as there exist simple closed form expressions for the evolution of the deformation gradient tensor over time. This is most clearly exemplified for the case of 1D linear viscoelasticity, where the stress $\sigma(t)$ is given by a convolution over the shear modulus $G(t)$ and the strain history $\varepsilon(t)$ (an invariant of $\mathbf{C}$(t)) as
%\begin{equation}
%\sigma(t)=\varepsilon(0)G(t)+\int_0^t G(t-s)\frac{d\varepsilon}{ds}ds,\label{eqn:linearvis} 
%\end{equation}
%where the differential $d\varepsilon/dt$ can be expressed explicitly in terms of the components of $\boldsymbol\epsilon(t)^\prime$, leading to efficient methods to solve (\ref{eqn:linearvis}).

\section{3D Deformation Gradient Tensor in Protean Coordinates}\label{sec:3D}

\subsection{Coordinate Reorientation in 3D Steady Flow}

In contrast to steady 2D flow, the additional degree of freedom associated with steady 3D flow admits the possibility of chaotic Lagrangian dynamics. Specifically,  steady flows with non-zero helicity density $h$ may exhibit chaotic dynamics and exponential fluid stretching due to relaxation of this topological constraint~\cite{Moffatt:1969aa,Sposito:2001aa}. In the Protean coordinate system, the base vector $\mathbf{e}_1^\prime$ aligns with the velocity vector $\mathbf{v}$, such that $\mathbf{v}^\prime=\{v,0,0\}$, however the transverse vectors $\mathbf{e}_2^\prime$, $\mathbf{e}_3^\prime$ are arbitrary up to a rotation about $\mathbf{e}_1^\prime$. As these base vectors are not necessary material coordinates, this gauge freedom does not impact the governing dynamics.

As per Section~\ref{sec:2D}, it is desirable to reorient the rate of strain tensor $\boldsymbol\epsilon^\prime(t)$ such that it is upper triangular, yielding explicit closed-form solution for the deformation gradient tensor $\mathbf{F}^\prime(t)$ and elucidating the deformation dynamics. Whilst all square matrices are unitarily similar to an upper triangular matrix, it is unclear in general whether such a similarity transform is orthogonal, or corresponds to a reorientation of frame, or furthermore is a reorientation into streamline coordinates. This condition is satisfied for all steady 2D flows, but this is an open question for steady 3D flows. %We shall explore this question throughout this Section. %However, there exist a subset of 3D steady flows for which $\boldsymbol\epsilon$ is rendered upper triangular via reorientation into streamline coordinates, as discussed below.

Following the procedure for 2D flows, we begin by considering the 3D rotation matrix $\mathbf{Q}_1(t)$ which reorients the Eulerian coordinate $\mathbf{e}_1$ to the corresponding Protean coordinate $\mathbf{e}^\prime_1$ which is tangent to the velocity vector $\mathbf{v}$. This reorientation is defined in terms of the rotation axis $\mathbf{q}$ and angle $\theta$ which are given in terms of the local velocity vector $\mathbf{v}$ and $\mathbf{e}_1$ as
\begin{align}
&\mathbf{q}=\frac{\mathbf{e}_1\times\mathbf{v}}{||\mathbf{e}_1\times\mathbf{v}||}=\frac{1}{\sqrt{v_2^2+v_3^2}}\{0,v_3,-v_2\},\\
&\cos\theta=\frac{\mathbf{e}_1\cdot\mathbf{v}}{||\mathbf{e}_1\cdot\mathbf{v}||}=\frac{v_1}{v},
\end{align}
and so
\begin{equation}
\begin{split}
\mathbf{Q}_1(t)
&=\cos\theta\mathbf{I}+\sin\theta (\mathbf{q})_{\times}^{T}+(1-\cos\theta)\mathbf{q}\otimes\mathbf{q},
%\\
%&=\frac{1}{v}
%\left(
%  \begin{array}{ccc}
 %   v_1 & -v_2 & -v_3 \\
 %   v_2 & a_1 & a_3 \\
 %   v_3 & a_3 & a_2 \\
%  \end{array}
%\right),
\label{eqn:A1}
\end{split}
\end{equation}
where $(\mathbf{q})_{\times}$ denotes the cross product matrix of $\mathbf{q}$. % and $a_1$, $a_2$, $a_3$ are
%\begin{align}
%a_1&=\frac{v_1 v_2^2+ v v_3^2}{\sqrt{v_2^2+v_3^2}},\\
%a_2&=\frac{v_1 v_3^2+ v v_2^2}{\sqrt{v_2^2+v_3^2}},\\
%a_3&=v_2 v_3\frac{v_1-v}{\sqrt{v_2^2+v_3^2}}.
%\end{align}
Whilst reorientation via $\mathbf{Q}_1(t)$ ensures the 1-coordinate in the Protean frame is always tangent to the velocity vector $\mathbf{v}$ along a streamline, there exists a degree of freedom regarding orientation of the $2-$, $3-$coordinates in 3D flows. As such, we consider a further reorientation by an arbitrary angle $\alpha$ about $\mathbf{e}_1^\prime$, such that the effective reorientation may be expressed as the composite 
\begin{equation}
\mathbf{Q}(t):=\mathbf{Q}_1(t)\cdot\mathbf{Q}_2(t),\label{eqn:composite_rotn}
\end{equation}
and $\mathbf{Q}_2(t)$ is 
\begin{equation}
\mathbf{Q}_2(t)=
\left(
  \begin{array}{ccc}
    1 & 0 & 0 \\
    0 & \cos\alpha & -\sin\alpha \\
    0 & \sin\alpha & \cos\alpha \\
  \end{array}
\right).\label{eqn:A2}
\end{equation}
%Physical space in this reoriented and moving Protean frame is again denoted by $\mathbf{x}^\prime$, and the transform between frames is given by the transform (\ref{eqn:objective_xform}). Likewise 
The basis vectors in this reoriented frame are then
\begin{equation}
\mathbf{Q}(t)=\{\mathbf{e}_1^\prime,\mathbf{e}_2^\prime,\mathbf{e}_3^\prime\}^T,\label{eqn:Ae1e2e3}
\end{equation}
and so the 3D velocity gradient $\boldsymbol\epsilon(t)$ transforms as
\begin{equation}
\begin{split}
\boldsymbol\epsilon^\prime(t)
:=&\mathbf{Q}^{T}(t)\cdot\boldsymbol\epsilon(t)\cdot\mathbf{Q}(t)+\mathbf{A}(t),\\
=&\tilde{\boldsymbol\epsilon}(t)+\mathbf{A}(t),\\
%=&\mathbf{Q}_2^{T}(t)\cdot\mathbf{Q}_1^{T}(t)\cdot\boldsymbol\epsilon(t)\cdot\mathbf{Q}_1(t)\cdot\mathbf{Q}_2(t)+\mathbf{A}(t),\\
=&\mathbf{Q}_2^{T}(t)\cdot\boldsymbol\epsilon^{(1)}(t)\cdot\mathbf{Q}_2(t)+\mathbf{A}(t).\label{eqn:epsilondefs}
\end{split}
\end{equation}
%where $\boldsymbol\epsilon^{(1)}(t):=\mathbf{Q}_1^{T}(t)\cdot\boldsymbol\epsilon(t)\cdot\mathbf{Q}_1(t)$.
Following (\ref{eqn:Ae1e2e3}), the moving frame contribution $\mathbf{A}(t)$ to $\boldsymbol\epsilon^\prime(t)$ is then
\begin{equation}
\begin{split}
\mathbf{A}(t)=\dot{\mathbf{Q}}^{T}(t)\cdot\mathbf{Q}(t)=\left(
   \begin{array}{ccc}
     0 & \mathbf{e}_2^\prime\cdot\dot{\mathbf{e}}_1^\prime & \mathbf{e}_3^\prime\cdot\dot{\mathbf{e}}_1^\prime \\
     -\mathbf{e}_2^\prime\cdot\dot{\mathbf{e}}_1^\prime & 0 & \mathbf{e}_3^\prime\cdot\dot{\mathbf{e}}_2^\prime \\
     -\mathbf{e}_3^\prime\cdot\dot{\mathbf{e}}_1^\prime & -\mathbf{e}_3^\prime\cdot\dot{\mathbf{e}}_2^\prime & 0 \\
   \end{array}
 \right),
\end{split}
\end{equation}
and these terms may be related to $\tilde{\boldsymbol\epsilon}(t)$ via the relations $\dot{\mathbf{v}}=\boldsymbol\epsilon(t)\cdot\mathbf{v}$, $\mathbf{e}_1^\prime=\mathbf{v}/v$, $\tilde\epsilon_{ij}=\mathbf{e}^\prime_i\cdot\boldsymbol\epsilon(t)\cdot\mathbf{e}^\prime_j$, yielding
\begin{align}
&\mathbf{e}_2^\prime\cdot\dot{\mathbf{e}}_1^\prime=\mathbf{e}_2^\prime\cdot\boldsymbol\epsilon(t)\cdot\mathbf{e}_1^\prime=\tilde{\epsilon}_{21},\label{eqn:e21}\\
&\mathbf{e}_3^\prime\cdot\dot{\mathbf{e}}_1^\prime=\mathbf{e}_3^\prime\cdot\boldsymbol\epsilon(t)\cdot\mathbf{e}_1^\prime=\tilde{\epsilon}_{31},\label{eqn:e31}\\
&\mathbf{e}_3^\prime\cdot\dot{\mathbf{e}}_2^\prime=v\mathbf{e}_3^\prime\cdot\frac{\partial\mathbf{e}_2^\prime}{\partial\mathbf{v}}\cdot\boldsymbol\epsilon(t)\cdot\mathbf{e}_1^\prime+\mathbf{e}^\prime_3\cdot\frac{\partial\mathbf{e}^\prime_2}{\partial\alpha}\frac{d\alpha}{dt},
%&\mathbf{e}_2^\prime\cdot\dot{\mathbf{e}}_3^\prime=-v\mathbf{e}_2^\prime\cdot\frac{d\mathbf{e}_3^\prime}{dv}\boldsymbol\epsilon\mathbf{e}_1^\prime,
\end{align}
where from (\ref{eqn:A1}), (\ref{eqn:A2})
\begin{equation}
\begin{split}
v\mathbf{e}_3^\prime\cdot&\frac{\partial\mathbf{e}_2^\prime}{\partial\mathbf{v}}
=\frac{1}{v+v_1}\{0,v_3,-v_2\},\\
=&\frac{v_3\cos\alpha-v_2\sin\alpha}{v+v_1}\mathbf{e}_2^\prime-\frac{v_2\cos\alpha+v_3\sin\alpha}{v+v_1}\mathbf{e}_3^\prime,
\end{split}
\end{equation}
and
\begin{align}
\mathbf{e}_3^\prime\cdot\frac{\partial\mathbf{e}_2^\prime}{\partial\alpha}=-1.
%v\mathbf{e}_2^\prime\cdot\frac{d\mathbf{e}_3^\prime}{dv}=\frac{1}{v+v_1}\{0,-v_3,v_2\}=\frac{-v_3\cos\alpha+v_2\sin\alpha}{v+v_1}\mathbf{e}_2^\prime+\frac{v_2\cos\alpha+v_3\sin\alpha}{v+v_1}\mathbf{e}_3^\prime,
\end{align}
As such, the $(2,3)$ component of $\mathbf{A}(t)$ is
\begin{equation}
\begin{split}
&A_{23}=\mathbf{e}_3^\prime\cdot\dot{\mathbf{e}}_2^\prime,\\
&=\frac{v_3\cos\alpha-v_2\sin\alpha}{v+v_1}\tilde{\epsilon}_{21}-\frac{v_2\cos\alpha+v_3\sin\alpha}{v+v_1}\tilde{\epsilon}_{31}-\frac{d\alpha}{dt},
\end{split}
\end{equation}
and the rotated velocity gradient tensor $\boldsymbol\epsilon^\prime(t)$ is then
\begin{align}
%\mathbf{A}&=\left(
%   \begin{array}{ccc}
%     0 & \tilde{\epsilon}_{21} & \tilde{\epsilon}_{31} \\
%     -\tilde{\epsilon}_{21} & 0 & \frac{-v_3}{v+v_1}\tilde{\epsilon}_{21}+\frac{v_2}{v+v_1}\tilde{\epsilon}_{31} \\
%     -\tilde{\epsilon}_{31} & \frac{v_3}{v+v_1}\tilde{\epsilon}_{21}+\frac{-v_2}{v+v_1}\tilde{\epsilon}_{31} & 0 \\
%   \end{array}
% \right),\\
 \boldsymbol\epsilon^\prime(t)&=\left(
   \begin{array}{ccc}
     \tilde{\epsilon}_{11} & \tilde{\epsilon}_{12}+\tilde{\epsilon}_{21} & \tilde{\epsilon}_{13}+\tilde{\epsilon}_{31} \\
     0 & \tilde{\epsilon}_{22} & \tilde{\epsilon}_{23}+A_{23}\\
     0 & \tilde{\epsilon}_{32}-A_{23} & -\tilde{\epsilon}_{22}-\tilde{\epsilon}_{11} \\
   \end{array}
 \right).\label{eqn:epsilonprime}
\end{align}
Similar to 2D steady flow $\epsilon^\prime_{21}$ and $\epsilon^\prime_{31}$  are both zero due to reorientation into the streamline frame, however in 3D flow $\epsilon^\prime_{32}\neq 0$ in general. However, (\ref{eqn:epsilonprime}) shows that manipulation of the transverse orientation angle $\alpha$ such that $A_{23}=\tilde{\epsilon}_{32}$ renders $\boldsymbol\epsilon^\prime(t)$ upper triangular, under the condition $A_{23}=\tilde{\epsilon}_{32}$. From (\ref{eqn:epsilondefs}) the components of $\tilde{\boldsymbol\epsilon}(t)$ and $\boldsymbol\epsilon^{(1)}(t)$ are related as
\begin{align}
&\tilde\epsilon_{21}=\epsilon^{(1)}_{21}\cos\alpha+\epsilon^{(1)}_{31}\sin\alpha,\\
&\tilde\epsilon_{31}=\epsilon^{(1)}_{31}\cos\alpha-\epsilon^{(1)}_{21}\sin\alpha,\\
&\tilde\epsilon_{32}=\epsilon^{(1)}_{32}\cos^2\alpha-\epsilon^{(1)}_{23}\sin^2\alpha+(\epsilon^{(1)}_{33}-\epsilon^{(1)}_{22})\cos\alpha\sin\alpha,
\end{align}
and so the condition $A_{23}=\tilde{\epsilon}_{32}$ is then
\begin{equation}
\begin{split}
\frac{d\alpha}{dt}&=g(\alpha,t)\\
&=a(t)\cos^2\alpha+b(t)\sin^2\alpha+c(t)\cos\alpha\sin\alpha,\label{eqn:alpha}
\end{split}
\end{equation}
where
\begin{align}
&a(t)=-\epsilon^{(1)}_{32}-\frac{v_2}{v+v_1}\epsilon^{(1)}_{31}-\frac{v_3}{v+v_1}\epsilon^{(1)}_{21},\label{eqn:a}\\
&b(t)=\epsilon^{(1)}_{23}+\frac{v_2}{v+v_1}\epsilon^{(1)}_{31}+\frac{v_3}{v+v_1}\epsilon^{(1)}_{21},\label{eqn:b}\\
&c(t)=\epsilon^{(1)}_{22}-\epsilon^{(1)}_{33}+\frac{2v_2}{v+v_1}\epsilon^{(1)}_{21}-\frac{2v_3}{v+v_1}\epsilon^{(1)}_{31}.\label{eqn:c}
\end{align}

\subsection{Evolution of Protean Orientation Angle $\alpha$}

Equation (\ref{eqn:alpha}) describes a 1st-order ODE for the transverse orientation angle $\alpha(t)$ along a streamline which renders $\boldsymbol\epsilon^\prime(t)$ upper triangular, where the temporal derivative for $\alpha(t)$ is associated with both change in flow structure along a streamline and impact of a moving coordinate transform as encoded by $\mathbf{A}(t)$. This is analogous to the ODE system (\ref{eqn:QRaux}) for the continuous QR method, with the important difference that the Protean reorientation gives a closed form solution for $\mathbf{e}_1=\mathbf{v}/v$ which constrains two degrees of freedom (d.o.f's) for the Protean frame and an ODE for the remaining d.o.f characterised by $\alpha(t)$, whereas the continuous QR method involves solution of the 3 degree of freedom  ODE system (\ref{eqn:QRaux}), and requires unitary integrators to preserve orthogonality of $\mathcal{Q}$. Similar to the 2D case, these methods differ with respect to the initial conditions $\mathcal{Q}(0)=\mathbf{I}$, $\mathbf{Q}(0)=\mathbf{Q}_1(0)\cdot\mathbf{Q}_2(0)$, where $\mathbf{Q}_1(0)$ is given explicitly by (\ref{eqn:A1}), and $\mathbf{Q}_2(0)$ is dependant upon $\alpha_0$ as per (\ref{eqn:A2}), (\ref{eqn:alpha}).

\begin{figure}
\begin{centering}
\begin{tabular}{cc}
\includegraphics[width=0.45\columnwidth]{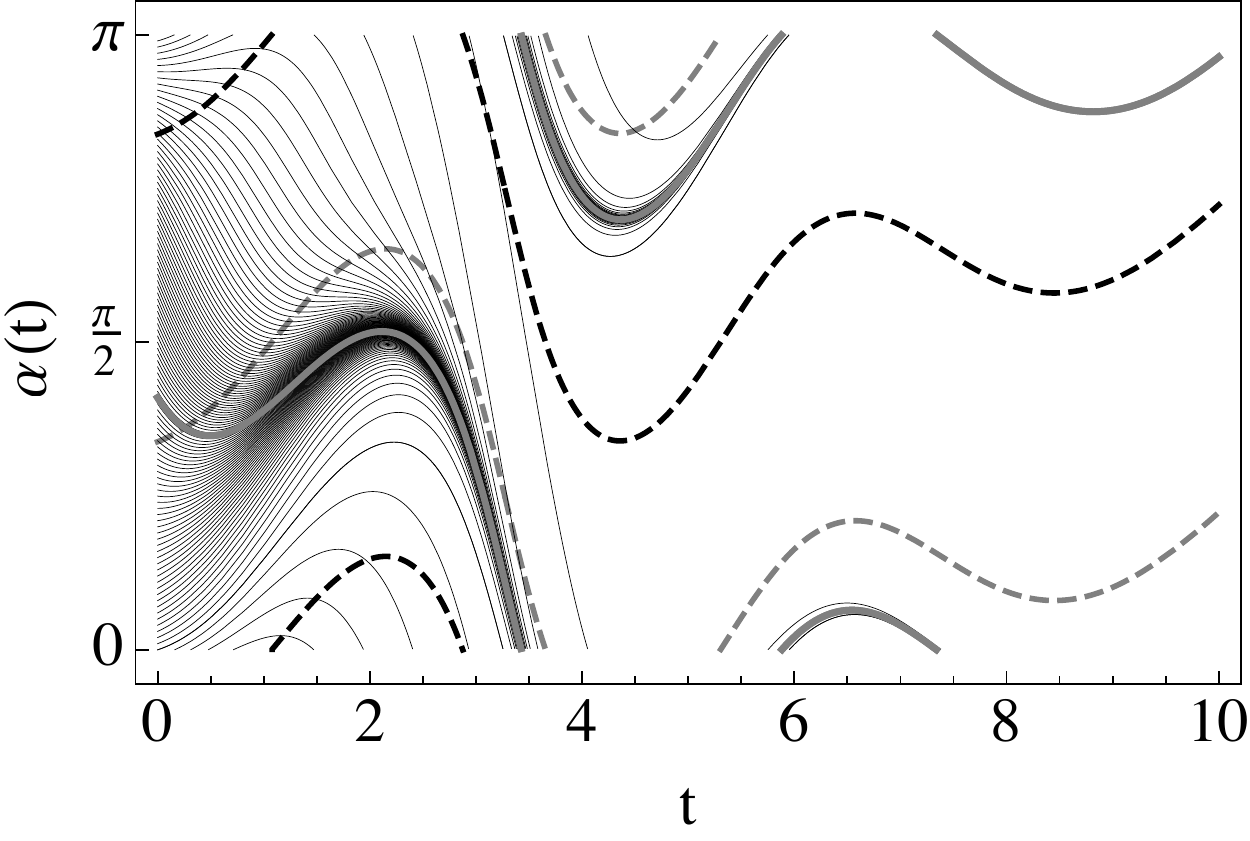}&
\includegraphics[width=0.45\columnwidth]{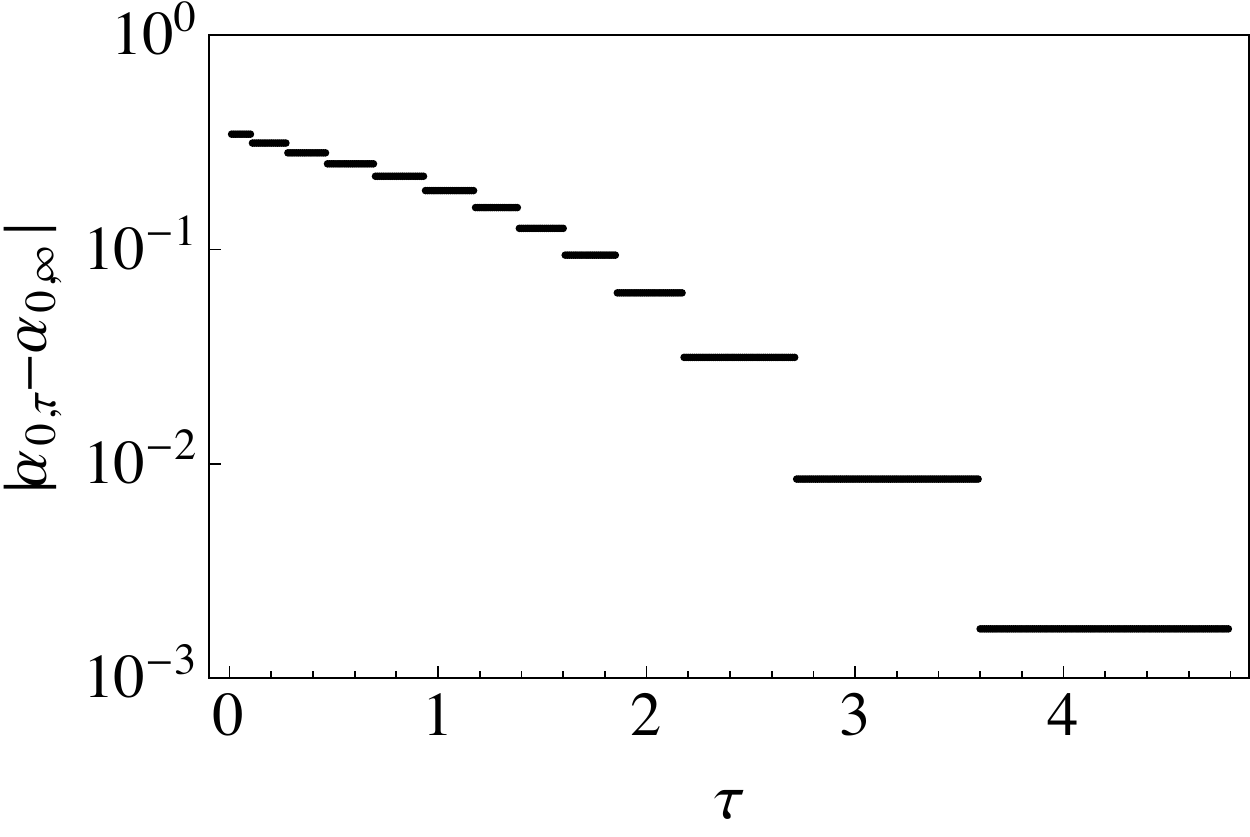}\\
(a) & (b)
\end{tabular}
\end{centering}
\caption{(a) Convergence of the transverse orientation angle $\alpha(t)$ for different initial conditions $\alpha(0)=\alpha_0$ (thin black lines) toward the inertial manifold $\mathcal{M}(t)=\alpha(t;\alpha_{0,\infty})$ (thick gray line) for the ABC flow. Note the correspondence between $\mathcal{M}(t)$ and the transverse angle $\alpha^\star(t)$ associated with minimum (gray dashed line) divergence $\partial g/\partial\alpha(\alpha,t)$ responsible for creation of the inertial manifold. The maximum divergence is also shown (black dashed line). (b) Convergence of the initial angle associated with the approximate inertial manifold $\alpha_{0,\tau}$ toward the infinite time limit $\alpha_{0,\infty}$.}
\label{fig:alpha}
\end{figure}

Whilst the ODE (\ref{eqn:alpha}) admits an arbitrary initial condition $\alpha(0)=\alpha_0$ which results in non-uniqueness of $\boldsymbol\epsilon^\prime(t)$, this non-autonomous ODE is locally dissipative as reflected by the divergence
\begin{equation}
\frac{\partial g}{\partial\alpha}=(b(t)-a(t))\sin 2\alpha+c(t)\cos 2\alpha,
\end{equation}
which admit maxima and minima of magnitude $\pm c\sqrt{1+(b-a/c)^2}$ respectively at
\begin{equation}
 \alpha^\star(t)=\frac{1}{2}\arctan\left(\frac{b-a}{c}\right)+\frac{\pi}{2}\frac{\text{sgn}\,c\mp 1}{2}.
\end{equation}
Hence for $c(t)\neq 0$, the ODE (\ref{eqn:alpha}) admits an non-autonomous inertial manifold~\cite{Potzsche:2009aa} $\mathcal{M}(t)$, which exponentially attracts solutions from all initial conditions $\alpha_0$ as illustrated in Fig.~\ref{fig:alpha}(a). As the inertial manifold is also a solution of (\ref{eqn:alpha}), it may be expressed as $\mathcal{M}(t)=\alpha(t;\alpha_{0,\infty})$, where $\alpha_{0,\infty}$ is the initial condition on $\mathcal{M}(t)$. Whilst all solutions of (\ref{eqn:alpha}) render $\boldsymbol\epsilon^\prime(t)$ upper triangular, only solutions along the inertial manifold $\mathcal{M}(t)$ represent asymptotic dynamics independent of the initial condition $\alpha_0$, hence define the Protean frame as that which corresponds to the unique inertial solution $\alpha(t;\alpha_{0,\infty})$.

As the inertial manifold $\mathcal{M}(t)$ maximises dissipation $-\partial g/\partial\alpha$ over long times, the associated initial condition $\alpha_{0,\infty}$ is quantified by the limit
\begin{align}
&\alpha_{0,\infty}:=\lim_{\tau\rightarrow\infty}\alpha_{0,\tau},
\end{align}
where
\begin{align}
&\alpha_{0,\tau}(t):=\arg\min_{\alpha_0} \int_0^\tau \frac{\partial g}{\partial\alpha}(\alpha(t^\prime;\alpha_0), t^\prime)dt^\prime.\label{eqn:finitetimeapprox}
\end{align}
Whilst the inertial manifold can be identified by evolving (\ref{eqn:alpha}) until acceptable convergence is determined, $\mathcal{M}(t)$ may be identified at shorter times $\tau\sim 1/|c|$ via the approximation (\ref{eqn:finitetimeapprox}), as shown in Fig.~\ref{fig:alpha}(b). This approach is particularly useful when Lagrangian data is only available over short time-frames, and moreover explicitly identifies the inertial initial orientation angle $\alpha_{0,\infty}$, allowing the Protean transform to be applied from $t=0$ rather than the finite convergence time $\tau$.

\subsection{Protean Deformation in 3D Steady Flow}

Given appropriate reorientation of the Protean frame given by (\ref{eqn:alpha}) such that $A_{23}=\tilde{\epsilon}^{(1)}_{32}$, the streamline velocity gradient tensor $\boldsymbol\epsilon^\prime(t)$ is upper triangular
%\begin{align}
%\boldsymbol\epsilon^\prime(t)&=\left(
 %  \begin{array}{ccc}
 %    \epsilon^\prime_{11} & \epsilon^\prime_{12} & \epsilon^\prime_{13} \\
  %   0 & \epsilon^\prime_{22} & \epsilon^\prime_{23} \\
  %   0 & 0 & \epsilon^\prime_{33} \\
  % \end{array}
 %\right),
%\end{align}
and from (\ref{eqn:rotF}) $\mathbf{F}^\prime(t)$ is also upper triangular, with components
\begin{align}
&F_{ij}^\prime(t)=0\,\,\text{for}\,\,i>j\label{eqn:Fij},\\
&F_{11}^\prime(t)=\frac{v(t)}{v(0)},\\
&F_{ii}^\prime(t)=\exp\left(\int_0^t dt^\prime \epsilon^\prime_{ii}(t^\prime)\right),\,\,i=(1,2,3)\label{eqn:Fii}\\
&F_{12}^\prime(t)=v(t)\int_0^t dt^\prime\frac{\epsilon^\prime_{12}(t^\prime) F^\prime_{22}(t^\prime)}{v(t^\prime)},\label{eqn:F12}\\
&F_{23}^\prime(t)=F^\prime_{22}(t)\int_0^t dt^\prime\frac{\epsilon^\prime_{23}(t^\prime) F^\prime_{33}(t^\prime)}{F^\prime_{22}(t^\prime)},\label{eqn:F23}\\
&F_{13}^\prime(t)=v(t)\int_0^t dt^\prime\frac{\epsilon^\prime_{12}(t^\prime)F^\prime_{23}(t^\prime)+\epsilon^\prime_{13}(t^\prime)F^\prime_{33}(t^\prime)}{v(t^\prime)}.\label{eqn:F13}
\end{align}
The upper triangular form of $\boldsymbol\epsilon^\prime(t)$ simplfies solution of the components of the deformation tensor $\mathbf{F}^\prime(t)$ in that the integrals (\ref{eqn:Fii})-(\ref{eqn:F13}) can solved sequentially in a manner analogous to Gaussian elimination. Note that similar to 2D flow the streamwise deformation component simply oscillates as $F^\prime_{11}(t)=v(t)/v(0)$, and cannot grow over time. These dynamics are a direct consequence of the steady nature of the flow, whereas for unsteady flow the ensemble average of this component can change over time. 
%As such, this form elucidates the deformation structure of the flow and furthermore provides a simple basis for the construction of stochastic models for deformation evolution.

\subsection{Protean Deformation in 3D Steady Zero Helicity Flow}

For 3D steady zero helicity flows, constraints upon the flow structure further simplify both the dynamics of both the orientation angle $\alpha$ and the deformation gradient evolution. Finnigan~\cite{Finnigan:1983aa} demonstrates that all 3D flows with zero helicity density $h=0$ (or \emph{complex lamellar} flows) may be expressed in the general form of an isotropic Darcy flow
\begin{equation}
\mathbf{v}(\mathbf{x})=-k(\mathbf{x})\nabla\phi=\nabla\psi\times\nabla\zeta,
\end{equation}
with $\nabla\psi\cdot\nabla\zeta=0$, where $k(\mathbf{x})$, $\phi$, $\psi$ are smooth continuous functions which respectively represent hydraulic conductivity, velocity potential and streamfunction in isotropic Darcy flow. Such flow fields admit an orthogonal streamline coordinate system which comprise of material surfaces of constant streamfunction $\psi$, isopotential surfaces of constant $\psi$, and an additional material surface of constant $\zeta$ normal to both of these. The velocity $\mathbf{v}$, vorticity $\boldsymbol\omega$ and Lamb $\mathbf{v}\times\boldsymbol\omega$ vectors are then normal to the isosurfaces of $\phi$, $\zeta$, $\psi$ respectively.

Sposito~\cite{Sposito:2001aa, Sposito:1997aa} identifies the stream surfaces of constant $\psi$ as \emph{Lamb surfaces}~\cite{Lamb:1932aa} of the flow which are smooth non-intersecting 2D manifolds which are spanned by both streamlines and vorticity lines of the flow. These Lamb surfaces are both geometrically flat and topologically simple, comprising of either topological cylinders or tori~\cite{Sposito:2001aa,Sposito:1997aa}. As such, the Poincar\'{e}-Bendixson theorem applies to streamlines, all of which are confined to these Lamb surfaces and so fluid deformation is restricted to grow algebraically in time. Conversely, streamlines in steady flows with non-zero helicity density do not conform to Lamb surfaces, and so may wander freely throughout the 3D flow domain and so may exhibit exponential fluid deformation.

In the material frame $(\phi,\psi,\zeta)$ the velocity gradient tensor $\boldsymbol\epsilon^\prime(t)$ takes on a particularly simple form, and so we define the Protean coordinate frame for zero helicity density flows as this coordinate frame, such that the $x_1^\prime$, $x_2^\prime$ coordinates align with the velocity and vorticity vectors respectively. Similar to 2D flow, the coordinate transform between the Cartesian and Protean frames is now explicit, consisting of the primary (\ref{eqn:A1}) and secondary (\ref{eqn:A2}) rotations, and $\alpha$ is explicitly
\begin{equation}
\alpha=\arctan\left(\frac{\mathbf{e}_3^{(1)}\cdot\boldsymbol\omega}{\mathbf{e}_2^{(1)}\cdot\boldsymbol\omega}\right),\label{eqn:alpha_zero_helicity}
\end{equation}
where $\mathbf{e}_j^{(1)}$ is the $j$-th column vector of $\mathbf{Q}_1$. Analysis of the differential geometry of this coordinate system in Appendix~\ref{App:zero_helicity} shows that the material nature of this frame automatically renders the Protean velocity gradient tensor $\boldsymbol\epsilon^\prime$ upper triangular. The analysis in Appendix~\ref{App:zero_helicity} also shows that the constraints associated with zero helicity density flows render the transverse shear and vorticity components of the Protean velocity gradient tensor (\ref{eqn:vgrad_zero_helicity}) to be zero
\begin{equation}
\epsilon_{23}^\prime=\epsilon^\prime_{32}=0,
\end{equation}
and so the $(2,3)$ components of the Protean deformation tensor $\mathbf{F}^\prime(t)$ then decouple as
\begin{equation}
\mathbf{F}^\prime(t)=
\left(
   \begin{array}{ccc}
     F_{11}^\prime & F_{12}^\prime & F_{13}^\prime \\
     0 & F_{22}^\prime & 0 \\
     0 & 0 & F_{33}^\prime \\
   \end{array}
 \right).\label{eqn:F_zero_helicity}
\end{equation}
As such the $(1,2)$ and $(1,3$) components of the deformation tensor then simplify to
\begin{align}
&F_{12}^\prime(t)=v(t)\int_0^t dt^\prime\frac{\epsilon^\prime_{12}(t^\prime) F^\prime_{22}(t^\prime)}{v(t^\prime)},\label{eqn:F12zh}\\
&F_{13}^\prime(t)=v(t)\int_0^t dt^\prime\frac{\epsilon^\prime_{13}(t^\prime)F^\prime_{33}(t^\prime)}{v(t^\prime)},\label{eqn:F13zh}
\end{align}
where $\epsilon^\prime_{12}$, $\epsilon^\prime_{13}$ represent longitudinal shear along surfaces of constant $\psi$, $\zeta$ respectively. Hence deformation in 3D zero helicity density flow evolves in a similar manner to that of 2D steady flow, where the only persistent deformation arises from shear and vorticity. This may be conceptualised as deformation due to 2D flow within a Lamb surface ($\epsilon^\prime_{12}$), as well as contributions due to shear between Lamb surfaces ($\epsilon^\prime_{13}$). This particularly simple deformation structure means that stochastic models of fluid deformation in steady 2D flows \cite{Dentz:2015aa,Dentz:2015ab} may be readily extended to steady 3D zero helicity flows.

%These separating surfaces partition the flow domain and for zero helicity density flows, the streamlines and vorticity lines intersect perpendicularly, as can be shown in streamline coordinate frame as
%\begin{equation}
%\begin{split}
%\boldsymbol\omega=&\boldsymbol\omega^\prime+\boldsymbol\varpi^{ijk}:\mathbf{A},\\
%=&\boldsymbol\omega^\prime-2A_{23}\mathbf{e}_1^\prime,\\
%=&\nabla^\prime\times\mathbf{v}^\prime(\mathbf{x}^\prime)-2A_{23}\mathbf{e}_1^\prime,\\
%=&\boldsymbol\varpi^{ijk}:\boldsymbol\epsilon^\prime(t)-2A_{23}\mathbf{e}_1^\prime,\\
%=&\{0,\tilde\epsilon_{13}+\tilde\epsilon_{31},-\tilde\epsilon_{12}-\tilde\epsilon_{21}\}.\label{eqn:vorticity}
%\end{split}
%\end{equation}
%Hence vorticity lines and streamlines form orthogonal curves (which are geodesics~\cite{Sposito:2001aa}) on Lamb surfaces, and (\ref{eqn:vorticity}) indicates that Lamb surfaces are oriented normal to direction of maximum shear deformation and streamline curvature.

%Such topological freedom admits chaotic Lagrangian dynamics and persistent exponential stretching of fluid elements, as reflected in evolution of the deformation gradient tensor $\mathbf{F}(t)$. As such, topological measures of the flow field such as the local helicity density $h$, equation (\ref{eqn:helicity}) (or the global equivalent $H:=\int_{\mathcal{D}}h(\mathbf{x})d^n\mathbf{x}$ which measures knottedness of vorticity lines) have a direct impact upon the deformation dynamics in steady 3D flows.

\section{Application to 3D Steady Flow}\label{sec:examples}

To illustrate utility of the method and uncover the deformation structure over a range of flows, we solve fluid deformation in the Protean frame over several classes of steady 3D flow summarised in Table~\ref{tab:flows}. For all flows, particle trajectories from a random initial position are calculated to precision $10^{-14}$ over the time period $t=[0,1000]$ via an implicit Gauss-Legendre method. The associated deformation tensor in the Eulerian frame is calculated via solution of (\ref{eqn:deform}) to the same precision via the discrete QR decomposition method due to the large associated deformations. The Protean rate of strain tensor $\boldsymbol\epsilon^\prime(t)$ is then determined along these trajectories by solution of the inertial manifold $\mathcal{M}(t)=\alpha(t;\alpha_{0,\infty})$ in (\ref{eqn:alpha}) via an explicit Runge-Kutta method over fixed time-steps $\Delta t=10^{-3}$ (unless specified otherwise) and the Protean deformation gradient tensor $\mathbf{F}^\prime(t)$ is determined by direct evaluation of the integrals (\ref{eqn:Fij})-(\ref{eqn:F12}). We analyse the errors between the deformation tensor in the Cartesian and Protean frames, and calculate relevant measures of fluid deformation. For the deterministic ABC flow, statistics for the components of $\epsilon^\prime(t)$ are computed over 1,000 particle trajectories from random initial locations within a chaotic region of the flow domain, whereas for the random flows the same statistics are gathered from a single trajectory over 1,000 realisations of the flow field.

\begin{table}
\begin{centering}
\begin{tabular}{c c c c}
\hline
Flow/Property & Random & Helical & Compressible \\
\hline
ABC Flow 				& O	& X 	& O \\
Kraichnan Flow 			& X	& X 	& O \\
Dual Streamfunction Flow 	& X 	& X 	& O \\
Potential Flow 			& X 	& O 	& X \\
\hline \label{tab:flows}
\end{tabular}
\caption{Example flow types and properties, where X denotes the flow field fulfils the property, O otherwise.}
\end{centering}
\end{table}

\subsection{Arnol'd-Beltrami-Childress (ABC) Flow}

\begin{figure}
\begin{centering}
\begin{tabular}{cc}
\includegraphics[width=0.45\columnwidth]{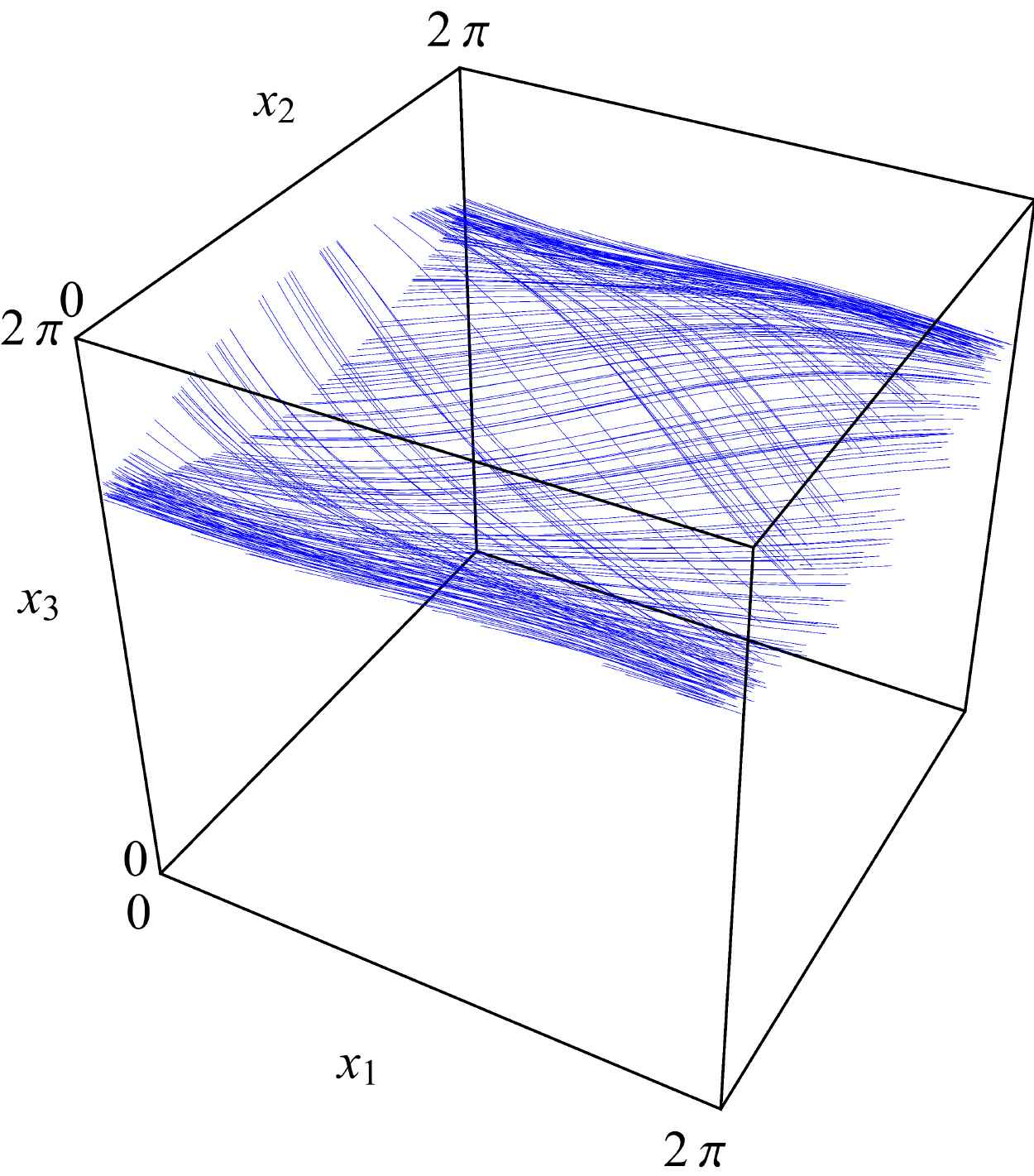}&
\includegraphics[width=0.5\columnwidth]{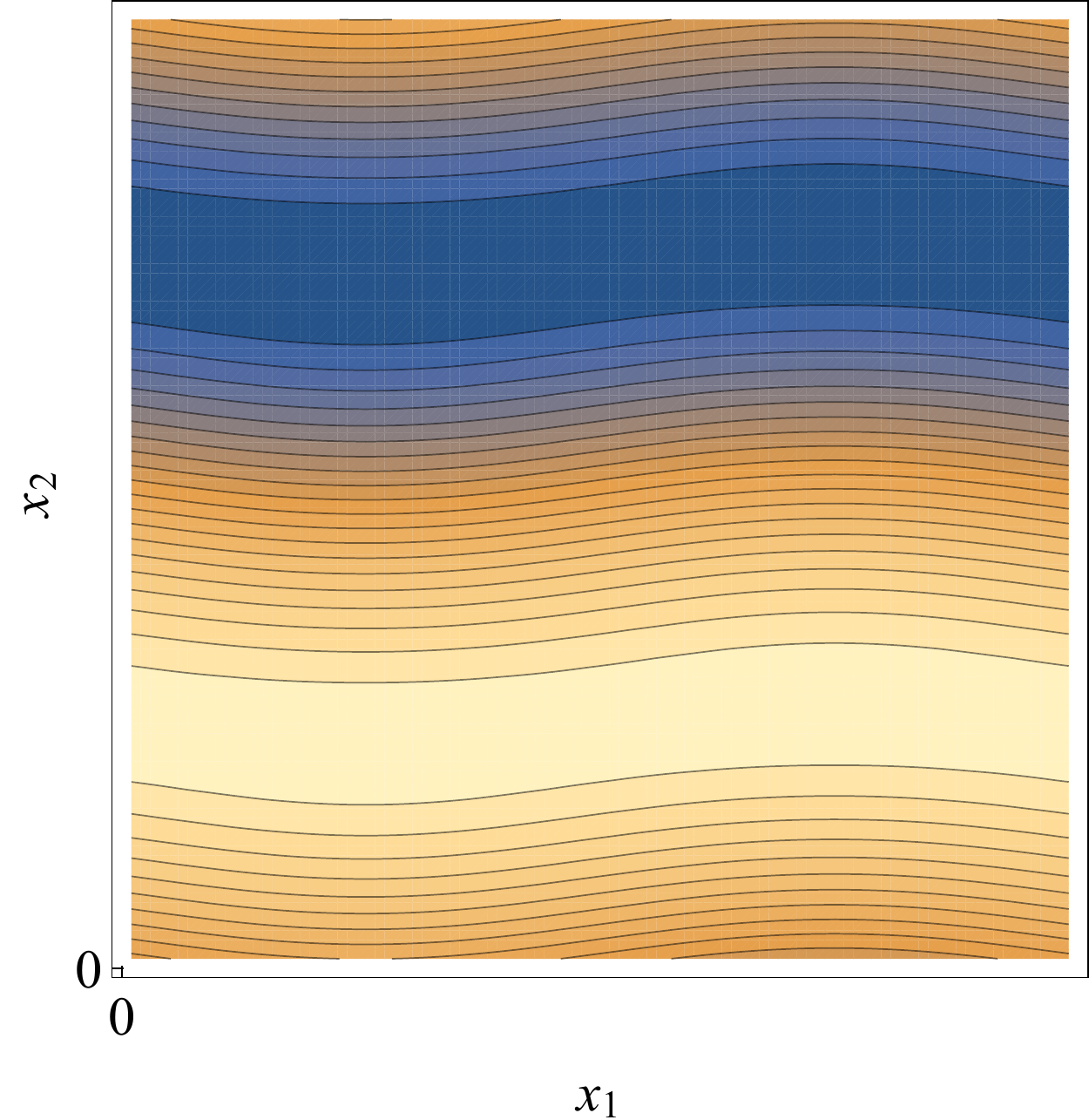}\\
(a) & (b)
\end{tabular}
\end{centering}
\caption{(a) Typical particle trajectory in the ergodic region of the Arnol'd-Beltrami-Childress flow, (b) contour plot of velocity magnitude distribution in $x_3=0$ plane.}\label{fig:ABC}
\end{figure}

We first apply the method to the Arnol'd-Beltrami-Childress (ABC) flow, an incompressible 3D Euler flow which is well-studied in the context of chaotic Lagrangian dynamics~\cite{DombreEA:1986} . The ABC flow is a member of the class of Beltrami flows, with the property $\mathbf{v}\propto \boldsymbol{\omega}$, hence the helicity density $h$ is non-zero throughout the triply-periodic flow domain $\mathcal{D}:\{x,y,z\}=[0,2\pi]\times[0,2\pi]\times[0,2\pi]$, as given by the velocity field
\begin{equation}
\begin{split}
\mathbf{v}(x,y,z)=&(A \sin z + C\cos y)\mathbf{e}_x\\
+&(B \sin x + A \cos z)\mathbf{e}_y\\
+&(C \sin y + B \cos x)\mathbf{e}_z.\label{eqn:ABC}
\end{split}
\end{equation}
We consider the parameter values $A=1.5$, $B=0.2$, $C=0.1$ used by Feingold~\cite{Feingold:1988aa} which generates chaotic trajectories over a subdomain of the global flow domain $\mathcal{D}$ as illustrated by the space-filling particle trajectory shown in Fig.~\ref{fig:ABC}(a). Whilst the entire flow domain is not globally chaotic, we restrict attention to the deformation dynamics in the ergodic subdomain shown in Fig.~\ref{fig:ABC}. As the Lagrangian domain of chaotic flows are partitioned into topologically distinct chaotic and regular subdomains, relevant arguments regarding deformation dynamics in ergodic flows apply within each distinct chaotic (mixing) region. To study the deformation dynamics of this region and test the Protean method, we solve 1,000 particle trajectories over the period $t\in[0,1000]$ in this region and solve the Cartesian deformation tensor $\mathbf{F}(t)$ in (\ref{eqn:deform}) using a discrete QR decomposition method~\cite{Dieci:1997aa,Dieci:2008aa}, and compare these results with the Protean deformation tensor $\mathbf{F}^\prime(t)$ calculated directly from the integrals (\ref{eqn:Fii})-(\ref{eqn:F13}).

\begin{figure}
\begin{centering}
\begin{tabular}{c c}
\includegraphics[width=0.45\columnwidth]{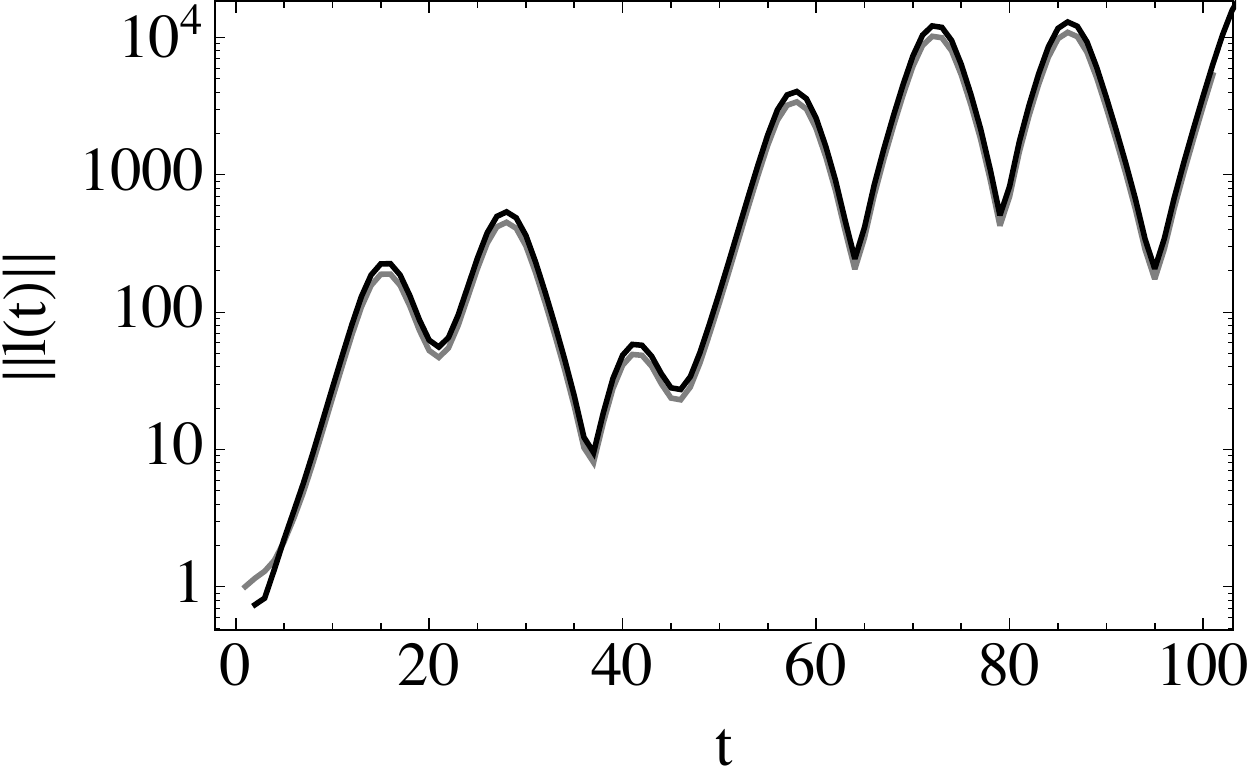}&
\includegraphics[width=0.45\columnwidth]{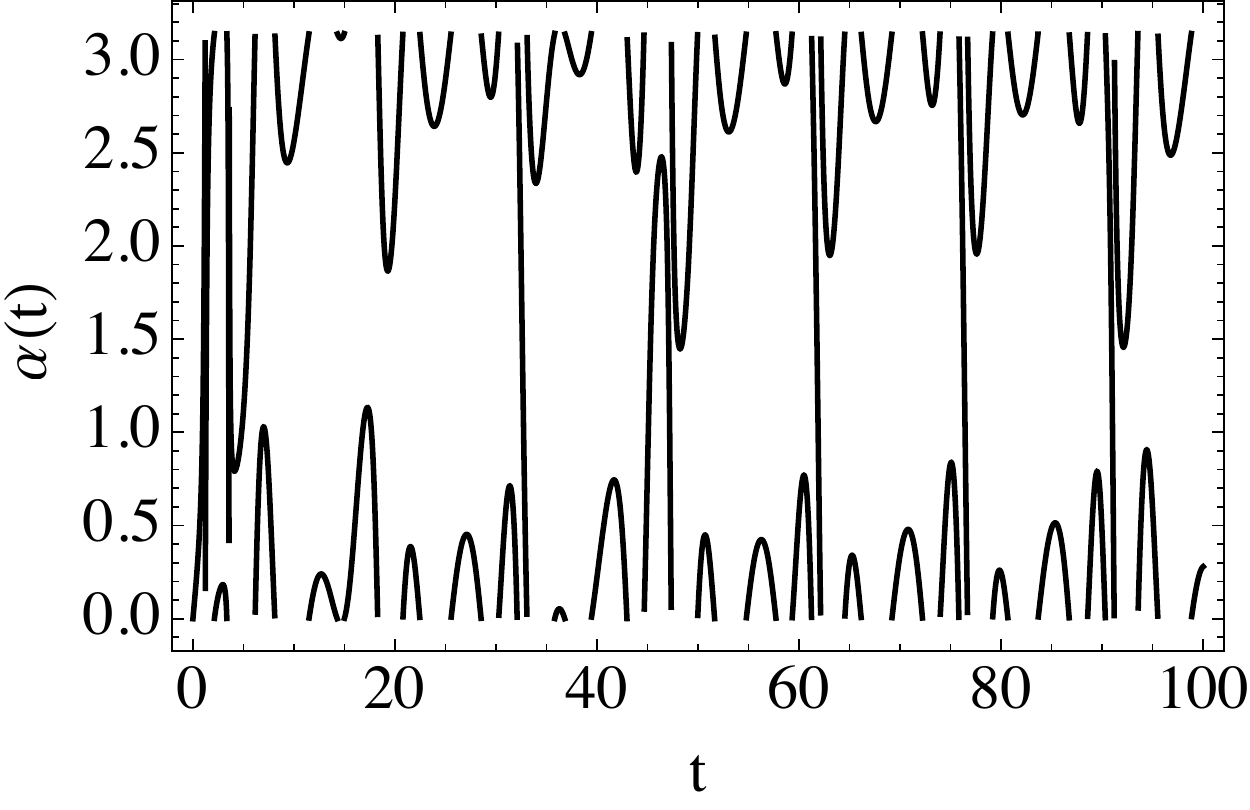}\\
(a) & (b)\\
\includegraphics[width=0.45\columnwidth]{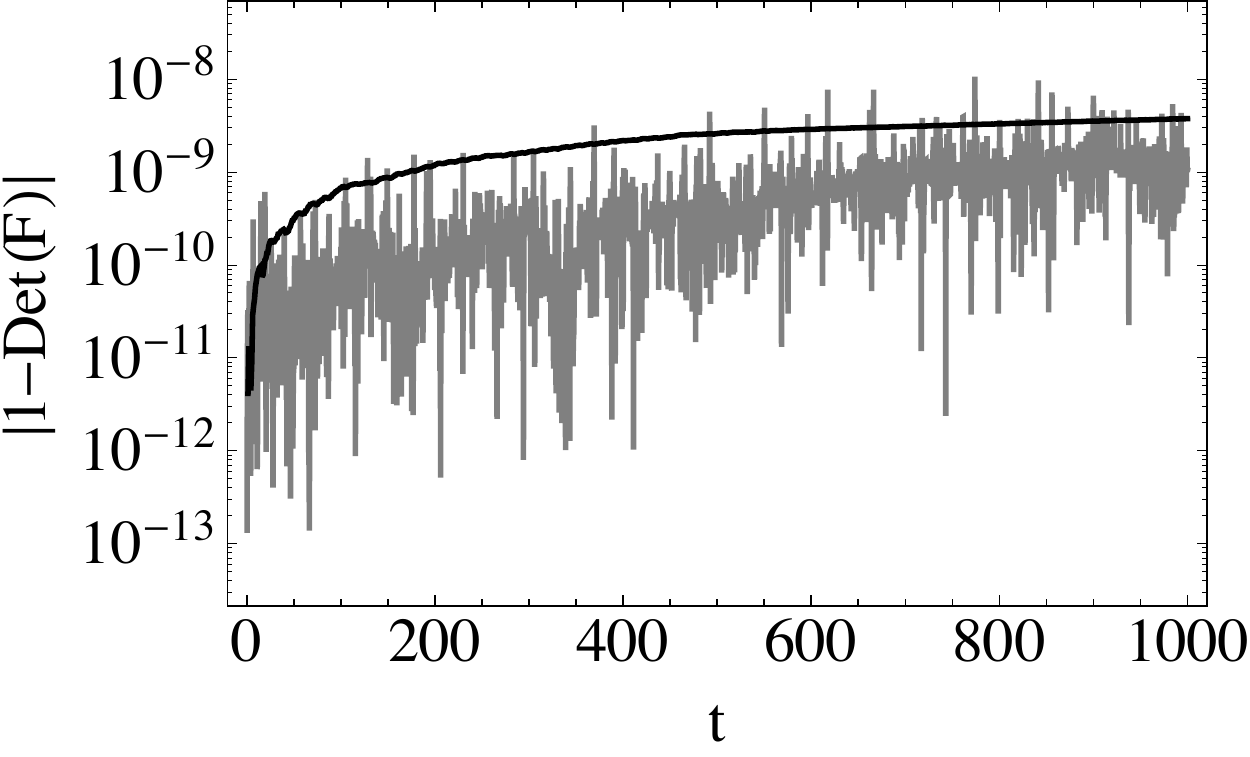}&
\includegraphics[width=0.45\columnwidth]{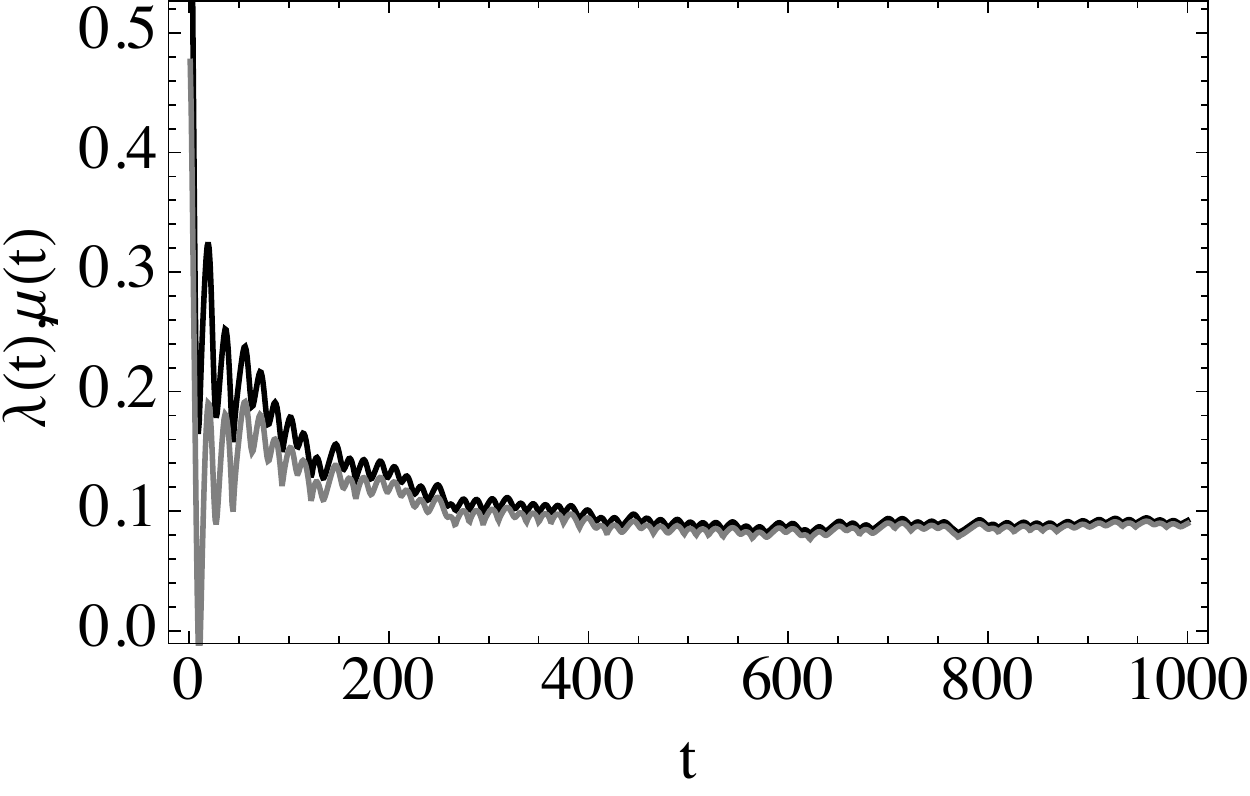}\\
(c) & (d)
\end{tabular}
\end{centering}
\caption{(a) Relative growth of the length $|\mathbf{l}(t)|$ of an infinitesimal material line along a single trajectory in the ABC flow (\ref{eqn:ABC}) calculated by (grey) particle tracking and (black) from $\mathbf{l}(t)=\mathbf{F}^\prime(t)\cdot\mathbf{l}(0)$. (b) Solution of the orientation angle $\alpha(t)$ along the inertial manifold $\mathcal{M}$. (c) Determinant error for the (grey) Cartesian $\mathbf{F}(t)$ and (black) Protean $\mathbf{F}^\prime(t)$ deformation tensors. (d) Convergence of the principal stretching exponent $\lambda(t,\mathbf{X})$ to the FTLE $\mu(t)$.
}\label{fig:plot1_ABC}
\end{figure}

The impact of chaotic mixing along the trajectory in Fig.~\ref{fig:ABC} is reflected by the rapid growth of the relative length $|\mathbf{l}(t)|/|\mathbf{l}(0)|$ of an infinitesimal ($|\mathbf{l}(0)|=10^{-16}$) material line calculated via particle tracking shown in Fig.~\ref{fig:plot1_ABC}(a), which matches that calculated from the Protean deformation tensor as $\mathbf{l}(t)=\mathbf{F}^\prime(t)\cdot\mathbf{l}(0)$. Fig.~\ref{fig:plot1_ABC}(b) shows that the non-zero helicity density of the ABC flow imparts significant rotation to the transverse orientation angle $\alpha(t)$ along a particle trajectory. As the ABC flow is incompressible, the determinant of $\mathbf{F}(t)$ should equal unity for this flow, and the associated errors for the deformation gradient tensors are shown in Fig.~\ref{fig:plot1_ABC}(c)). As indicated by Fig.~\ref{fig:plot1_ABC}(d)), convergence between the principal stretching rate $\lambda(t,\mathbf{X})$ and the FTLE $\mu(t,\mathbf{X})$ (\ref{eqn:Lyapunov}) is fairly rapid for this flow, however the FTLE approximation given by (\ref{eqn:FTLEapprox}) converges significantly faster (as indicated by Figure~\ref{fig:FTLEapprox}).

To ascertain the accuracy of the coordinate transform as a function of the timestep $\Delta t$, the quantities in Figure~\ref{fig:plot1_ABC} are calculated over a range of time steps $\Delta t=10^{-3}$, $10^{-2}$, $10^{-1}$, $10^0$, and the associated errors for the deformation tensor $\mathbf{F}^\prime(t)$ for these time steps are summarized in Figure~\ref{fig:plot1_Dt} below. For all cases, the errors for $\Delta t\leqslant 10^{-1}$ are remarkably insensitive to the size of the time-step, indicating the coordinate transform is quite robust for all but the coarsest of time step $\Delta t= 10^0$. This timestep corresponds to spatial increments along particle trajectories which are of the order of the length-scale of the flow field, hence significant errors arising from cubic interpolation which impact estimation of both $\boldsymbol\epsilon^\prime(t)$ and $\mathbf{F}^\prime(t)$. Hence the coordinate transform appears to be very stable so long as the spatial discretisation is sufficient to resolve the underlying flow features.

\begin{figure}
\begin{centering}
\begin{tabular}{c c}
\includegraphics[width=0.45\columnwidth]{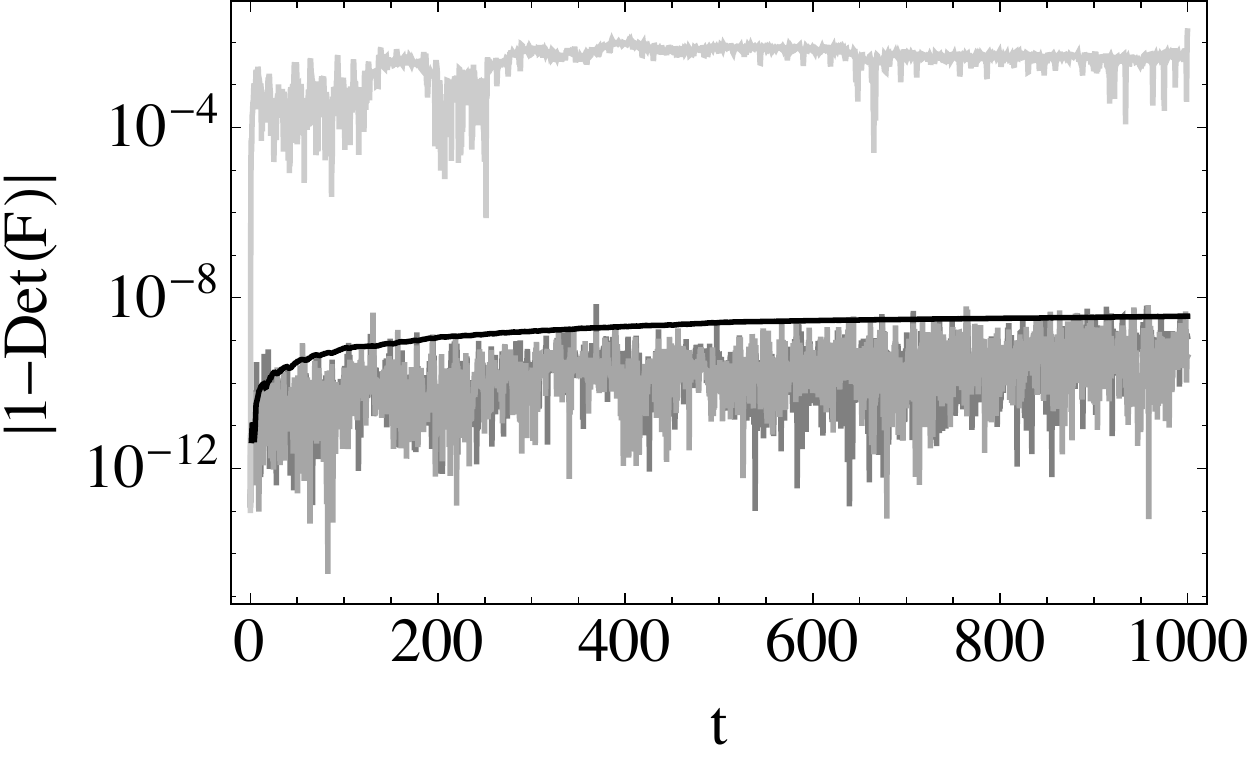}&
\includegraphics[width=0.45\columnwidth]{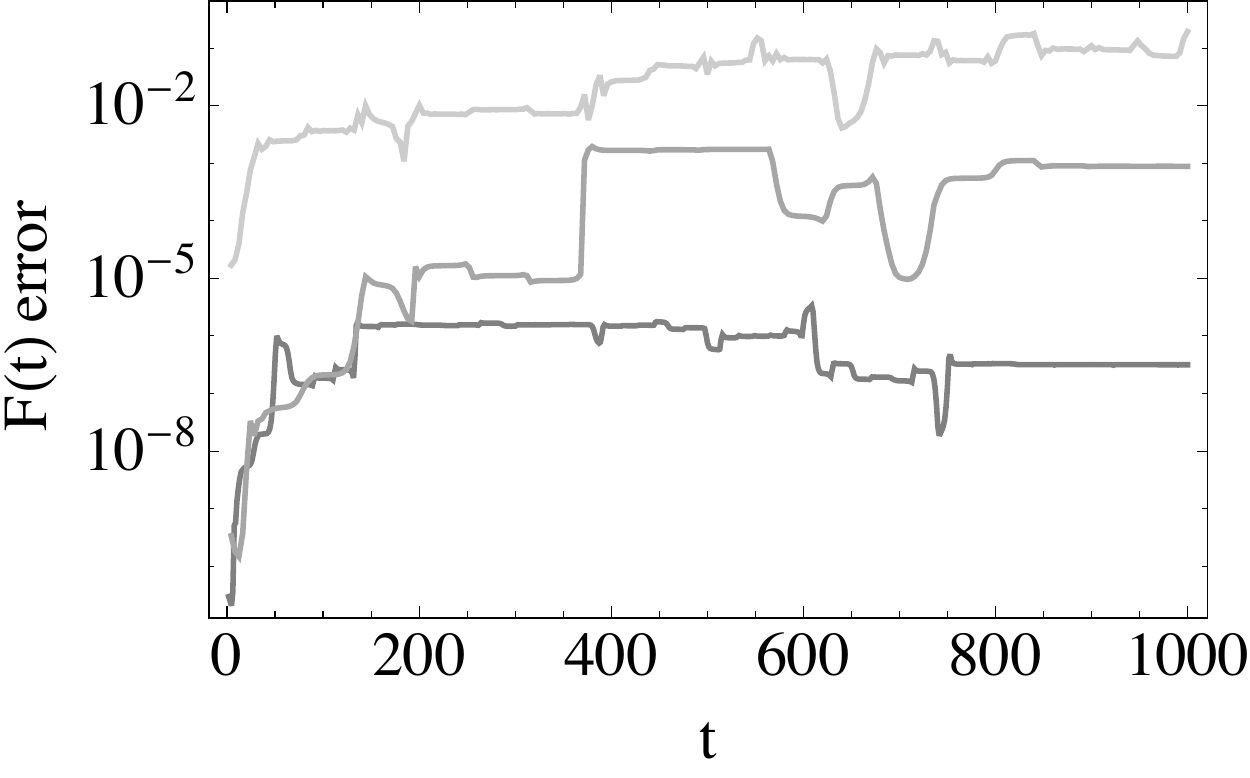}\\
(a) & (b)
\\
\includegraphics[width=0.45\columnwidth]{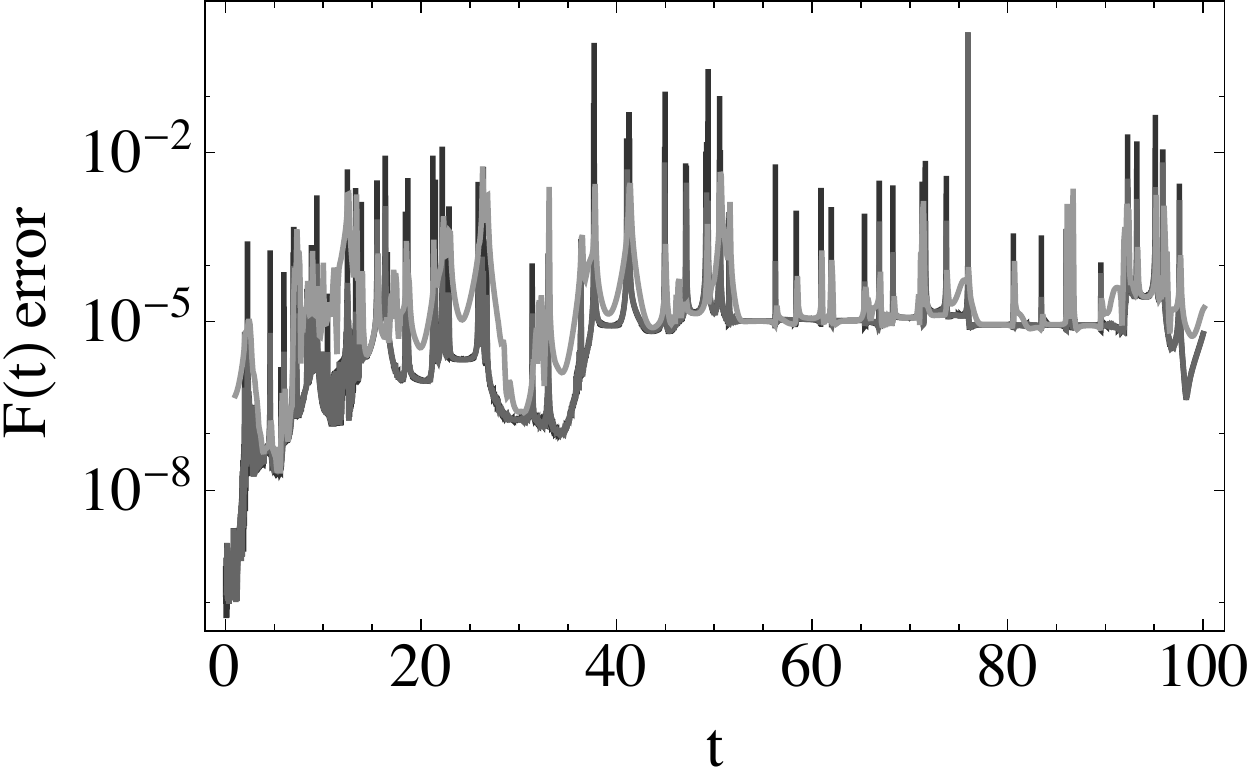}&
\includegraphics[width=0.45\columnwidth]{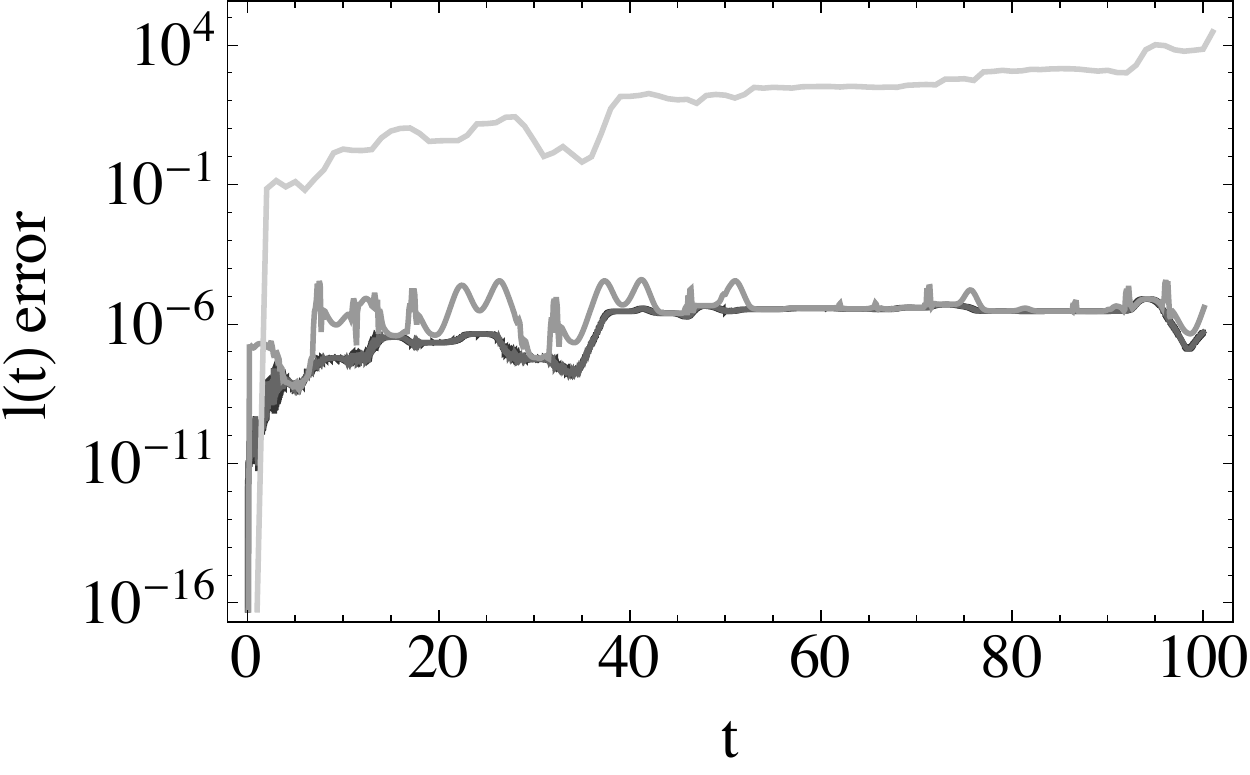}\\
(c) & (d)
\end{tabular}
\end{centering}
\caption{Errors between Cartesian and Protean deformation tensors for time steps $\Delta t=10^{-3}$ (black), $10^{-2}$ (dark grey), $10^{-1}$ (medium grey), $10^0$ (light grey) in terms of (a) determinant of the deformation tensor $\mathbf{F}(t)$, (b) eigenvalues of right Cauchy-Green tensor $\mathbf{C}$, (c) norm of error between deformation tensors, (d) error in material line length $l(t)$.}\label{fig:plot1_Dt}
\end{figure}

\begin{figure}
\begin{centering}
\begin{tabular}{c c}
\includegraphics[width=0.45\columnwidth]{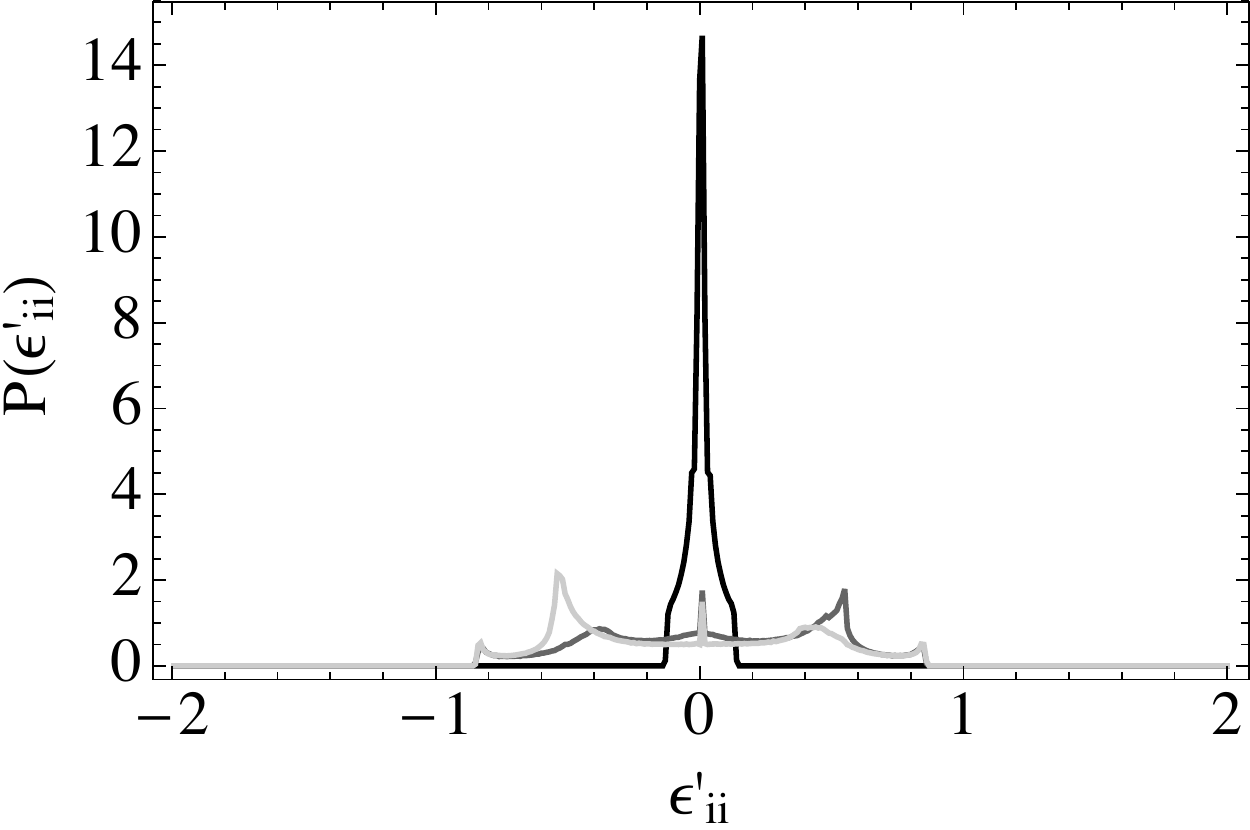}&
\includegraphics[width=0.45\columnwidth]{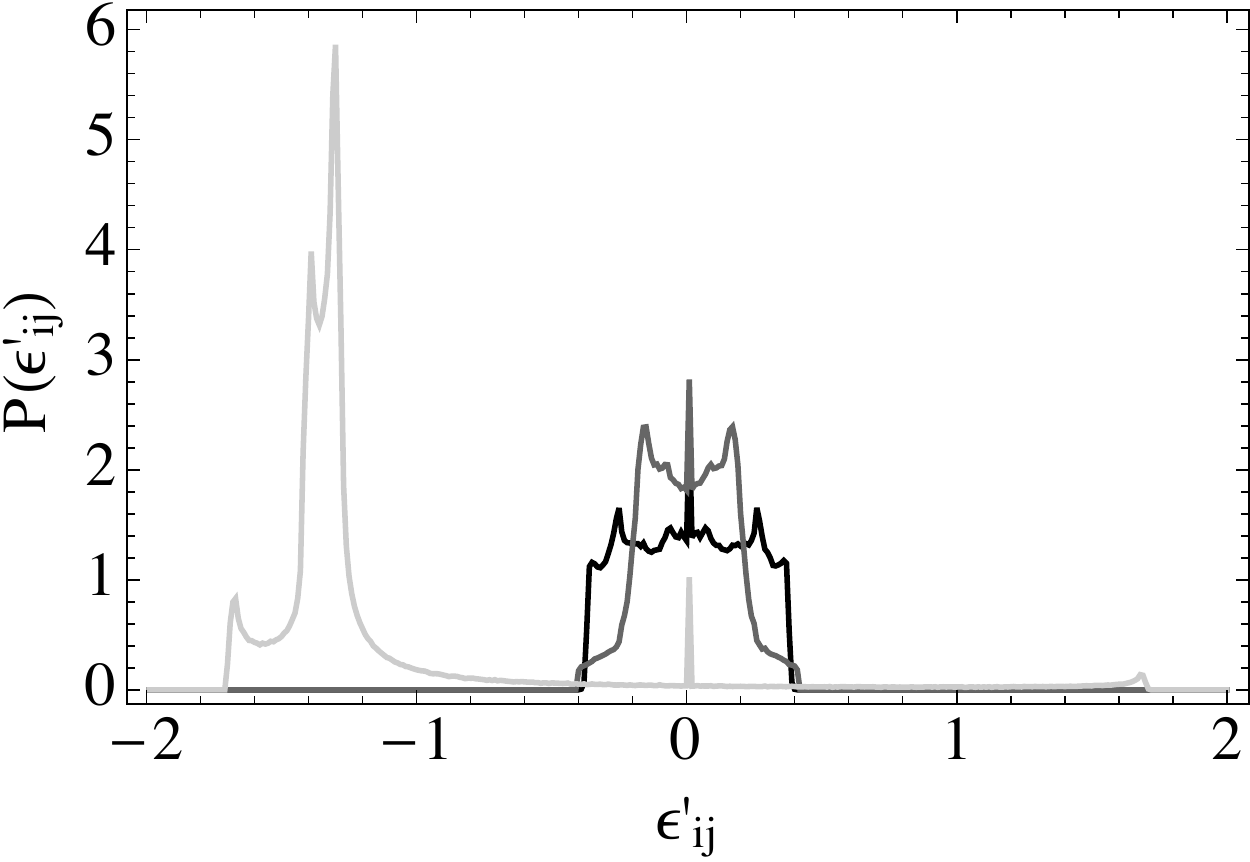}\\
(a) & (b)
\end{tabular}
\end{centering}
\caption{(a) Distribution of diagonal reoriented rate of strain components $\boldsymbol\epsilon^\prime_{ii}(t)$ for the ABC flow, dark grey $\epsilon^\prime_{11}$, medium grey $\epsilon^\prime_{22}$, light grey $\epsilon^\prime_{33}$, (b) distribution of off-diagonal reoriented rate of strain components $\boldsymbol\epsilon^\prime_{ij}(t)$ for the ABC flow, dark grey $\epsilon^\prime_{12}$, medium grey $\epsilon^\prime_{13}$, light grey $\epsilon^\prime_{23}$.}\label{fig:ABCpdf}
\end{figure}

Probability density functions (PDFs) of the non-zero components of $\boldsymbol\epsilon^\prime(t)$ over 10,000 streamlines in the ergodic region of the ABC flow are shown in Figure~\ref{fig:ABCpdf}, these correspond to spatial distributions throughout this region due to ergodicity. To within numerical precision, the diagonal components $\epsilon^\prime_{ii}(t)$ of this incompressible flow have mean stretching rates of $\lambda_{\infty,i}=\{0,\lambda,-\lambda\}$, $\lambda\approx 0.0543$ which agrees very favourably with the infinite-time Lyapunov exponent computed as $\mu_{\infty}\approx 0.0548$. 

The deformation structure in Figure~\ref{fig:ABCpdf}(a) indicates that the velocity fluctuations associated with $\epsilon^\prime_{11}(t)$ is highly peaked around zero, suggesting weak acceleration of the flow. Conversely, the transverse deformations $\epsilon^\prime_{22}(t)$,  $\epsilon^\prime_{33}(t)$, are more broadly distributed throughout the ergodic region. Distributions of the off-diagonal components $\epsilon^\prime_{ij}(t)$ in Figure~\ref{fig:ABCpdf}(b) indicate the stream-wise components $\epsilon^\prime_{12}(t)$, $\epsilon^\prime_{13}(t)$ have zero mean, as is expected due to ergodicity. Conversely, the transverse component $\epsilon^\prime_{23}(t)$ is consistently negative with mean -1.166, reflecting the strong helical component of Beltrami flows such as the ABC flow. In contrast with random ergodic flows, where $\epsilon^\prime_{23}(t)$ will typically have zero mean due to stationary, this component may have non-zero mean as this chaotic flow is deterministic yet gives rise to ergodic trajectories.

We find the components of $\boldsymbol\epsilon^\prime(t)$ are all uncorrelated, with the exception of diagonal elements which are correlated as $r(\epsilon^\prime_{11},\epsilon^\prime_{22})=0.453$, $r(\epsilon^\prime_{11},\epsilon^\prime_{33})=-0.542$, $r(\epsilon^\prime_{22},\epsilon^\prime_{33})=-0.993$, a consequence of the volume-preserving nature of the ABC flow. The simplicity of the deformation structure of the ABC flow indicates the feasibility of developing stochastic models for evolution of the deformation tensor and FTLE PDFs. Whilst this development is beyond the scope of this study, these distributions above form the building blocks of such models, and clearly illustrate how the Eulerian flow features govern Lagrangian fluid deformation. %This approach also clearly elucidates the how physical flow properties and topological constraints such as incompressibility, non-zero helicity, ergodicity and stationarity directly impact fluid deformation and restrict the classes of deformation evolution.

\subsection{3D Kraichnan Flow}

\begin{figure}
\begin{centering}
\begin{tabular}{cc}
\includegraphics[width=0.45\columnwidth]{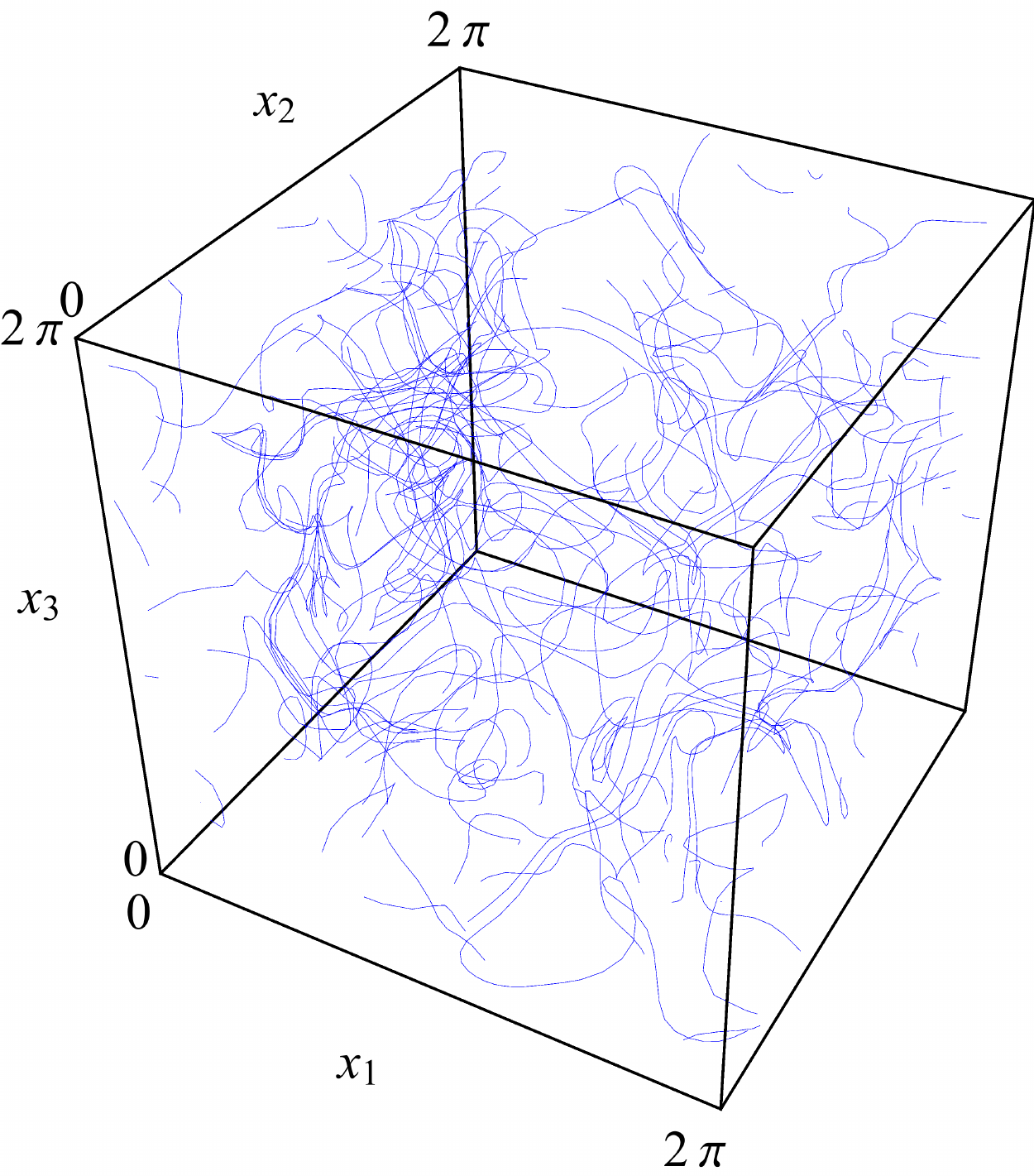}&
\includegraphics[width=0.5\columnwidth]{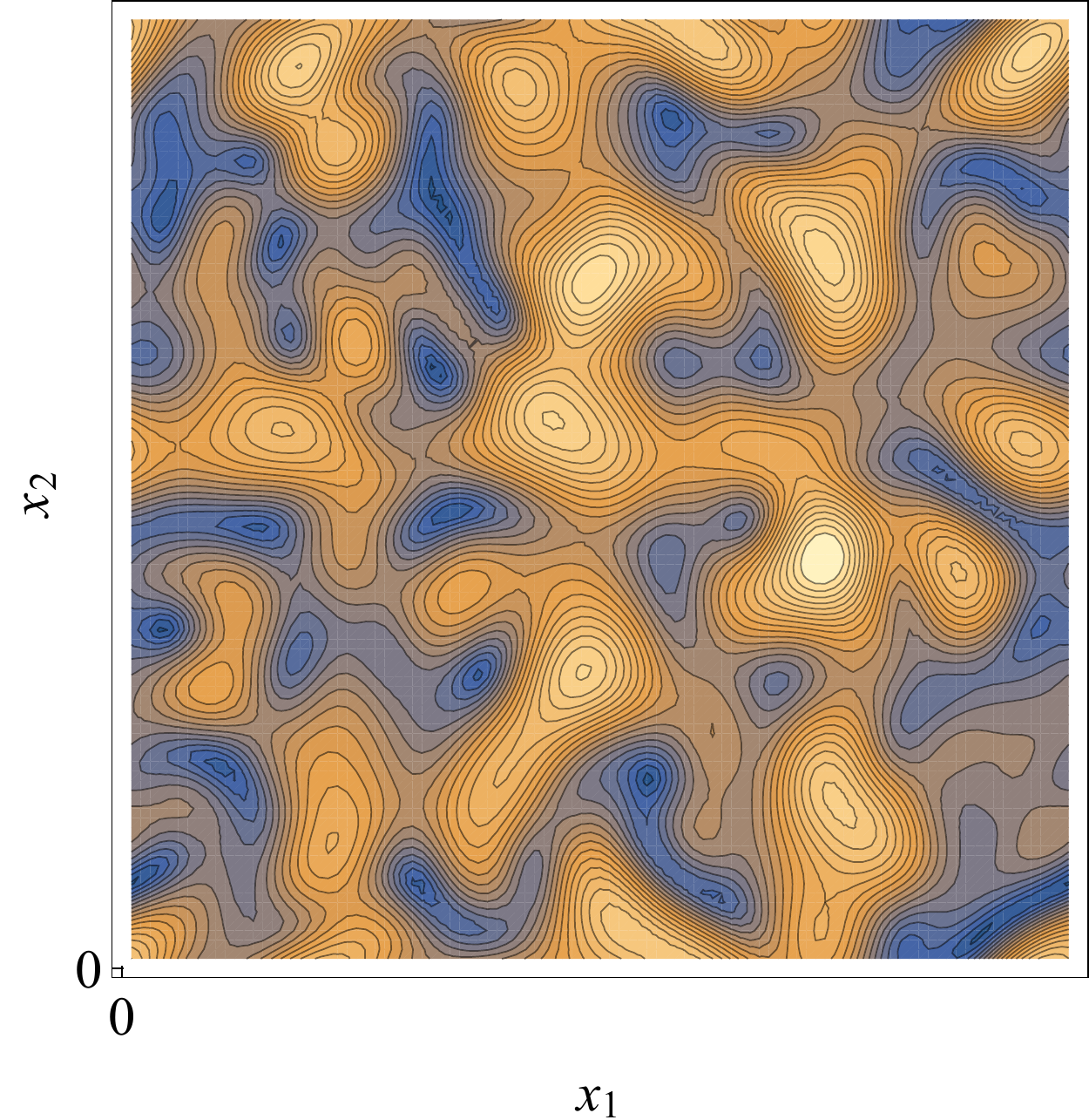}\\
(a) & (b)
\end{tabular}
\end{centering}
\caption{(a) Typical particle trajectory in a realisation of the Kraichnan flow, (b) contour plot of velocity magnitude distribution in $x_3=0$ plane.}\label{fig:Kraichnan}
\end{figure}

\begin{figure}
\begin{centering}
\begin{tabular}{c c}
\includegraphics[width=0.45\columnwidth]{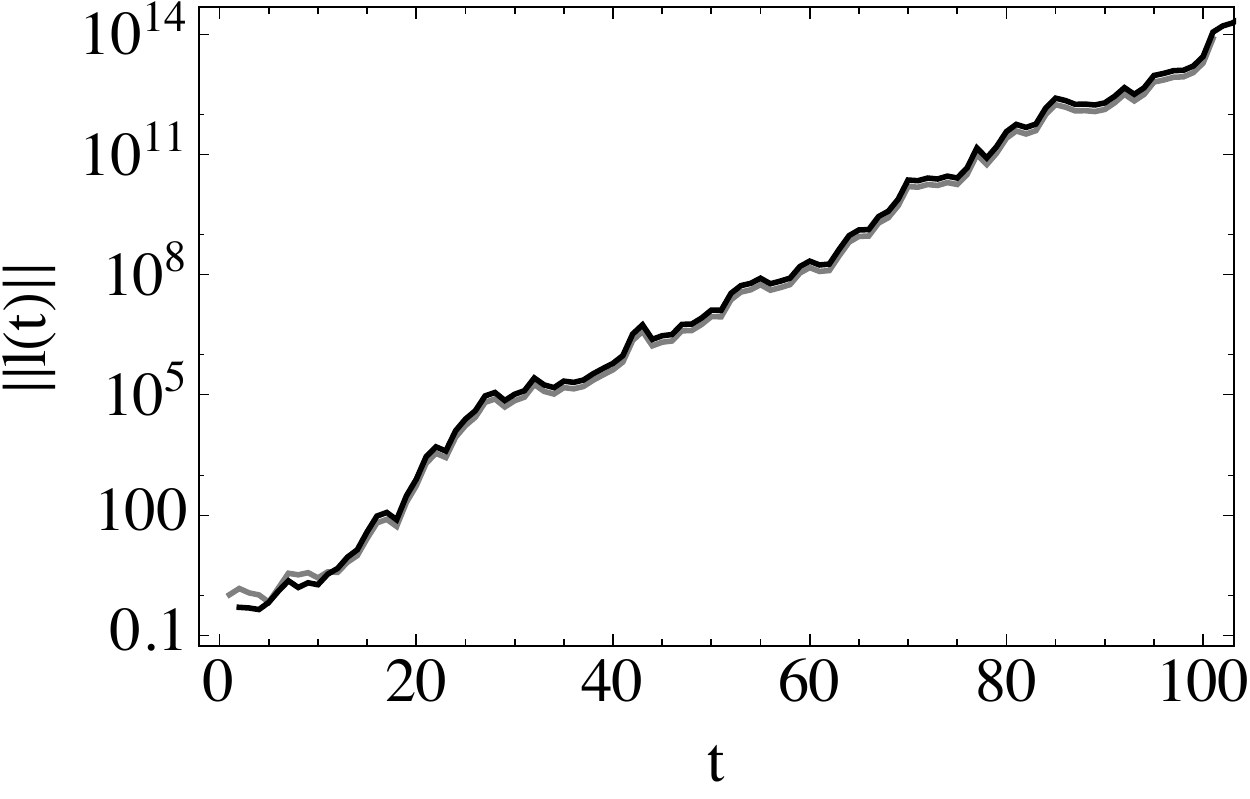}&
\includegraphics[width=0.45\columnwidth]{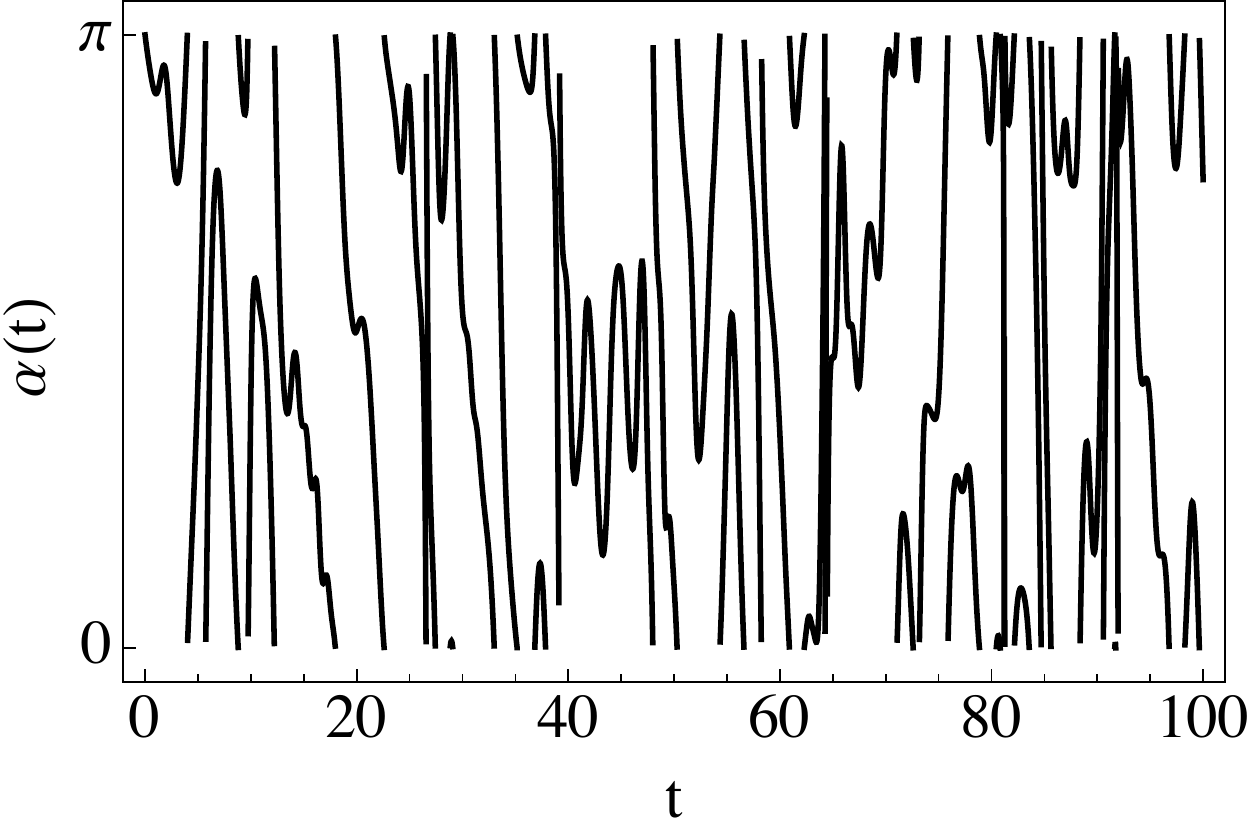}\\
(a) & (b)%\\
%\includegraphics[width=0.45\columnwidth]{detF_Kraichnan_inertial_pdfs.pdf}&
%\includegraphics[width=0.45\columnwidth]{lyapunov_Kraichnan_inertial.pdf}\\
%(c) & (d)
\end{tabular}
\end{centering}
\caption{(a) Relative growth of the length $|\mathbf{l}(t)|$ of an infinitesimal material line along a single trajectory in the Kraichnan flow (\ref{eqn:Kraichnanfield}) calculated by (grey) particle tracking and (black) from $\mathbf{l}(t)=\mathbf{F}^\prime(t)\cdot\mathbf{l}(0)$. %(b) Solution of the orientation angle $\alpha(t)$ along the inertial manifold $\mathcal{M}$. (c) Determinant error for the (grey) Cartesian $\mathbf{F}(t)$ and (black) Protean $\mathbf{F}^\prime(t)$ deformation tensors. (d) Convergence of the principal stretching exponent $\lambda(t,\mathbf{X})$ to the FTLE $\mu(t)$.
}\label{fig:plot1_Kraichnan}
\end{figure}

We also apply the method to a 3D Kraichnan flow, a random model flow used extensively in the study of transport in homogeneous isotropic turbulence~\cite{Kraichnan:1970aa}. A variant of this incompressible flow with non-zero helicity density is given by the velocity field
\begin{align}
\mathbf{v}(x,y,z)=\nabla\times\psi_1(x,y)+\nabla\times\psi_2(x,z)+\nabla\times\psi_3(y,z),\label{eqn:Kraichnanfield}
\end{align}
where the 2D streamfunction is given by
\begin{equation}
\begin{split}
&\psi_{i}(x_1,x_2)=\\
&\sum_{m=1}^M\sum_{n=1}^N\frac{A^i_{m,n}}{m^2+n^2}\cos m(x_1+\phi^{1,i}_{m,n})\cos n(x_2+\phi^{2,i}_{l,m,n}),
\end{split}
\end{equation}
and $A^i_{m,n}$, $\phi^i_{m,n}$ are uniformly distributed independent random variables in $[-1,1]$ and $[0,2\pi]$ respectively and modes up to $M$=$N$=5 are used. A typical particle trajectory and the basic flow structure is shown in Figure~\ref{fig:Kraichnan}. The Kraichnan flow possess non-zero helicity density and significant fluid stretching as per Figure~\ref{fig:plot1_Kraichnan}(a), which consists of punctuated stretching events which lead to minimal persistent stretching. Rapid reorientation of the transverse angle $\alpha(t)$ driven by non-zero helicity density is reflected in Figure~\ref{fig:plot1_Kraichnan}(b). 
%For this incompressible flow, the volumetric errors for the Cartesian and Protean deformation tensors are both small. The strong exponential stretching exhibited by this flow leads to rapid convergence between the principal stretching rate $\lambda(t,\mathbf{X})$ and the FTLE $\mu(t,\mathbf{X})$ along a particle trajectory.

\begin{figure}
\begin{centering}
\begin{tabular}{c c}
\includegraphics[width=0.45\columnwidth]{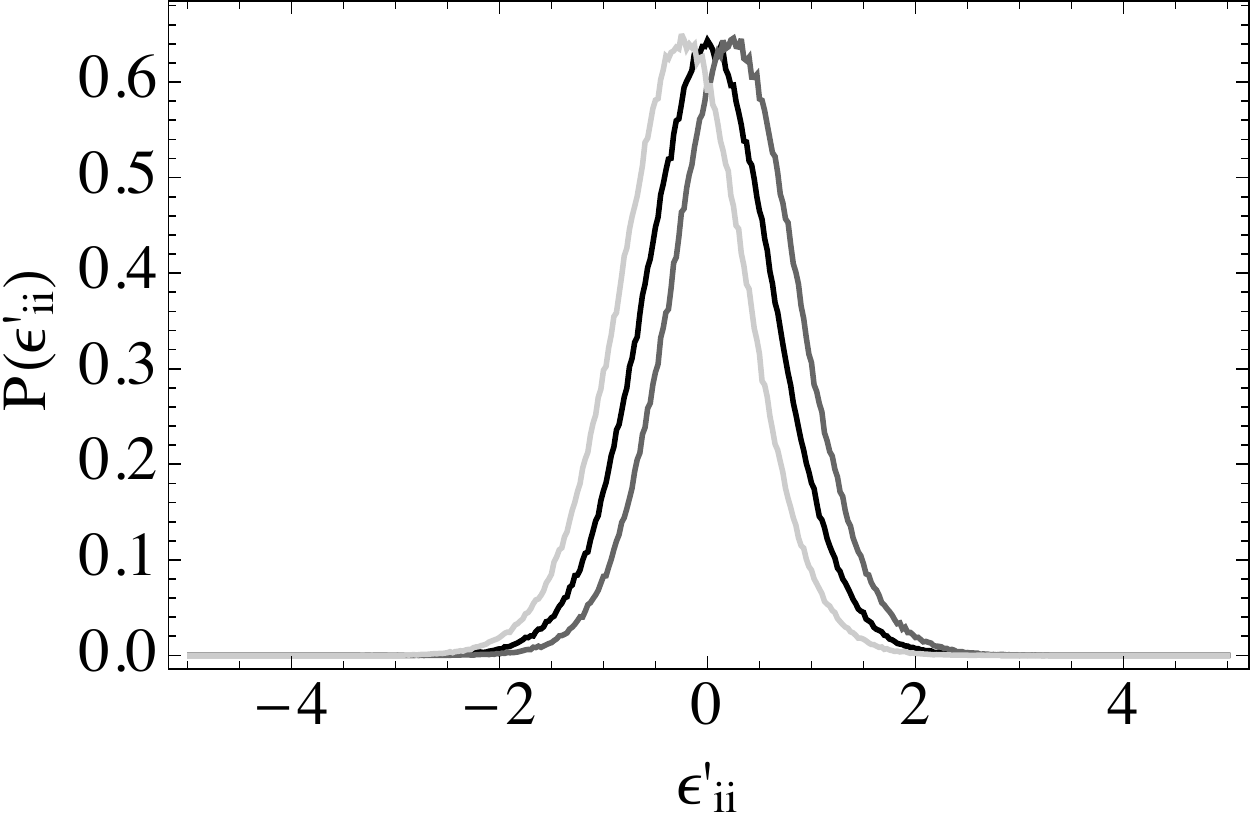}&
\includegraphics[width=0.45\columnwidth]{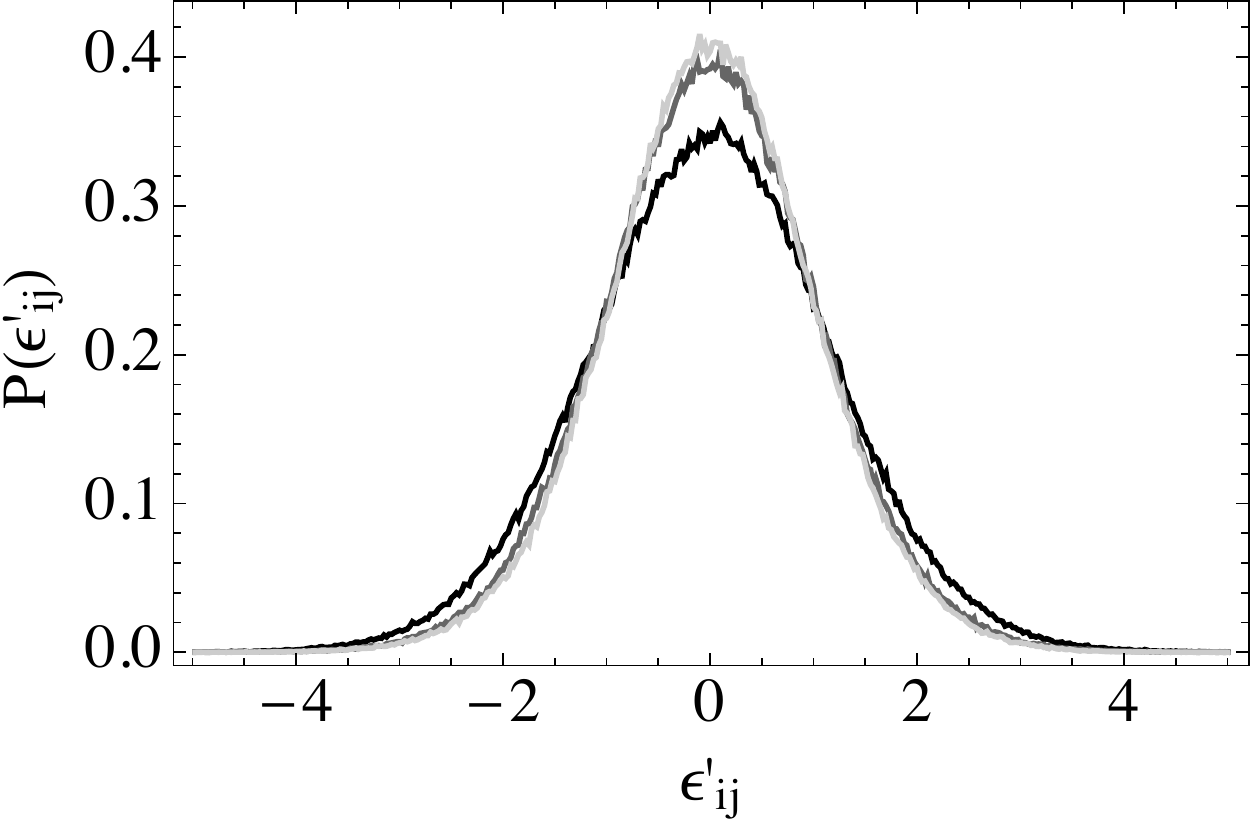}\\
(a) & (b)
\end{tabular}
\end{centering}
\caption{(a) Distribution of diagonal reoriented velocity gradient components $\boldsymbol\epsilon^\prime_{ii}(t)$ for the Kraichnan flow, dark grey $\epsilon^\prime_{11}$, medium grey $\epsilon^\prime_{22}$, light grey $\epsilon^\prime_{33}$, (b) distribution of off-diagonal reoriented rate of strain components $\boldsymbol\epsilon^\prime_{ij}(t)$ for the Kraichnan flow, dark grey $\epsilon^\prime_{12}$, medium grey $\epsilon^\prime_{13}$, light grey $\epsilon^\prime_{23}$.}\label{fig:plot2_Kraichnan}
\end{figure}

As shown in Fig~\ref{fig:plot2_Kraichnan}, the deformation structure of the Kraichnan flow is particularly simple as all of the Protean rate of strain components are Gaussian distributed as a consequence of ergodicity and the Central Limit Theorem. Analysis of the normalized moments of all of these distributions indicate that all odd moments are statistically insignificant, and the 4th and 6th moments agree with those of a Gaussian distrubtion to within 10\% and 20\% respectively. The diagonal components $\epsilon^\prime_{ii}(t)$ and the off-diagonal components $\epsilon^\prime_{ij}(t)$ all appear to have the same variance, and the diagonal components have mean $\lambda_{\infty,i}=\{0,\lambda,-\lambda\}$ with $\lambda\approx 0.23$, indicating significant exponential fluid stretching. Conversely, the off-diagonal components have zero mean, hence net fluid stretching arises from combination of these principal deformations and non-trivial correlations between the diagonal and off-diagonal stretching components. Similar to the ABC flow, all the components of $\boldsymbol\epsilon^\prime(t)$ are uncorrelated except for the diagonal components which are coupled as $r(\epsilon^\prime_{11},\epsilon^\prime_{22})=-0.503$, $r(\epsilon^\prime_{11},\epsilon^\prime_{33})=-0.503$, $r(\epsilon^\prime_{22},\epsilon^\prime_{33})=-0.493$ as consequence of incompressibility. This remarkably simple deformation structure of the 3D Kraichnan flow in Protean coordinates indicates that deformation in this flow can be fully characterised in terms of only a handful of parameters, hence it appears that prediction of the deformation tensor PDF is feasible via random walk models.

\subsection{Dual Stream Function Flow}

\begin{figure}
\begin{centering}
\begin{tabular}{cc}
\includegraphics[width=0.44\columnwidth]{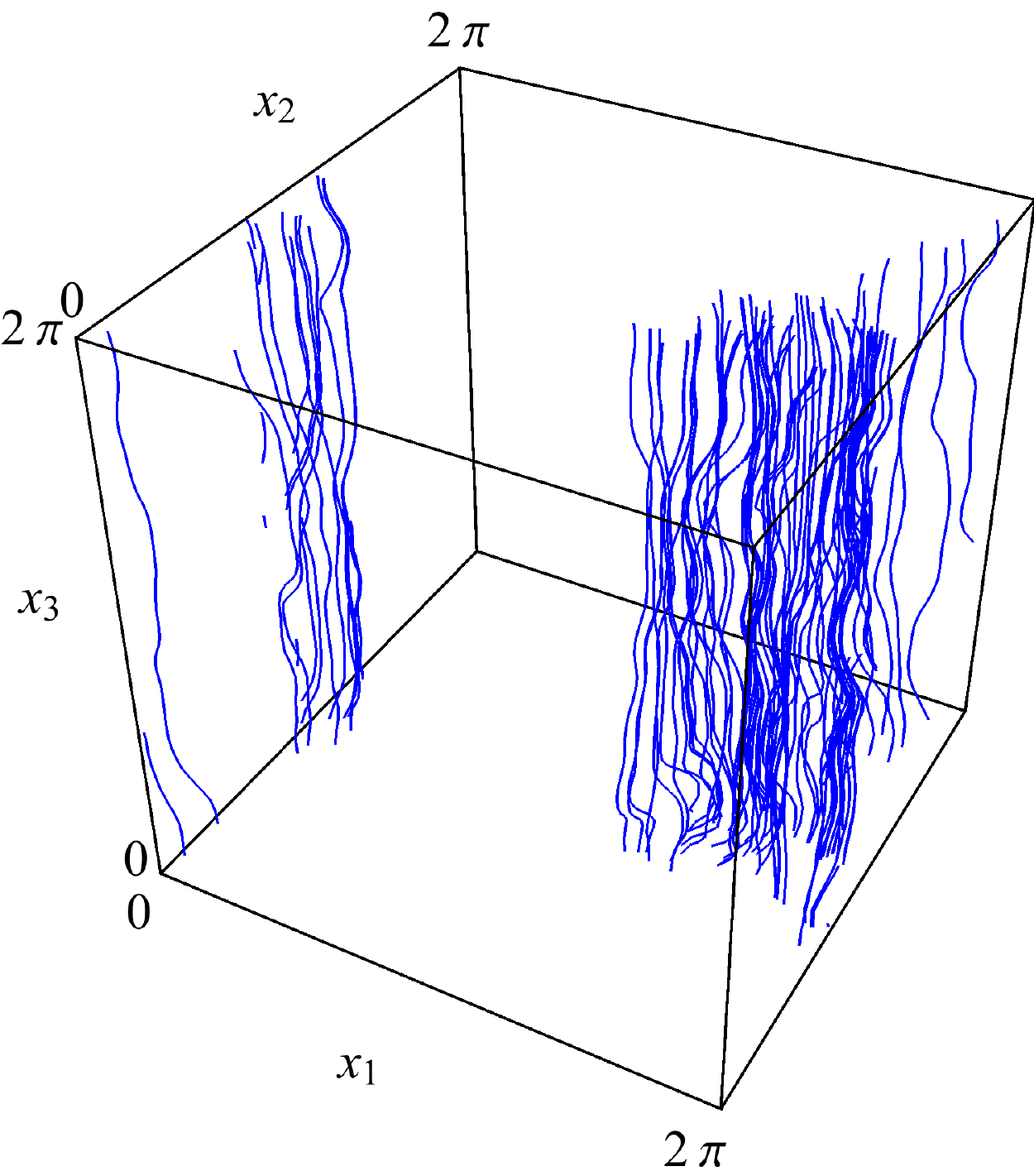}&
\includegraphics[width=0.5\columnwidth]{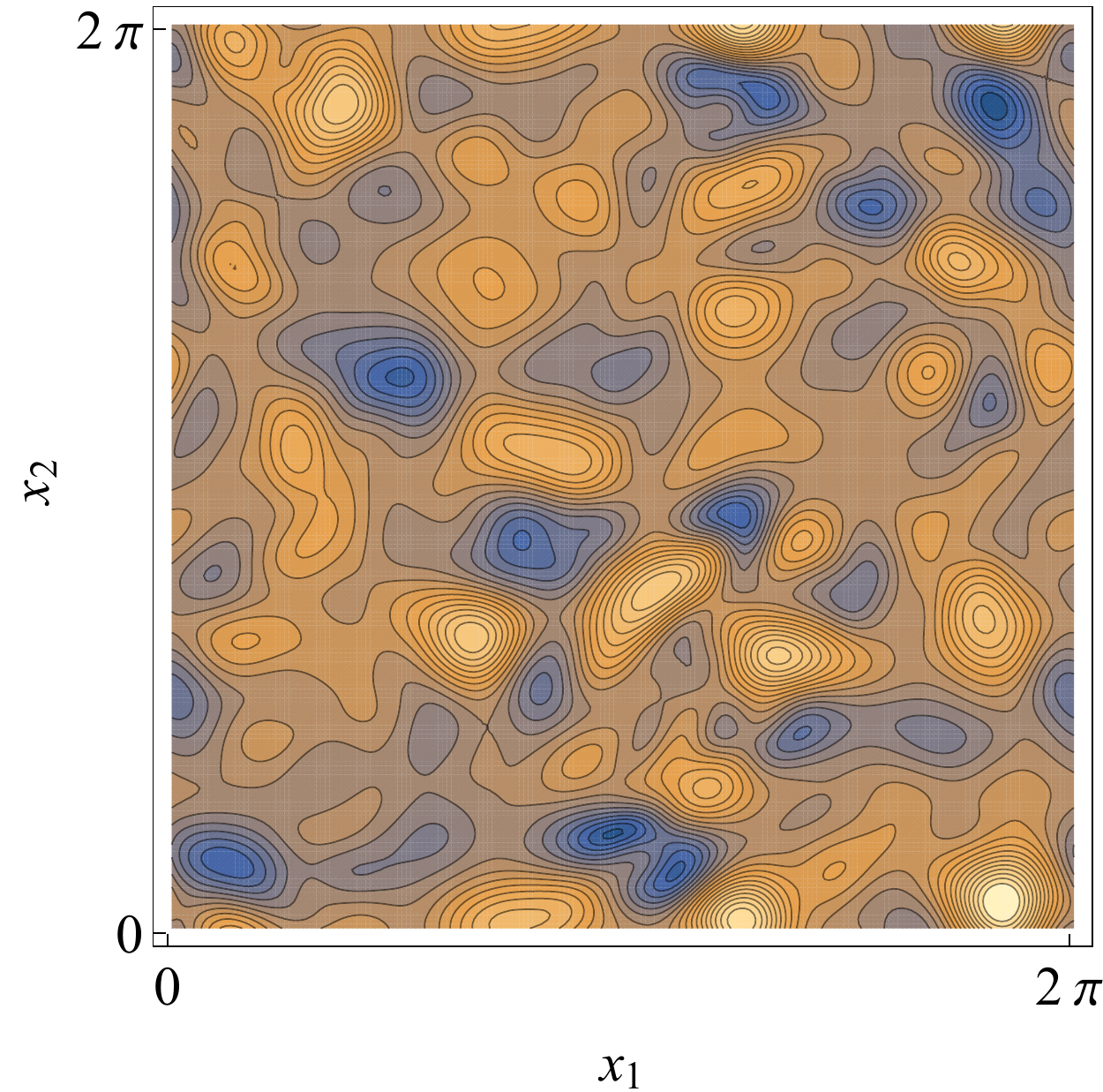}\\
(a) & (b)
\end{tabular}
\end{centering}
\caption{(a) Typical particle trajectory in a realisation of the After flow, (b) contour plot of velocity magnitude distribution in $x_3=0$ plane.}\label{fig:Arter}
\end{figure}

\begin{figure}
\begin{centering}
\begin{tabular}{c c}
\includegraphics[width=0.45\columnwidth]{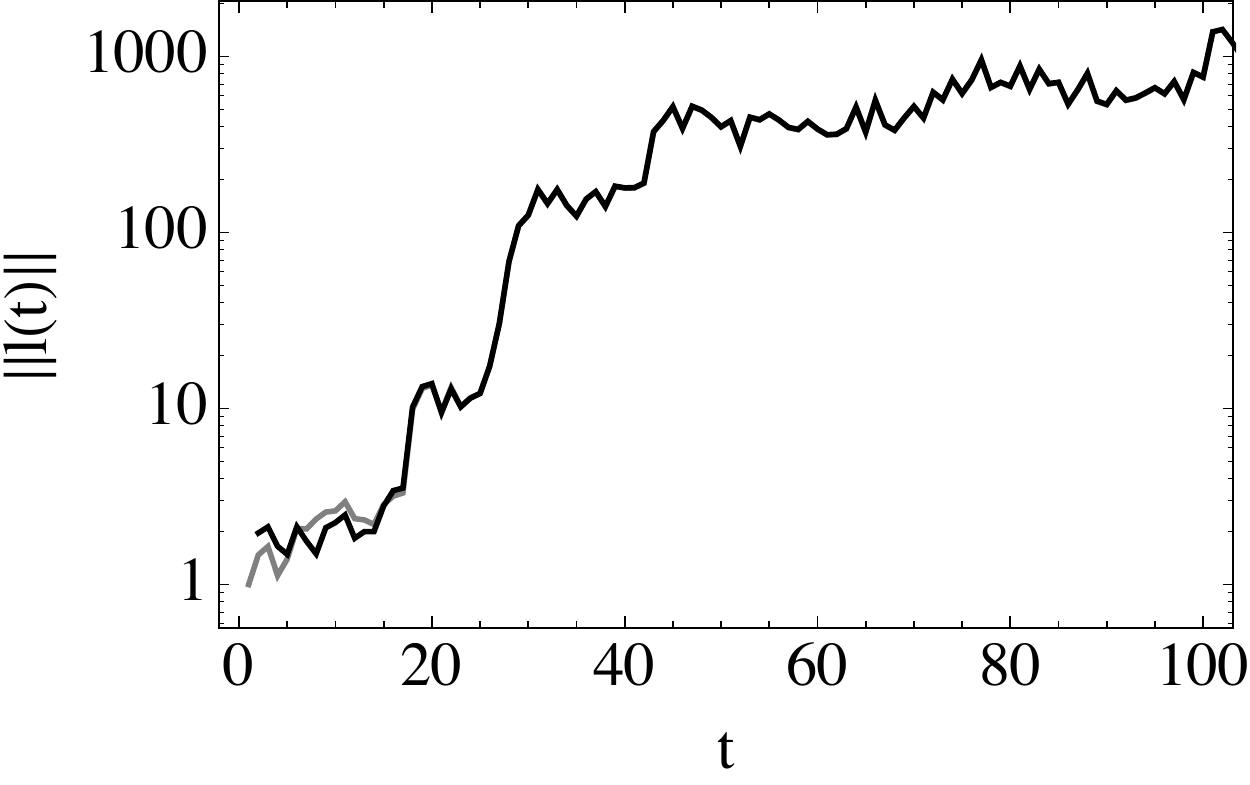}&
\includegraphics[width=0.45\columnwidth]{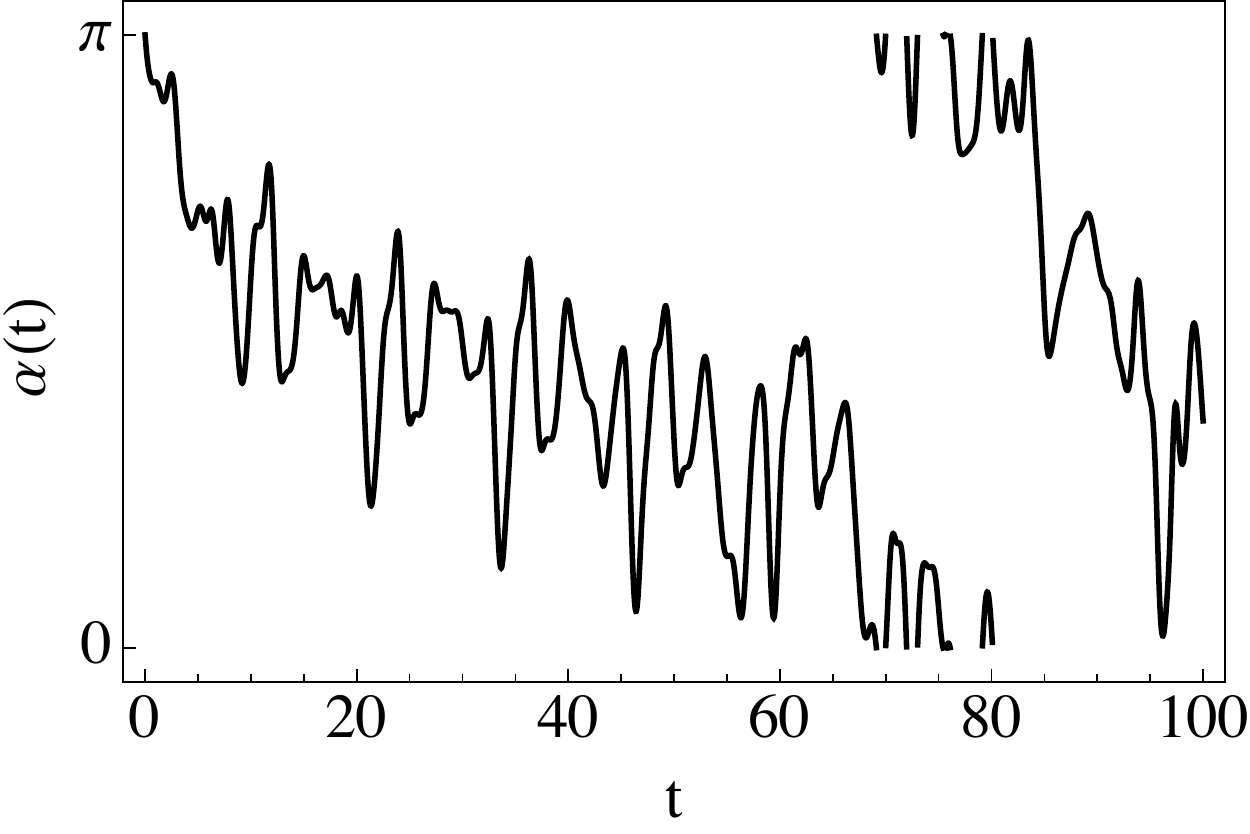}\\
(a) & (b)
%\\
%\includegraphics[width=0.45\columnwidth]{detF_Dual_inertial_pdfs.pdf}&
%\includegraphics[width=0.45\columnwidth]{lyapunov_Dual_inertial.pdf}\\
%(c) & (d)
\end{tabular}
\end{centering}
\caption{(a) Relative growth of the length $|\mathbf{l}(t)|$ of an infinitesimal material line along a single trajectory in the dual streamfunction flow (\ref{eqn:Arterflow}) calculated by (grey) particle tracking and (black) from $\mathbf{l}(t)=\mathbf{F}^\prime(t)\cdot\mathbf{l}(0)$. (b) Solution of the orientation angle $\alpha(t)$ along the inertial manifold $\mathcal{M}$. 
%(c) Determinant error for the (grey) Cartesian $\mathbf{F}(t)$ and (black) Protean $\mathbf{F}^\prime(t)$ deformation tensors. (d) Convergence of the principal stretching exponent $\lambda(t,\mathbf{X})$ to the FTLE $\mu(t)$.
}\label{fig:plot1_Arter}
\end{figure}

Here we apply the Protean transform to a random 3D flow described by two stream functions. This incompressible flow with non-zero helicity density is given by the velocity field
\begin{align}
\mathbf{v}(x,y,z)=\nabla\psi_1\times\nabla\psi_2,\label{eqn:Arterflow}
\end{align}
where
\begin{equation}
\begin{split}
&\psi_{i}(x,y,z)=\sum_{l=1}^L\sum_{m=1}^M\sum_{n=1}^N\frac{A^i_{l,m,n}}{l^2+m^2+n^2}\cos l(x+\phi^{x,i}_{l,m,n})\\
&\cos m(y+\phi^{y,i}_{l,m,n})\cos n(z+\phi^{z,i}_{l,m,n}),\,\,i=(1,2),
\end{split}\label{eqn:psi}
\end{equation}
and $A^i_{l,m,n}$, $\phi^i_{l,m,n}$ are uniformly distributed independent random variables in $[-1, 1]$ and $[0,2\pi]$ respectively, and modes up to $L$=$M$=$N$=5 are used. A typical particle trajectory and the basic flow structure is shown in Figure~\ref{fig:Arter}. As the stream functions $\psi_1$ and $\psi_2$ are invariants of the flow, streamlines are given by the intersection of level surfaces of $\psi_1$, $\psi_2$. Although this flow is helical and chaotic, fluid stretching is much smaller than the ABC or Kraichnan flows as per Figure~\ref{fig:plot1_Arter}(a), consisting of punctuated stretching events which lead to minimal persistent stretching. The rapid reorientation of transverse angle $\alpha(t)$ driven by non-zero helicity density is reflected in Figure~\ref{fig:plot1_Arter}(b). 
%For this incompressible flow, the volumetric errors for the Eulerian and Protean deformation tensors are both small, and the relative error for Cauchy-Green eigenvalue between these frames are also small. Similarly, whilst errors for the material line length $l(t)$ and reoriented deformation tensor $\mathbf{F}(t)$ initially grow rapidly, long-term growth is limited.

Similar to the Kraichnan flow, the deformation structure of the dual streamfunction flow is particularly simple, consisting of Gaussian distributed Protean rate of strain components with the same variance amongst the diagonal and off-diagonal components. The off-diagonal components all have zero mean, with $\epsilon_{12}^\prime$ and $\epsilon_{13}^\prime$ having zero mean due to ergodicity, whilst $\epsilon_{23}^\prime$ has zero mean due to the random structure of the dual streamfunction flow. The diagonal components of $\boldsymbol\epsilon^\prime(t)$ have mean the form $\lambda_{\infty,i}=\{0,\lambda,-\lambda\}$ with significantly weaker exponential stretching exhibited as $\lambda\approx 0.0816$, again of similar magnitude to the Lyapunov exponent $\mu\approx 0.0813$ of the flow. As per the ABC and 3D Kraichnan flows, only the diagonal components of $\boldsymbol\epsilon^\prime(t)$ for the dual streamfunction flow are correlated due to incompressibility as $r(\epsilon^\prime_{11},\epsilon^\prime_{22})=-0.520$, $r(\epsilon^\prime_{11},\epsilon^\prime_{33})=-0.528$, $r(\epsilon^\prime_{22},\epsilon^\prime_{33})=-0.450$.

\begin{figure}
\begin{centering}
\begin{tabular}{c c}
\includegraphics[width=0.45\columnwidth]{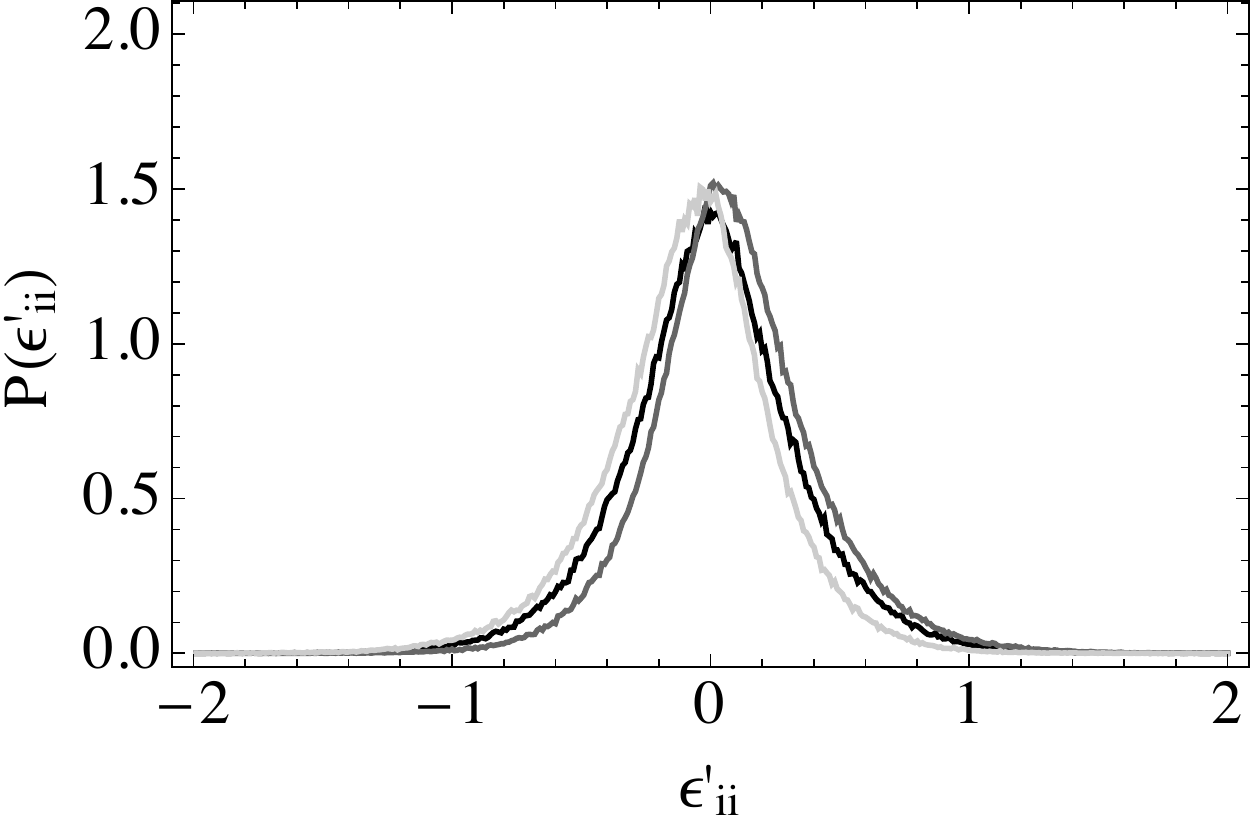}&
\includegraphics[width=0.45\columnwidth]{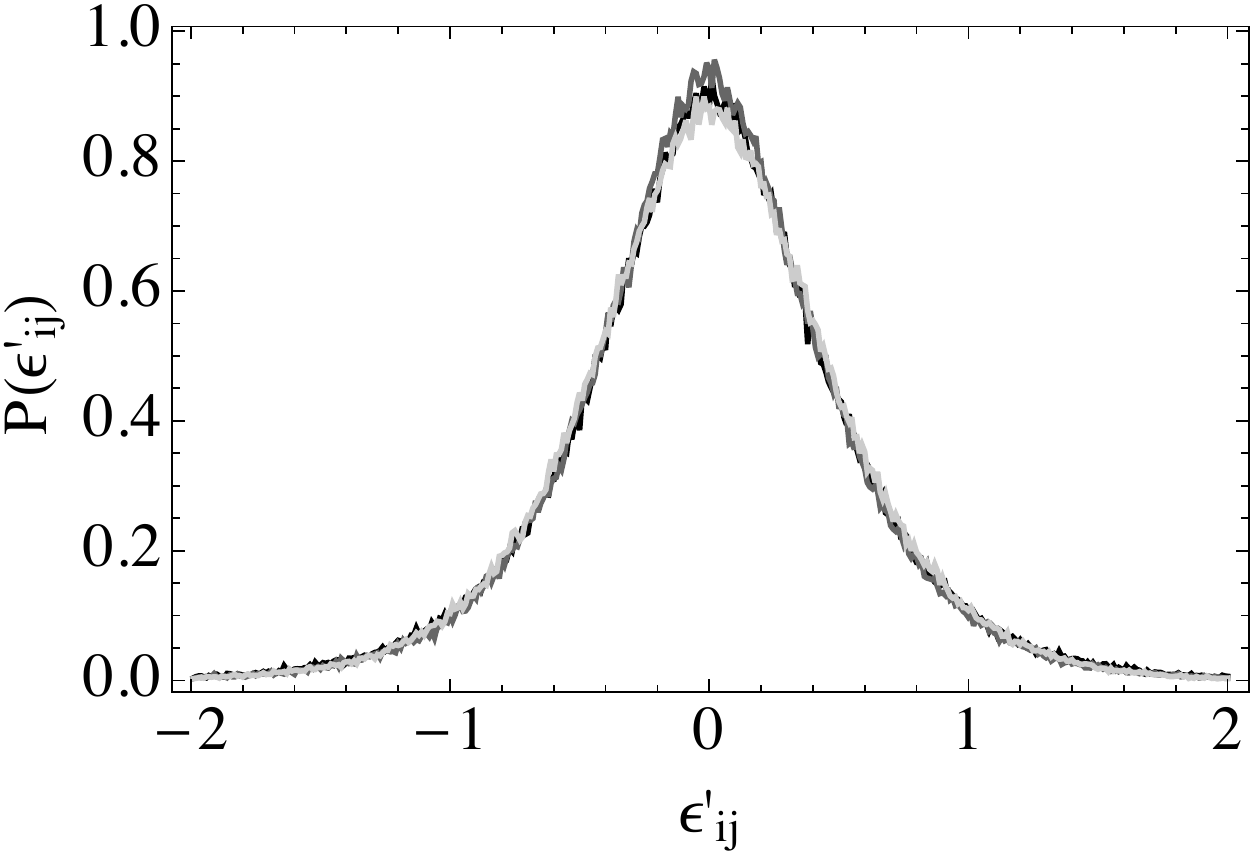}\\
(a) & (b)
\end{tabular}
\end{centering}
\caption{(a) Distribution of diagonal reoriented rate of strain components $\boldsymbol\epsilon^\prime_{ii}(t)$ for the dual streamfunction flow, dark grey $\epsilon^\prime_{11}$, medium grey $\epsilon^\prime_{22}$, light grey $\epsilon^\prime_{33}$, (b) distribution of off-diagonal reoriented rate of strain components $\boldsymbol\epsilon^\prime_{ij}(t)$ for the dual streamfunction flow, dark grey $\epsilon^\prime_{12}$, medium grey $\epsilon^\prime_{13}$, light grey $\epsilon^\prime_{23}$.}\label{fig:plot2_Arter}
\end{figure}

\subsection{Random Potential Flow}

\begin{figure}
\begin{centering}
\begin{tabular}{cc}
\includegraphics[width=0.45\columnwidth]{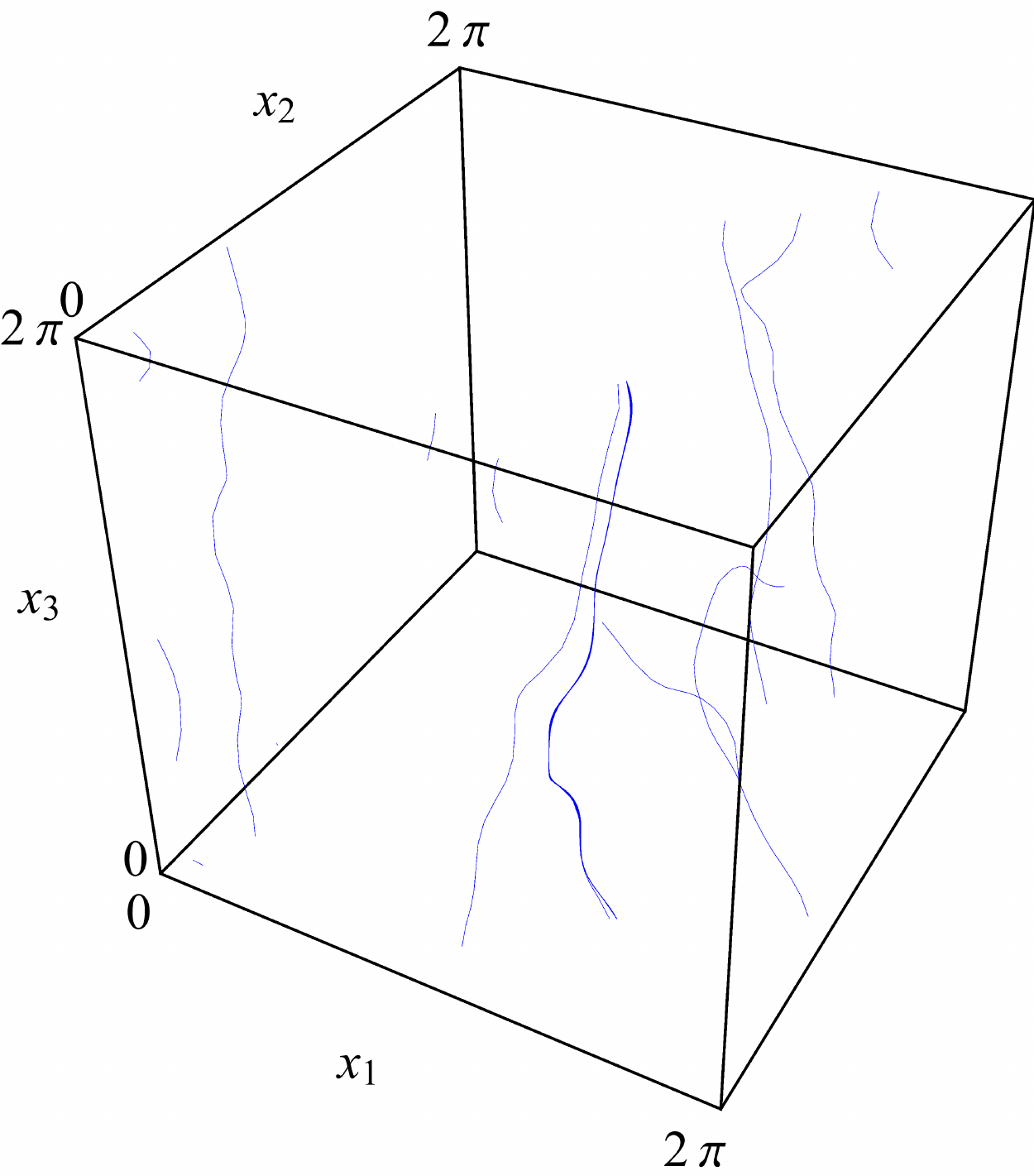}&
\includegraphics[width=0.5\columnwidth]{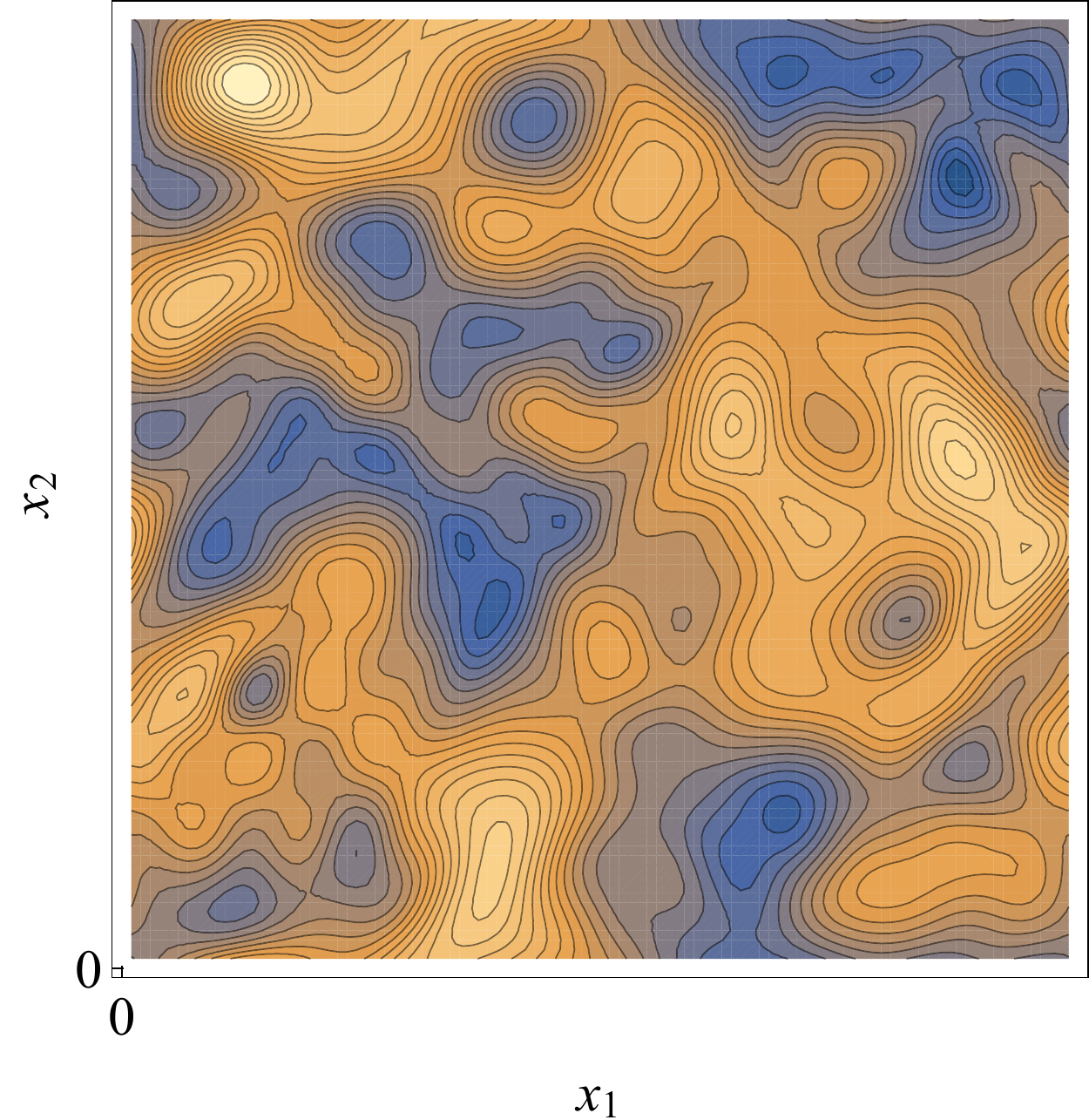}\\
(a) & (b)
\end{tabular}
\end{centering}
\caption{(a) Typical particle trajectory in a realisation of the 3D random potential flow, (b) contour plot of velocity magnitude distribution in $x_3=0$ plane.}\label{fig:Potential}
\end{figure}

\begin{figure}
\begin{centering}
\begin{tabular}{c c}
\includegraphics[width=0.45\columnwidth]{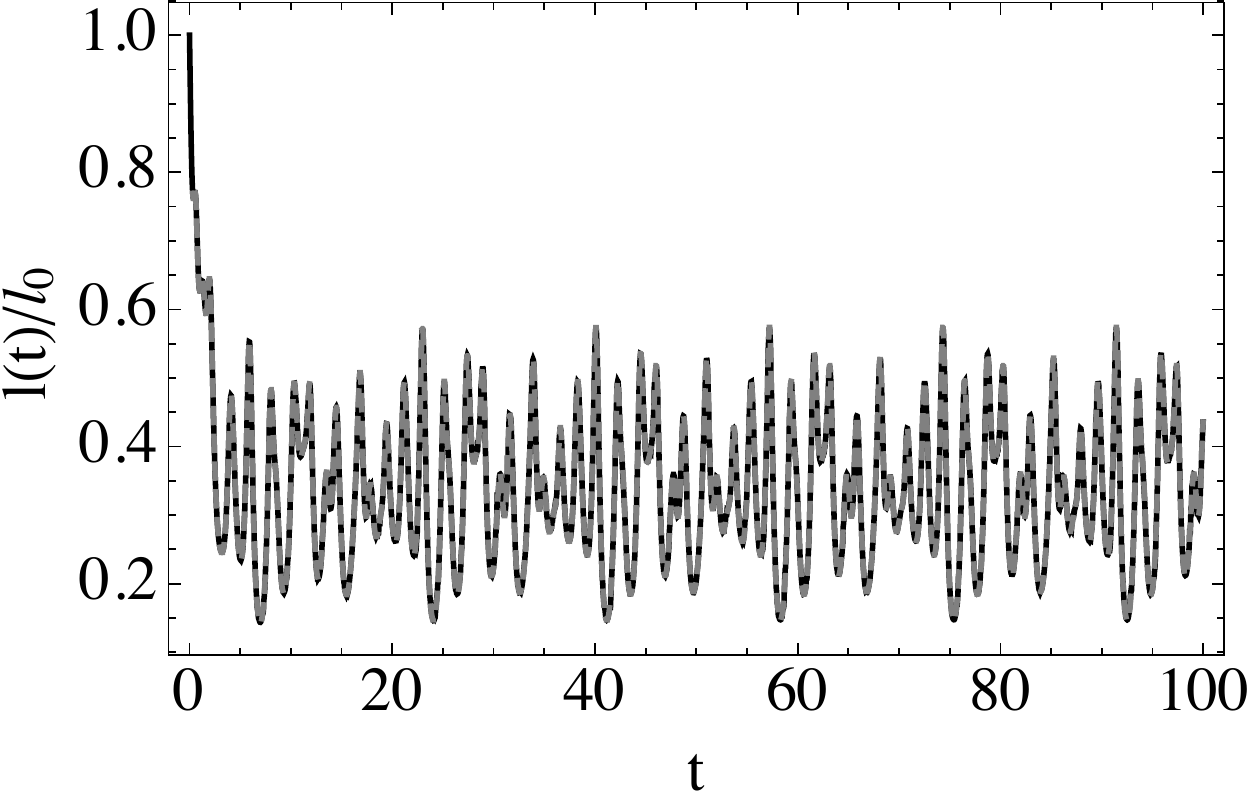}&
\includegraphics[width=0.45\columnwidth]{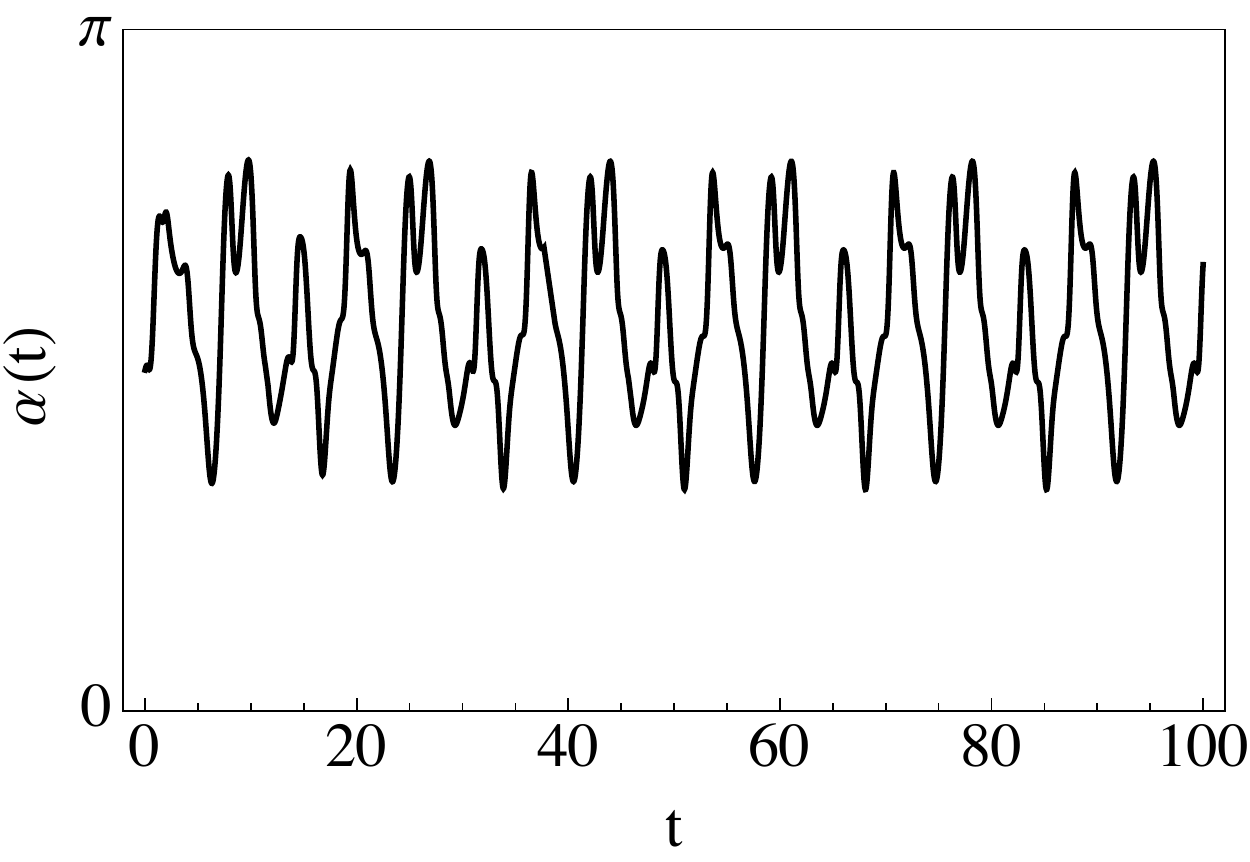}\\
(a) & (b)
%\\
%\includegraphics[width=0.45\columnwidth]{DetF_3D_potential2.pdf}&
%\includegraphics[width=0.45\columnwidth]{F_t__err_3D_potential2.pdf}\\
%(c) & (d)
\end{tabular}
\end{centering}
\caption{(a) Relative growth of the length $|\mathbf{l}(t)|$ of an infinitesimal material line along a single trajectory in the 3D potential flow (\ref{eqn:Arterflow}) calculated by (grey) particle tracking and (black) from $\mathbf{l}(t)=\mathbf{F}^\prime(t)\cdot\mathbf{l}(0)$. (b) Solution of the orientation angle $\alpha(t)$ along the inertial manifold $\mathcal{M}$. 
%(c) Determinant error for the (grey) Cartesian $\mathbf{F}(t)$ and (black) Protean $\mathbf{F}^\prime(t)$ deformation tensors. (d) Error between the Cartesian $\mathbf{F}(t)$ and Protean $\mathbf{F}^\prime(t)$ deformation tensors.
}\label{fig:plot1_potential}
\end{figure}

Finally, we apply the Protean transform to the 3D random potential flow which is both compressible and irrotational, and so possesses identically zero helicity density. This flow is given by the superposed unidirectional and random velocity fields
\begin{align}
\mathbf{v}(x,y,z)=V_0 z + \nabla\Phi,\label{eqn:potential_flow}
\end{align}
where
\begin{equation}
\begin{split}
&\Phi_(x,y,z)=\sum_{l=1}^L\sum_{m=1}^M\sum_{n=1}^N\frac{A_{l,m,n}}{l^2+m^2+n^2}\cos l(x+\phi^{x}_{l,m,n})\\
&\cos m(y+\phi^{y}_{l,m,n})\cos n(z+\phi^{z}_{l,m,n}),\,\,i=(1,2),
\end{split}\label{eqn:psi2}
\end{equation}
and $A_{l,m,n}$, $\phi_{l,m,n}$ are uniformly distributed independent random variables in $[-1, 1]$ and $[0,2\pi]$ respectively, and modes up to $L$=$M$=$N$=5 are used. The unidirectional background flow magnitude is set as $V_0=1$. As per Figure~\ref{fig:plot1_potential}(a), zero helicity density imparts much weaker algebraic fluid stretching. Note that whilst this flow is non-helical, as it is irrotational it does not admit Lamb surfaces, and so does not possess an orthogonal material coordinate system such as (\ref{eqn:Darcycross}), and so the reoriented Protean coordinate system given by (\ref{fig:alpha}) for helical flows is used for this flow. Whilst the divergence of this flow has zero mean, particle trajectories are attracted to high density regions (where $\nabla\cdot\mathbf{v}<0$), and likewise are repelled from low density regions. This bias causes particle trajectories to experience net compression in the Lagrangian frame, although the density- and flux-weighted average has zero mean by continuity. The non-helical nature of the flow imparts slow reorientation of the transverse angle $\alpha(t)$ as per Figure~\ref{fig:plot1_potential}(b), which now only evolves in response to changes in the fluid structure. %For this incompressible flow, the volumetric errors for the Cartesian and Protean deformation tensors are both small, as well as the overall error in $\mathbf{F}(t)$ between these frames shown in Figure~\ref{fig:plot1_potential}(d). 

Similar to the other random flows, the deformation structure of the 3D potential flow is particularly simple, consisting of Gaussian distributed Protean velocity gradient components with the same variance amongst the diagonal and off-diagonal components. Due to stationarity all of the off-diagonal components have zero mean, and zero helicity density everywhere ensures the diagonal components also all have zero mean. As such, the PDFs of the potential flow are completely characterised by two parameters; the variance of the diagonal and off-diagonal components. Similar to 2D steady random flows~\cite{Dentz:2015aa}, persistent fluid stretching arises from correlations between the diagonal and off-diagonal components of $\boldsymbol\epsilon^\prime(t)$, leading to algebraic fluid stretching. Again only the diagonal components of $\boldsymbol\epsilon^\prime(t)$ are correlated, however more weakly so due to the compressible nature of the flow: $r(\epsilon^\prime_{11},\epsilon^\prime_{22})=0.335$, $r(\epsilon^\prime_{11},\epsilon^\prime_{33})=0.105$, $r(\epsilon^\prime_{22},\epsilon^\prime_{33})=0.273$.

\begin{figure}
\begin{centering}
\begin{tabular}{c c}
\includegraphics[width=0.45\columnwidth]{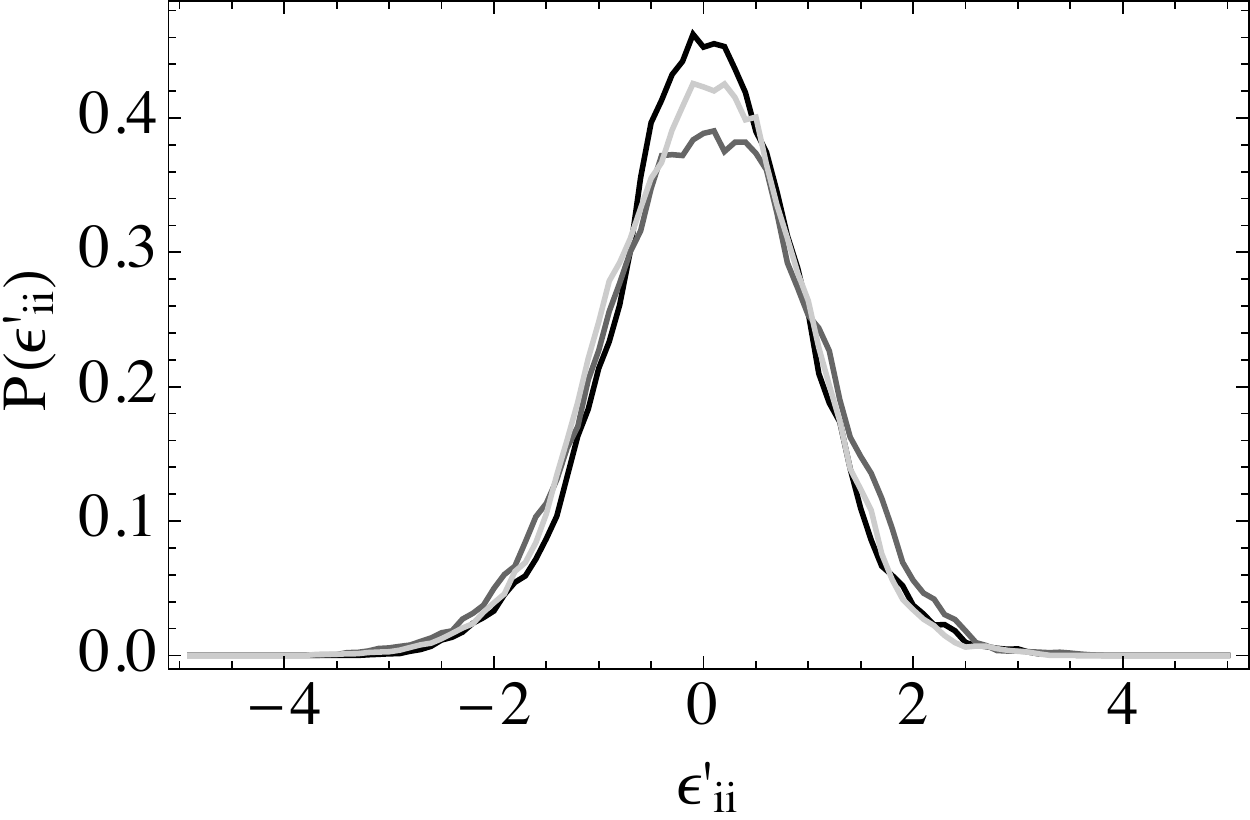}&
\includegraphics[width=0.45\columnwidth]{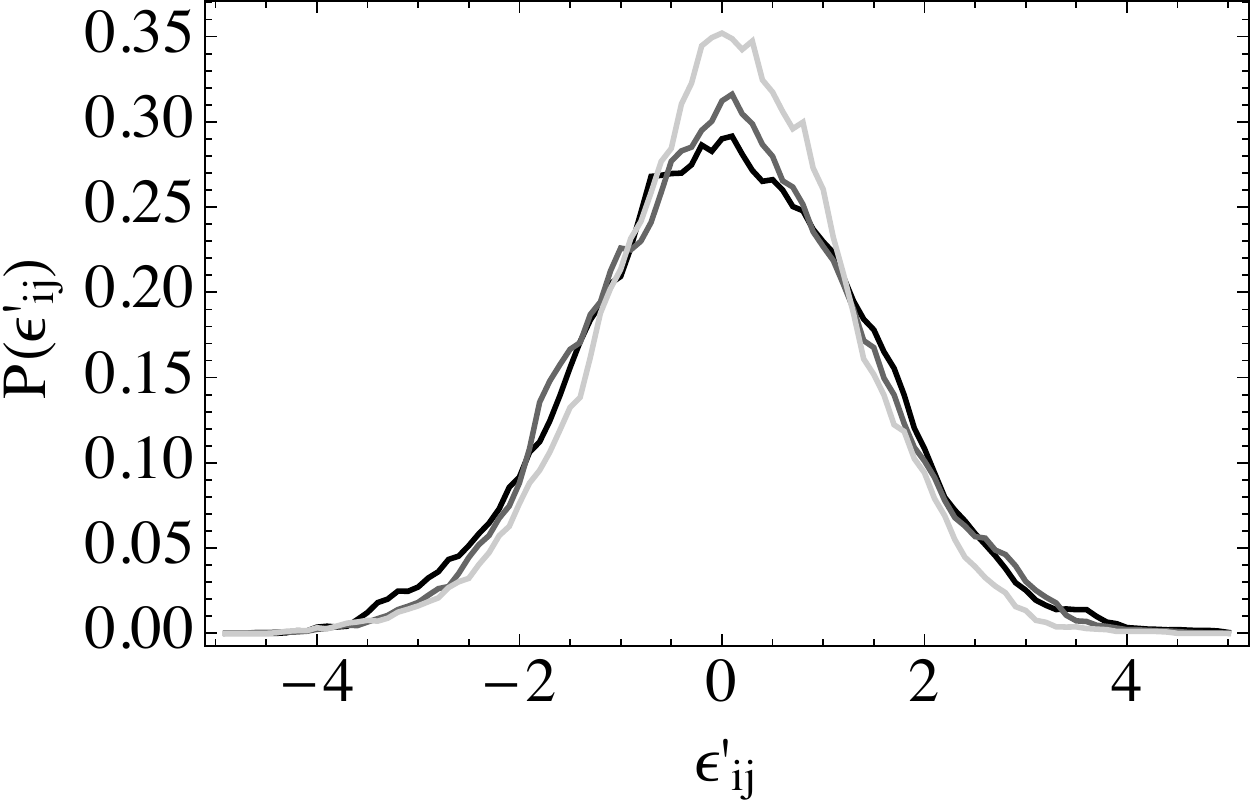}\\
(a) & (b)
\end{tabular}
\end{centering}
\caption{(a) Distribution of diagonal reoriented rate of strain components $\boldsymbol\epsilon^\prime_{ii}(t)$ for the 3D potential flow, dark grey $\epsilon^\prime_{11}$, medium grey $\epsilon^\prime_{22}$, light grey $\epsilon^\prime_{33}$, (b) distribution of off-diagonal reoriented rate of strain components $\boldsymbol\epsilon^\prime_{ij}(t)$ for the 3D potential flow, dark grey $\epsilon^\prime_{12}$, medium grey $\epsilon^\prime_{13}$, light grey $\epsilon^\prime_{23}$.}\label{fig:plot2_Potential}
\end{figure}

\section{Discussion}

Similar to 2D steady flows, reorientation into streamline coordinates yields explicit expressions for the deformation tensor $\mathbf{F}^\prime(t)$ as the series of integrals (\ref{eqn:Fii})-(\ref{eqn:F13}), and the eigenvalues of $\mathbf{F}^\prime(t)$ are simply the diagonal components $F^\prime_{ii}(t)$. The explicit solution $F^\prime_{11}(t)=v(t)/v(0)$ as $\epsilon_{11}(t)=\partial v/\partial x_1^\prime$ is a direct consequence of the steady nature of the flow field. We define the temporal average of the principal deformation along a streamline with Lagrangian coordinate
$\mathbf{X}$ as
\begin{equation}
\lambda_i(t,\mathbf{X}):=\frac{1}{t}\int_0^t dt^\prime \epsilon^\prime_{ii}(t^\prime,\mathbf{X}),\label{eqn:expstretch}
\end{equation}
and the asymptotic limit as $\hat{\lambda}_{\infty,i}(\mathbf{X}):=\lim_{t\rightarrow\infty}\lambda_i(t,\mathbf{X})$. The eigenvalues of $\mathbf{F}^\prime(t)$ are then $F^\prime_{ii}(t)$=$\exp(\lambda_i(t,\mathbf{X}) t)$ and for divergence-free flows $\sum_i\lambda_i(t,\mathbf{X})=0$. Due to conservation of mass, the expansion or compression of all fluid elements is bounded, hence $\hat{\lambda}_{\infty,1}(\mathbf{X})=0$ for both compressible and incompressible flows, with the exception of streamlines which terminate at a stagnation point. As the $2$-direction aligns with the mean fluid stretching then $\hat{\lambda}_{\infty,2}(\mathbf{X})\geqslant 0$, and conversely in the 3-direction $\hat{\lambda}_{\infty,3}(\mathbf{X})\leqslant 0$ due to conservation of mass. For ergodic (mixing) flows such as turbulent, chaotic or random flows, the temporal average $\hat{\lambda}_{\infty,i}(\mathbf{X})$ and the ensemble average $\langle\epsilon^\prime_{ii}\rangle$ both converge to a unique global average $\lambda^i_{\infty}$ over the flow domain. Hence for ergodic steady 3D flows the set of average principal deformations may then be characterised in terms of a single parameter $\lambda\geqslant 0$, where  $\{\lambda^1_{\infty}, \lambda^2_{\infty}, \lambda^3_{\infty}\}$=$\{0,\lambda,-\lambda\}$.

%\subsection{Deformation in zero helicity density 3D flow}

As steady zero helicity density flows (but not necessarily flows with zero total helicity $H$~\cite{Holm:1991aa}) preclude chaotic dynamics~\cite{Sposito:2001aa}, $\lambda_i(t,\mathbf{X})\rightarrow 0$ along all streamlines (irrespective of ergodicity), and the diagonal components $F^\prime_{ii}$ grow sub-exponentially in time, $\lambda=0$. The Protean frame corresponding to the material coordinates $(\phi,\psi,\zeta)$ naturally recovers this constraint of sub-exponential stretching as a direct consequence of the Poincar\'{e}-Bendixson theorem applied to Lamb surfaces of the flow, and the orientation angle $\alpha$ in given explicitly as (\ref{eqn:alpha_zero_helicity}). Furthermore, as shown in Appendix~\ref{App:zero_helicity} the transverse shear and vorticity in these flows is zero ($\epsilon_{23}^\prime=0$, and so fluid deformation evolves longitudinally due to $\epsilon_{12}^\prime$, $\epsilon_{13}^\prime$ in a manner analogous to 2D steady flow. This behaviour builds upon the insights discussed by Sposito~\cite{Sposito:1997aa,Sposito:2001aa} regarding the deformation structure of isotropic Darcy flow, and shows that stochastic models of fluid deformation in such flows can be developed as a simple extension of those~\cite{Dentz:2015aa,Dentz:2015ab} for steady 2D flows.

%\subsection{Deformation in non-zero helicity density 3D flow}

Steady 3D flows with non-zero helicity density exhibit Lagrangian chaos and exponential fluid stretching, and so $\lambda_i(t,\mathbf{X})$ converges to persistent non-zero values along streamlines, $\lambda>0$. Although the ensemble average of the Cartesian velocity gradient $\langle\boldsymbol\epsilon\rangle$ in random steady 3D flows converges toward zero due to statistical stationarity, persistent exponential stretching arises in these flows due to the asymmetry between fluid stretching and compression. Fluid stretching aligns material elements with the stretching direction, accentuating the stretching process, whereas compression aligns elements normal to the compression direction, retarding compression. This basic asymmetry leads to persistent exponential fluid stretching in random steady 3D flows, hence chaotic advection is the norm. This behaviour is not captured by the ensemble mean of the Cartesian velocity gradient $\langle\boldsymbol\epsilon\rangle$, whereas in the Protean frame exponential stretching is directly quantified by the diagonal elements of $\langle\boldsymbol\epsilon^\prime\rangle$. 

%\subsection{Off-diagonal components of Protean velocity gradient tensor}

Similar to the 2D case, in 3D ergodic flows the off-diagonal components of $\boldsymbol\epsilon^\prime(t)$ associated with the 1-coordinate $\epsilon^\prime_{1j}(t)$ have zero mean over long times due to cancellation of shear and streamline curvature over open particle trajectories. Conversely, the transverse off-diagonal component $\epsilon^\prime_{23}(t)$ may have non-zero mean in ergodic (mixing) flows, but in random flows this transverse component must be zero due to stationarity. The mechanism leading to persistent fluid stretching in random zero helicity density flows ($\lambda=0$) is the same as that for ergodic 2D flows described in Section~\ref{sec:2D}. 

%\subsection{Approximation of Protean deformation gradient tensor}

For non-zero helicity density flows, the integrals (\ref{eqn:Fii})-(\ref{eqn:F13}) are significantly simplified by exponential growth and decay respectively of $F_{22}^\prime(t)$, $F_{33}^\prime(t)$. As the integrand in (\ref{eqn:F23}) decays as $\exp(-2\lambda t)$, then these converge to
\begin{align}
&F^\prime_{23}(t)\stackrel{t}\longrightarrow K_1 F^\prime_{22}(t),\label{eqn:F23F22}\\
&F^\prime_{13}(t)\stackrel{t}\longrightarrow K_1 F^\prime_{12}(t),\label{eqn:F13F12}
\end{align}
where the constant $K_1$ given by the integral in (\ref{eqn:F23}) is approximated as
\begin{equation}
K_1\approx \frac{1}{2\lambda}\langle\epsilon_{23}\rangle.
\end{equation}
As $\epsilon_{12}^\prime(t)$ is bounded with zero mean, $F_{12}^\prime(t)$ may be expressed as
\begin{align}
F^\prime_{12}(t)=\tilde{m}(t) F^\prime_{22}(t)\approx K_2 F^\prime_{22}(t),\label{eqn:F12F22}
\end{align}
where $\tilde{m}(t)$ is an oscillatory function with non-zero mean $K_2$. Note that these relationships hold both over a single trajectory or over an ensemble of trajectories in terms of the averages $\langle F_{ij}^\prime\rangle$, $\langle K_l\rangle$ if the relevant quantities are uncorrelated. For such chaotic flows $\epsilon_{12}^\prime$ is the dominant off-diagonal component of $\boldsymbol\epsilon^\prime(t)$ with respect to evolution of $\mathbf{F}^\prime(t)$, where the full history of fluid shear and streamline curvature along a particle trajectory dictates $\tilde{m}(t)$, $K_2$ (which may be large in magnitude). In contrast $K_1$ is typically small and $\epsilon_{23}^\prime$ is only relevant at short times, and $\epsilon_{13}^\prime$ plays a negligible role in such flows. %From these results, the deformation tensor may be approximated as
%\begin{equation}
%\mathbf{F}^\prime(t)\approx
%\left(
 %  \begin{array}{ccc}
  %   v(t)/v(0) & F_{12}^\prime(t) & K_1 F_{12}^\prime(t) \\
   %  0 & F_{22}^\prime(t) & K_1F_{22}^\prime(t) \\
    % 0 & 0 & 1/F_{22}^\prime(t) \\
   %\end{array}
 %\right),
%\end{equation}

%\subsection{Approximation of Lyapunov exponents}

\begin{figure}
\begin{centering}
\begin{tabular}{cc}
\includegraphics[width=0.47\columnwidth]{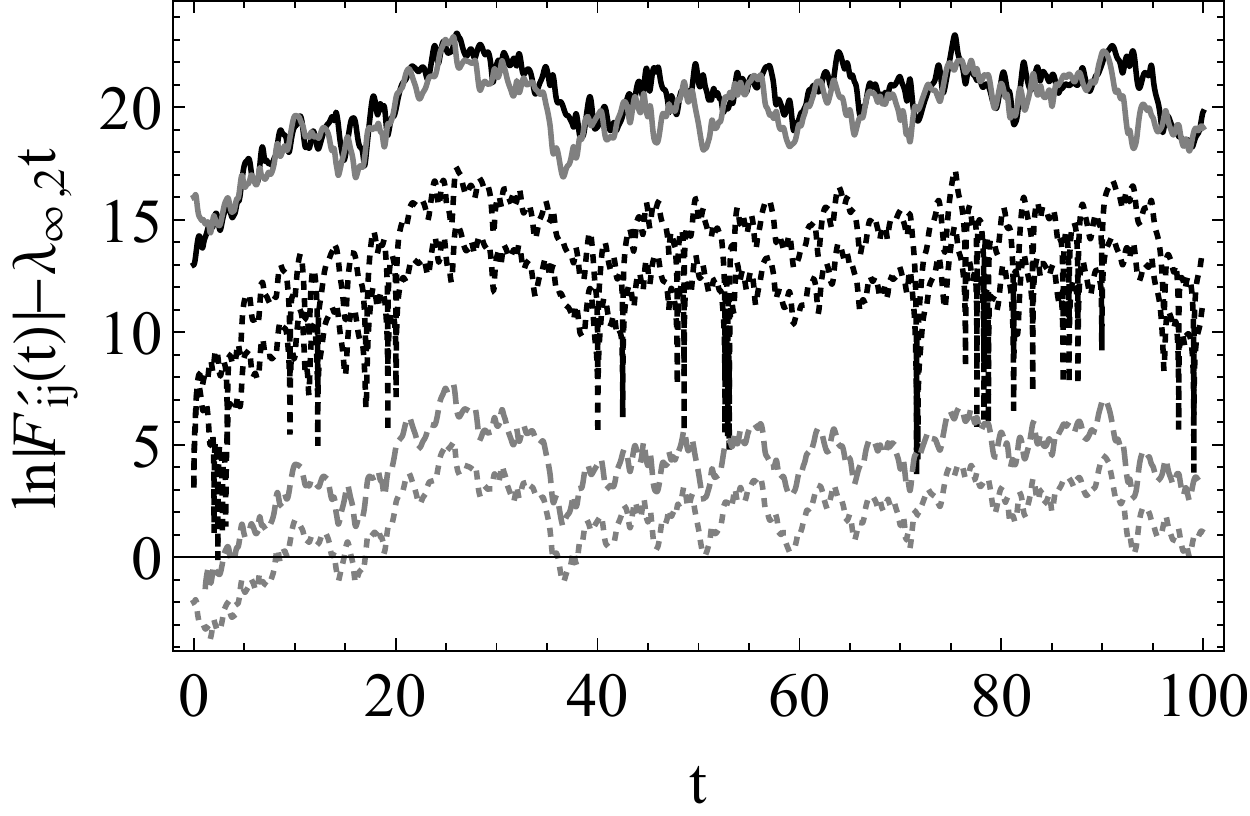}&
\includegraphics[width=0.48\columnwidth]{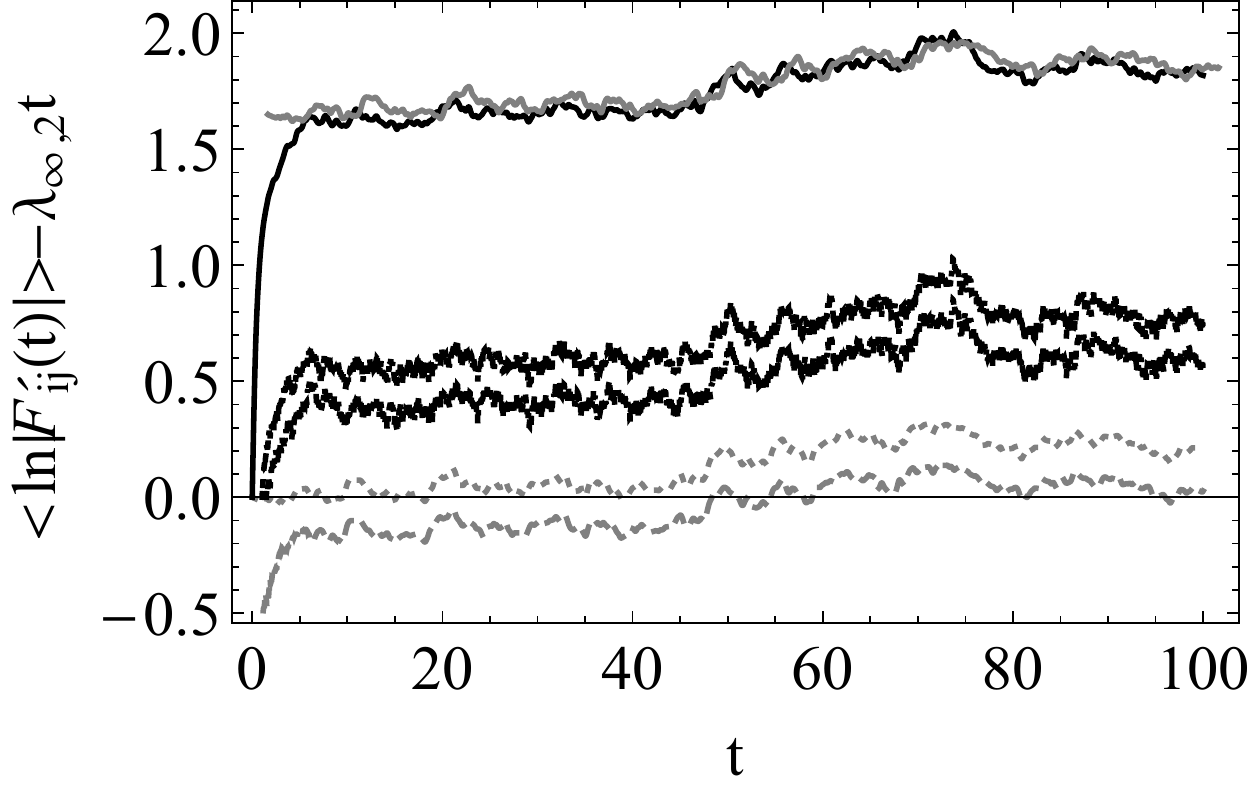}\\
(a) & (b)
\end{tabular}
\end{centering}
\caption{Comparison of the logarithm of the magnitude of the components of $\mathbf{F}^\prime(t)$ and FTLE $\mu(t)$ over (a) a single trajectory and (b) and the mean over an ensemble of 1,000 trajectories within the 3D Kraichnan flow (\ref{eqn:Kraichnanfield}). For clarity of comparison all plots are shown with the principal stretching rate $\lambda_{\infty,2} t$ subtracted. In both plots $\ln|F_{22}^\prime(t)|$ is shown by the grey dotted line, $\ln|F_{23}^\prime(t)|$ by the gray dashed line, $\ln|F_{12}^\prime(t)|$ by the black dotted line, and $\ln|F_{13}^\prime(t)|$ by the black dashed line. The offsets between these curves reflect the ratios $K_1$, $K_2$ between these components given by (\ref{eqn:F23F22})-(\ref{eqn:F12F22}). The solid black line shows the FTLE $\mu(t)$ over these trajectories compared with the estimate (solid grey) given by (\ref{eqn:FTLEapprox}).}
\label{fig:FTLEapprox}
\end{figure}

An important measure with respect to fluid mixing, deformation and the formation of Lagrangian coherent structures (LCSs) is the finite-time Lyapunov exponent (FTLE) $\mu(t,\mathbf{X})$, which measures the exponential rate of fluid stretching as
\begin{equation}
\mu(t,\mathbf{X}):=\frac{1}{2t}\ln[\nu_1(t,\mathbf{X})],\label{eqn:Lyapunov}
\end{equation}
where $\nu_j(t,\mathbf{X})$ is the $j$-th largest eigenvalue of the Cauchy-Green tensor $\mathbf{C}(t,\mathbf{X})$ along streamline with Lagrangian coordinate $\mathbf{X}$. For ergodic flows the corresponding infinite-time Lyapunov exponent
\begin{equation}
\mu_\infty:=\lim_{t\rightarrow\infty}\mu(t,\mathbf{X}),
\end{equation}
also converges to the ensemble average of $\langle\nu_1\rangle$ due to ergodicity. In the presence of exponential fluid stretching, $\lambda>0$, in the asymptotic limit all the components of $\mathbf{F}^\prime(t)$ grow no faster than $\exp(\lambda t)$, and so the Lyapunov exponent $\mu_\infty$ is given the ensemble average of $\epsilon^\prime_{22}$
\begin{equation}
\mu_\infty=\lambda=\langle\epsilon^\prime_{22}\rangle,\label{eqn:ITLE}
\end{equation}
providing a means to compute Lagrangian deformation from the Eulerian velocity gradient. From (\ref{eqn:ITLE}), the principal stretching rate converges toward the FTLE along a trajectory as
\begin{equation}
\lambda_2(t,\mathbf{X})\stackrel{t}\longrightarrow\mu(t,\mathbf{X}).\label{eqn:mu_lambda_converge}
\end{equation}
A more accurate approximation for the FTLE is provided by consideration of the full deformation tensor $\mathbf{F}^\prime$. As the sum of the singular values of $\mathbf{F}^\prime(t)$ (given by the eigenvalues $\nu_j$ of $\mathbf{C}$)  is equal to the Frobenius norm of $\mathbf{F}^\prime(t)$
\begin{equation}
||\mathbf{F}^\prime||_F:=\sqrt{\sum_{i,j}F_{ij}^{\prime 2}}=\sqrt{\text{tr}(\mathbf{C})}=\sqrt{\sum_j \nu_j(t)},\label{eqn:frobnorm}
\end{equation}
where for volume-preserving flows  $\prod_j \nu_j (t,\mathbf{X})=\prod_j \lambda_j(t,\mathbf{X})=1$, then for large deformations the FTLE is well-approximated as
\begin{equation}
\mu(t,\mathbf{X})\approx\frac{1}{2t}\ln||\mathbf{F}^\prime||_F.
\end{equation}
Hence the FTLE is well approximated by the dominant terms in the Frobenius norm $||\mathbf{F}^\prime||_F$ over a single trajectory (labelled with Lagrangian coordinate $\mathbf{X}$) or ensemble of trajectories respectively as 
\begin{equation}
\begin{split}
\mu(t,\mathbf{X})\approx &\frac{1}{t}\ln |F_{22}^{\prime}(t,\mathbf{X})|\\
+&\frac{1}{t}\ln\sqrt{1+K_1^2(\mathbf{X})}+\frac{1}{t}\ln\sqrt{1+K_2^2(\mathbf{X})},\label{eqn:FTLEapprox}
\end{split}
\end{equation}
\begin{equation}
\begin{split}
\langle\mu(t)\rangle\approx &\frac{1}{t}\langle\ln |F_{22}^{\prime}(t)|\rangle\\
+&\frac{1}{t}\langle\ln\sqrt{1+K_1^2}\rangle+\frac{1}{t}\langle\ln\sqrt{1+K_2^2}\rangle.
\label{eqn:FTLEapprox_ensemble}
\end{split}
\end{equation}
The accuracy of  this approximation over both a single trajectory and an ensemble of 1,000 realisations of the Kraichnan flow (\ref{eqn:Kraichnanfield}) described in $\S$\ref{sec:examples} is illustrated in Figure~\ref{fig:FTLEapprox}. These results indicate that for all but short times (\ref{eqn:FTLEapprox}), (\ref{eqn:FTLEapprox_ensemble}) accurately capture the finite-time stretching dynamics in the presence of significant exponential stretching. The offsets between the deformation components $\langle \ln|F_{ij}^\prime(t)|\rangle$ in Figure~\ref{fig:FTLEapprox} are given by ensemble averages $\langle\sqrt{1+K_1^2}\rangle$, $\langle\sqrt{1+K_2^2}\rangle$, and the dominant FTLE is well approximated by (\ref{eqn:FTLEapprox}), (\ref{eqn:FTLEapprox_ensemble}). As this approximation also holds over each trajectory, for ergodic flows (\ref{eqn:FTLEapprox}) also facilitates estimation of both the FTLE $\mu(t,\mathbf{X})$ and deformation tensor $\mathbf{F}^\prime(t)$ probability distribution function (PDF) given the statistics for $F_{22}(t)$, $K_1$, $K_2$.
 
 \subsection{Stochastic Modelling of Lagrangian Deformation}
 
It has been observed~\cite{Le-Borgne:2008aa,Le-Borgne:2008ab,Anna:2013aa} that the Lagrangian velocities in several random 2D flows follow a spatial Markov process, facilitating quantification of transport in these flows as a continuous time random walk (CTRW) model. This remarkable property appears to also extend to the deformation structure of random 2D flows, where Dentz et al~\cite{Dentz:2015aa} have recently developed a CTRW model in the Protean frame to predict evolution of the PDF of $\mathbf{F}^\prime(t)$ as a L\'{e}vy process. This approach differs from other stochastic models~\cite{Thalabard:2014aa, Villermaux:2003ab,Gonzalez:2009aa} for fluid deformation in that the Protean transform directly provides the kernels for the deformation CTRW from the velocity gradient along streamlines, and automatically recovers topological and kinematic constraints inherent to the flow. This approach clearly elucidates the link between flow structure and deformation evolution and facilitate the identification of flow features which govern deformation. As such, the Protean transform provides a means to quantify the full deformation structure of complex ergodic steady 3D flows. As these flows exhibit Gaussian  in terms of a small num

Extension of this CTRW deformation model to 3D steady flows facilitates prediction of the PDFs of Lagrangian deformation and the FTLE from the underlying Eulerian flow structure. We anticipate that this CTRW framework is applicable to a range of steady ergodic 3D flows, spanning non-zero helicity density flows such as chaotic flow fields, and zero helicity density flows such as heterogeneous Darcy flow. Such a development opens the door to prediction of fluid deformation from statistical flow features or the underlying material properties for flow in disordered media. Whilst establishment of spatial Markovianity and development of stochastic models is beyond the scope of this study, the simple deformation structure observed for the random flows in $\S$\ref{sec:examples} point to the feasibility of predict of the deformation PDF using such methods.

\section{Conclusions}\label{sec:conclusions}
We present an method to calculate fluid deformation in steady 3D flows using a Protean (streamline) coordinate system which renders the velocity gradient tensor $\boldsymbol\epsilon^\prime(t)$ upper triangular, facilitating explicit solution of the deformation gradient tensor $\mathbf{F}^\prime(t)$ in terms of definite integrals of the components of $\boldsymbol\epsilon^\prime(t)$. Reorientation into Protean coordinates greatly simplifies computation of material deformation along particle trajectories, and naturally recovers physical, topological and kinematic constraints associated with the underlying flow field. These constraints such as the Poincar\'{e}-Bendixson theorem directly impact the dynamics of 3D fluid deformation as reflected by evolution of the deformation gradient tensor. Whilst this method is analogous to the continuous QR method, the QR method does not necessarily align with streamline coordinates. Similarly streamlines coordinate do not necessarily render $\boldsymbol\epsilon^\prime$ upper triangular. Explicit reorientation in the Protean frame renders strict adherence to the above constraints, and moreover elucidates the underlying deformation structure and providing useful approximations for the finite- and infinite-time Lyapunov exponents, thus linking Lagrangian deformation with the Eulerian flow structure.

Although several models of fluid mixing require the mean (scalar) fluid deformation rate as an input, in general the full deformation tensor $\mathbf{F}(t)$ is required as this measure completely quantifies infinitesimal fluid deformation to first order at time $t$. For example, the evolution of a continuously injected dye plume from a point or line source evolves in a different manner to that of a temporal pulse or indeed the area of a dye plume illuminated by a laser sheet transverse to the mean flow direction. Depending upon the situation at hand, these structures are governed by either the full deformation tensor $\mathbf{F}(t)$ or a subset of this measure. As such, quantification of the evolution of the deformation tensor can be applied to all of these cases, whereas scalar deformation measures are case-specific.

With respect to modelling of viscoelastic flows, solution of materials stress from strain history using a memory integral constitutive model is simplified in the Protean frame as the evolution of the fluid deformation gradient tensor (and hence associated Cauchy-Green $\mathbf{C}$, Finger $\mathbf{F}$ or Hencky $\mathbf{H}$ tensors) is given in closed form as integrals of the rate of strain tensor. This simplification also extends to solution of memory integral constitutive models for determining stress evolution in non-Newtonian (viscoelastic) materials, where the convolution of the entire strain history with a viscoelastic kernel is also significantly simplified.

This approach uncovers the deformation structure of a wide class of steady 3D flows, such that the link between compressibility, helicity density and stationarity and the components of $\boldsymbol\epsilon^\prime(t)$ are clearly elucidated. For zero helicity flows the Protean frame aligns with an orthogonal material coordinate system which recovers the inherent constraints of sub-exponential fluid deformation and zero transverse shear and vorticity. The resultant deformation structure is particularly simple, and may be considered as a generalisation of  deformation in 2D shear flows.

For ergodic systems including random stationary flows or deterministic mixing flows, this approach points toward concrete methods to predict Lagrangian fluid deformation directly from the Eulerian flow structure whilst directly observing the constraints upon deformation evolution imposed by the underlying flow properties. This approach involves extension of a CTRW framework for 2D fluid deformation~\cite{Dentz:2015aa} to 3D flows, where the Eulerian PDF of $\boldsymbol\epsilon^\prime(t)$ and correlation between components serve as inputs to predict evolution of the PDF of the deformation tensor $\mathbf{F}^\prime(t)$. This approach is particularly useful for developing stochastic models of deformation evolution in a wide class of ergodic flows including deterministic chaotic flows, random turbulent flows, or flow in disordered media, and in the latter case this approach opens the door to linking medium properties and statistical controls to deformation evolution.

We apply the Protean transformation method to several ergodic steady 3D flows, including model chaotic flow (ABC flow), isotropic homogeneous turbulence (Kraichnan model) and random potential flow,  and demonstrate that the basic flow properties such as compressibility, helicity density and stationarity constrain the Protean rate of strain tensor as predicted and uncovers the underlying deformation structure of complex 3D flow fields, including prediction of the infinite-time Lyapunov exponent from finite-time data. The deformation structure of the random flows is remarkably simple (Gaussian) and completely uncorrelated apart from the diagonal components of $\boldsymbol\epsilon^\prime(t)$ which are correlated due to conservation of mass. These results provide the building blocks for the development of stochastic models of Lagrangian fluid deformation which observe relevant flow physics are relevant to fluid phenomena ranging from fluid mixing and transport through to particle dispersion and alignment. 

\appendix

\section{Analogy to Continuous QR Decomposition\label{App:QR}}

As the Protean coordinate frame renders the transformed velocity gradient
tensor $\boldsymbol\epsilon^\prime(t)$ upper triangular, this method is directly analogous to the continuous QR decomposition method for a $d$-dimensional autonomous nonlinear dynamical system
\begin{align}
\frac{d\mathbf{x}}{dt}=\mathbf{f}(\mathbf{x}),\label{eqn:fx}
\end{align}
which may be considered as the generalisation of the advection equation $\dot{\mathbf{x}}=\mathbf{v}(\mathbf{x})$ for a steady flow field. For a given solution trajectory $\mathbf{x}(t)$, the Lyapunov exponents of (\ref{eqn:fx}) are given by the eigenvalues of the fundamental solutions $\mathcal{Y}(t)$ of the linear variational equation
\begin{equation}
\frac{d \mathcal{Y}}{dt}=\mathcal{A}(t)\cdot\mathcal{Y}(t),\,\,Y(0)=\mathbf{I},
\end{equation}
which is the generalisation of (\ref{eqn:deform}), where $\mathcal{A}(t)=\partial\mathbf{f}/\partial\mathbf{x}$ is the Jacobian along the trajectory $\mathbf{x}(t)$. The continuous QR method considers the decomposition
\begin{equation}
\mathcal{Y}(t)=\mathcal{Q}(t)\cdot\mathcal{R}(t),
\end{equation}
where $\mathcal{Q}(t)$ is orthogonal and $\mathcal{R}(t)$ upper triangular, and satisfy the auxiliary equations
\begin{align}
&\frac{d\mathcal{R}}{dt}=\mathcal{A}^\prime\cdot\mathcal{R}(t),\,\,\mathcal{R}(0)=\mathbf{I},\label{eqn:QRauxR}\\
&\frac{d\mathcal{Q}}{dt}=\mathcal{Q}(t)\cdot\mathcal{H}(t),\,\,\mathcal{Q}(0)=\mathbf{I},\label{eqn:QRaux}
\end{align}
where, similar to (\ref{eqn:Ftransform})-(\ref{eqn:epsilon}),
\begin{align}
\mathcal{A}^\prime(t):=\mathcal{Q}^T(t)\cdot\mathcal{A}(t)\cdot\mathcal{Q}(t)-\mathcal{Q}^T(t)\cdot\dot{\mathcal{Q}},\label{eqn:Aaux}
\end{align}
\begin{equation}
\mathcal{H}(t):=\mathcal{Q}^T(t)\cdot\dot{\mathcal{Q}}=
\begin{cases}
[\mathcal{Q}^T(t)\cdot\mathcal{A}(t)\cdot\mathcal{Q}(t)]_{ij},&\quad i>j,\\
0,&\quad i=j,\\
-[\mathcal{Q}^T(t)\cdot\mathcal{A}(t)\cdot\mathcal{Q}(t)]_{ji},&\quad i<j.
\end{cases}
\end{equation}
Hence the evolution equations for the orthogonal and upper triangular matrices $\mathcal{Q}(t)$, $\mathcal{A}^\prime(t)$, $\mathcal{R}(t)$ for the continuous QR method are directly analogous to the reorientation operator $\mathbf{Q}(t)$, Protean rate of strain $\boldsymbol\epsilon^\prime(t)$ and Protean deformation gradient $\mathbf{F}^\prime(t)$ tensors for the Protean transform method. However, the actual values differ in that the initial condition for the QR method corresponds to the unrotated frame ($\mathcal{Q}(0)=\mathbf{I}$), whereas the Protean frame always aligns with the flow direction as per (\ref{eqn:Qrotn}). Due to the temporal derivative in the reoriented Jacobian (\ref{eqn:Aaux}) for the QR decomposition method, solutions to $\mathcal{Q}(t)$ which render $\mathcal{A}^\prime(t)$ upper triangular are not unique.

Whilst $\mathcal{Q}(t)$ asymptotically converges to the Protean coordinate frame (due to the dissipative nature of (\ref{eqn:QRaux})), for finite times the continuous QR method does not align with the streamlines of the flow and hence inferences regarding topological and kinematic constraints upon the dynamics do not universally hold. Moreover, for 2D steady flows (and analogous dynamical systems), the Protean transform provides a simple closed solution (\ref{eqn:Qrotn}) for $\mathbf{Q}(t)$, hence it is not necessary to explicitly solve the ODE system (\ref{eqn:QRaux}) or employ unitary integration routines~\cite{Dieci:1997aa} to preserve orthogonality of $\mathcal{Q}(t)$.

\section{Velocity Gradient Tensor in Zero Helicity Density Flow\label{App:zero_helicity}}

As shown in \cite{Finnigan:1983aa,Sposito:2001aa}, all 3D zero helicity density flows may be posed in the form of an isotropic Darcy flow
\begin{equation}
\mathbf{v}(\mathbf{x})=-k(\mathbf{x})\nabla\phi=\nabla\psi\times\nabla\zeta,\label{eqn:Darcycross}
\end{equation}
where $\phi$ is the flow potential, $\psi$ the streamfunction and $\zeta$ is a scalar, with $\nabla\psi\cdot\nabla\zeta=0$. The streamsurfaces of constant $\psi$ are the Lamb surfaces of the flow, material surfaces which are spanned by both the velocity and vorticity vectors. Such zero helicity density flows admit an orthogonal material coordinate system which aligns with the stream lines, vorticity lines and Lamb vector field lines of the flow. In this frame the velocity gradient tensor $\boldsymbol\epsilon^\prime$ takes on a particularly simple form, and so for zero helicity density flows we define this material frame as the Protean coordinate system. The covariant base vectors $\hat{\mathbf{e}}_i$ of this material coordinate system are then given by the normalised velocity, vorticity and Lamb vectors respectively as
\begin{align}
\hat{\mathbf{e}}_1=\frac{\mathbf{v}}{|\mathbf{v}|},\quad
\hat{\mathbf{e}}_2=\frac{\boldsymbol\omega}{|\boldsymbol\omega|},\quad
\hat{\mathbf{e}}_3=\frac{\mathbf{v}\times\boldsymbol\omega}{|\mathbf{v}\times\boldsymbol\omega|}.
\end{align}
Under this formulation the isopotential, stream and Lamb surfaces (respectively $\phi$=const., $\psi=$const., $\zeta=$const.) are all orthogonal and are
normal to $\mathbf{v}$, $\boldsymbol\omega$, $\mathbf{v}\times\boldsymbol\omega$ respectively. The orthogonal coordinates of the material coordinate system are then
\begin{align}
\xi^1&=\phi=\phi(x_1^\prime),\label{eqn:xi1}\\
\xi^2&=\psi=\psi(x_2^\prime),\label{eqn:xi2}\\
\xi^3&=\zeta=\zeta(x_3^\prime),\label{eqn:xi3}
\end{align}
where $x_i^\prime$ denote the distance along the coordinate $\xi^i$. The differential arc length $ds$ then satisfies $ds^2=g_{\alpha\beta}d\xi^\alpha d\xi^\beta=dx^\prime_\alpha dx^\prime_\beta$, and the metric tensor for the orthogonal coordinate system is then
\begin{equation}
g_{\alpha\beta}=
\left(
   \begin{array}{ccc}
     h_1^2 & 0 & 0 \\
     0 & h_2^2 & 0 \\
     0 & 0 & h_3^2 \\
   \end{array}
 \right).
 \end{equation}
Note that as the components of the covariant $g_{\alpha\beta}$ and contravariant $g^{\alpha\beta}$ metric tensors transform as
\begin{align}
&g^\prime_{ij}(\mathbf{x}^\prime)=\frac{\partial x^k}{\partial x^{\prime i}}\frac{\partial x^l}{\partial x^{\prime j}}g^\prime_{kl}(\mathbf{x}),\\
&g^{\prime ij}(\mathbf{x}^\prime)=\frac{\partial x^{\prime i}}{\partial x^k}\frac{\partial x^{\prime j}}{\partial x^l}g^{\prime kl}(\mathbf{x}),
\end{align}
and as $g_{\alpha\beta}^{-1}=g^{\alpha\beta}$, then the scale factors $h_i$ are then
\begin{align}
h_i=\sqrt{\left(\frac{\partial x_1}{\partial\xi^i}\right)^2+\left(\frac{\partial x_2}{\partial\xi^i}\right)^2+\left(\frac{\partial x_3}{\partial\xi^i}\right)^2}=\frac{1}{|\nabla\xi^i|}=\frac{\partial x_i^\prime}{\partial\xi^i}.
\end{align}
From the isotropic Darcy equation (\ref{eqn:Darcycross}) and coordinate definitions (\ref{eqn:xi1})-(\ref{eqn:xi3}) these scale factors are explicitly
\begin{align}
h_1=\frac{\partial x_1^\prime}{\partial\phi}=\frac{k}{v},\\
h_2=\frac{\partial x_2^\prime}{\partial\psi}=\frac{\rho}{v},\\
h_3=\frac{\partial x_3^\prime}{\partial\zeta}=\frac{1}{\rho},
\end{align}
where the local density of Lamb surfaces is defined as
\begin{equation}
\rho:=\frac{\partial\zeta}{\partial x_3^\prime}.
\end{equation}
Following Batchelor (1967), the components of the velocity gradient tensor $\boldsymbol\epsilon^\prime(t)$ are then
\begin{align}
&\epsilon_{ii}^\prime(t)=\frac{1}{h_i}\frac{\partial v_i}{\partial \xi^j}+\sum_{j}\frac{v_i}{h_i h_j}\frac{\partial h_i}{\partial\xi^j},\\
&\epsilon_{ij}^\prime(t)=\frac{h_i}{h_j}\frac{\partial}{\partial \xi^j}\left(\frac{v_i}{h_i}\right),
\end{align}
hence the velocity gradient is upper triangular, and the $(2,3)$ components fully decouple as
\begin{equation}
\boldsymbol\epsilon^\prime(t)=
\left(
   \begin{array}{ccc}
     \epsilon_{11}^\prime & \epsilon_{12}^\prime & \epsilon_{13}^\prime \\
     0 & \epsilon_{22}^\prime & 0 \\
     0 & 0 & \epsilon_{33}^\prime \\
   \end{array}
 \right),\label{eqn:vgrad_zero_helicity}
\end{equation}
where
\begin{align}
&\epsilon_{ii}^\prime=\frac{\partial v_i}{\partial x_i^\prime},\\
&\epsilon_{ij}^\prime=2\dot\gamma_1-\omega_i,
\end{align}
and the longitudinal shear rate $\dot\gamma_i$ and vorticity $\omega_i$ are respectively 
\begin{align}
&\dot\gamma_i=\frac{\partial v}{\partial x_i^\prime},\\
&\omega_i=v\frac{\partial \ln k}{\partial x_i^\prime}.
\end{align}
Hence fluid deformation in 3D steady flows with zero helicity density evolve in a similar manner to 2D steady flow due to longitudinal shear and vorticity within Lamb surfaces and stream surfaces. As these surfaces are material, there is no transverse shear in the $(2,3)$ directions, leading to decoupling as per (\ref{eqn:vgrad_zero_helicity}).
%
%\begin{align}
%&\epsilon^\prime_{11}=\frac{\partial v_1}{\partial x_1^\prime},\\
%&\epsilon^\prime_{22}=\frac{\partial v_2}{\partial x_2^\prime},\\
%&\epsilon^\prime_{33}=\frac{\partial v_3}{\partial x_3^\prime},\\
%&\epsilon^\prime_{12}=2\frac{\partial v}{\partial x_2^\prime}-v\frac{\partial \ln k}{\partial x_2^\prime},\\
%&\epsilon^\prime_{13}=2\frac{\partial v}{\partial x_3^\prime}-v\frac{\partial \ln k}{\partial x_3^\prime},\\
%\epsilon^\prime_{23}=\epsilon^\prime_{32}=\epsilon^\prime_{21}=\epsilon^\prime_{31}=0,
%\end{align}

%\bibliographystyle{plain}
%\bibliography{database.ref}

\end{document}